\documentclass[10pt]{article}
\usepackage[lmargin=1in,rmargin=1in,tmargin=1in,bmargin=1in]{geometry}
\usepackage[english]{babel} 
\usepackage[utf8]{inputenc}
\usepackage{lmodern}
\usepackage{xargs} 

\usepackage{parskip}
\usepackage[dvipsnames]{xcolor}

\usepackage{hyperref}
\hypersetup{
    colorlinks,
    urlcolor=black,
    linkcolor=black,
    citecolor=black
}

\usepackage[shortlabels]{enumitem} 
\setlist[enumerate]{itemsep=0.5pt plus 1.0 pt minus 0.5pt, topsep=0.1pt plus 1.0 pt minus 0.5pt} 
\setlist[itemize]{itemsep=0.5pt plus 0.5 pt minus 0.5pt, topsep=0.1pt plus 0.0 pt minus 0.0pt}

\usepackage{float} 
\usepackage{booktabs} 
\usepackage{xtab} 

\usepackage{graphicx} 
\usepackage{subcaption}

\usepackage{amsfonts}
\usepackage{amsmath}
\usepackage{amssymb}
\usepackage{amsthm}
\usepackage{array}
\usepackage{bigints} 
\usepackage{commath}
\usepackage[mathscr]{eucal}
\usepackage{latexsym, marvosym} 
\usepackage{mathrsfs}
\usepackage{mathtools}
\mathtoolsset{showonlyrefs=true} 
\usepackage{nicematrix}
\usepackage{slashed}
\usepackage{upgreek} 
\usepackage{harmony}

\newtheoremstyle{customthm}
  {5pt} 
  {5pt} 
  {\itshape} 
  {} 
  {\bfseries} 
  {.} 
  {.5em} 
  {} 

\newtheoremstyle{customdef}
  {5pt} 
  {5pt} 
  {\upshape} 
  {} 
  {\bfseries} 
  {.} 
  {.5em} 
  {} 

\newtheoremstyle{customrmk}
  {5pt} 
  {5pt} 
  {\upshape} 
  {} 
  {\slshape} 
  {.} 
  {.5em} 
  {} 

\numberwithin{equation}{section}

\theoremstyle{customdef}
\newtheorem{definition}{Definition}[section]

\theoremstyle{customrmk}
\newtheorem{remark}[definition]{Remark}

\theoremstyle{customthm}
\newtheorem{proposition}[definition]{Proposition}
\newtheorem{lemma}[definition]{Lemma}
\newtheorem{theorem}[definition]{Theorem}
\newtheorem*{conjecture}{Conjecture}
\newtheorem*{extension}{Generalised $C^k$ extension principle with $\Lambda > 0$}
\newtheorem{corollary}[definition]{Corollary}

\newtheorem{innercustomthm}{Theorem}
\newenvironment{customthm}[1]
  {\renewcommand\theinnercustomthm{#1}\innercustomthm}
  {\endinnercustomthm}

\usepackage{soul}
\usepackage{contour}
\usepackage[normalem]{ulem}

\contourlength{0.8pt}

\newcommand{\myuline}[1]{%
  \uline{\phantom{#1}}%
  \llap{\contour{white}{#1}}%
}

\usepackage{csquotes}
\usepackage[backend=bibtex, style=alphabetic, giveninits=true,
            doi=false, isbn=false, url=false, eprint=true, 
            useprefix=true,
            sorting=nyt, maxbibnames=99, maxalphanames=5, date=year]{biblatex}
\renewbibmacro{in:}{}
\title{Gravitational collapse to extremal black holes with $\Lambda>0$: counterexamples to the third law without charge}

\usepackage[affil-it]{authblk}

\author{Mónica Tapia del Moral\thanks{mt891@cam.ac.uk} \vspace{0.5em}}
\affil{\small University of Cambridge, Department of Applied Mathematics and Theoretical Physics, Wilberforce~Road,~Cambridge~CB3~0WA,~United Kingdom}

\usepackage{texdate}
\initcurrdate
\date{25 November 2026 \vspace{1em}}

\begin{document}

\maketitle

\begin{abstract}
    We study the dynamical formation of the extremal Schwarzschild--de Sitter solution, in both its time orientations, and the related Nariai solution. We prove that there exist spherically symmetric, regular Cauchy data for the Einstein-scalar field system with $\mathbb S^3$ topology, for any scalar field mass $\mathfrak{m}\geq 0$, that lead to the formation of an exactly extremal Schwarzschild--de Sitter black hole in finite time, and that moreover the apparent horizon of the solution is exactly isometric to that of the subextremal Schwarzschild--de Sitter solution at earlier times. We also show that if the scalar field mass satisfies $\mathfrak{m}>\sqrt{\Lambda}/2$, an analogous result holds having the extremal Schwarzschild--de Sitter white hole or the Nariai solution in place of the black hole. In the black hole case, the exactly extremal Schwarzschild--de Sitter region is bounded to the future by two event horizons and can be seen as the analogue of the exterior region in the asymptotically flat case. We moreover establish that these two black hole horizons that bound the exactly extremal Schwarzschild--de Sitter region are the only event horizons present in the spacetime, something non-trivial given the existence of multi-black hole configurations when $\Lambda > 0$. Consequently, the solutions arising from these data can be thought to constitute, in particular, regular counterexamples to the third law of black hole thermodynamics with $\Lambda>0$ arising from the collapse of physically plausible matter which is moreover uncharged.  Our approach to construct the Cauchy data relies on $C^k$ characteristic gluing results for the Einstein-scalar field system in spherical symmetry, in the spirit of \cite{Kehle_Unger_2025}. 
\end{abstract}

\newpage

\tableofcontents

\newpage

\section{Introduction}
\label{sec:introduction}

The construction of solutions to the Einstein equations in which gravitational collapse leads to black hole formation has long been a central problem in general relativity. In the asymptotically flat setting, it is well known that subextremal black holes can emerge as the final state of a gravitational collapse process for different matter models \cite{Lemaitre_1933,Oppenheimer_Snyder_1939} and even in vacuum \cite{Christodoulou_2009_blackholes,Li_Yu_2015,Kehle_Unger_2024}. Extremal black holes, on the other hand, have traditionally been regarded as exotic objects of theoretical interest but little physical relevance, with their dynamical formation in astrophysically plausible scenarios deemed unfeasible. Much of this widely held belief probably stems from the so-called \textit{third law of black hole thermodynamics}, originally postulated by Bardeen, Carter and Hawking \cite{Bardeen_Carter_Hawking_1973}, and later made precise by Israel \cite{Israel_1986, Israel_1992} in the form:

\vspace{0.0em}

\begin{center}
    \textit{``A subextremal black hole cannot become extremal in finite time by any continuous process,}\\ \textit{no matter how idealised, in which the spacetime and matter fields remain regular}\\  \textit{and obey the weak energy condition.''}
\end{center}

\vspace{0.5em}

The recent work of Kehle and Unger \cite{Kehle_Unger_2025} demonstrates that the above statement is false: there do exist asymptotically flat, regular initial data for the Einstein--Maxwell-charged scalar field system that undergo gravitational collapse to form first a subextremal Reissner--Nordström apparent horizon, and subsequently an exactly extremal Reissner--Nordström event horizon at a later advanced time. Kehle and Unger's work has helped shift the perception of extremal black holes as inaccessible through physically reasonable and sufficiently regular gravitational collapse processes, triggering a series of related works \cite{Marin_2024,Reall_2024,McSharry_Reall_2025, Gadioux_Reall_Santos_2026,Crump_Gadioux_Reall_Santos_2026,Dunajski_Szybka_2026,Schneider_2026, Mittal_Gundlach_Martel_2026, Lee_2026, McSharry_Reall_Santos_2026,Turiaci_Wu_2026, Hod_Piran_2026aa,Czimek_2026}. 

\vspace{0.1em}

The results in \cite{Kehle_Unger_2025} are proven for a charged massless scalar field, and by \cite[Remark 1.2]{Kehle_Unger_2025}, the proof extends to a charged massive scalar field with mass $\mathfrak{m}\in\mathbb R_{>0}$ and charge $\mathfrak{e}\in \mathbb R\backslash\{0\}$ satisfying $\frac{|\mathfrak{e}|}{\mathfrak{m}}\gg 1$. A natural question to ask then is whether third-law-violating solutions can be ruled out by requiring matter to satisfy a local mass-charge inequality, excluding ``supercharged'' matter in particular. 

\vspace{0.1em}

Indeed, this has been shown for supersymmetric black holes \cite{Reall_2024}: if the matter satisfies a local mass-charge inequality, which for a charged scalar field requires $\frac{|\mathfrak{e}|}{\mathfrak{m}}\hspace{-0.1em}\leq 1$, supersymmetric black holes cannot form via gravitational collapse in finite time. This specifically rules out the possibility of forming extremal Reissner--Nordström black holes by gravitational collapse of a non-supercharged scalar field in finite time. The same result, that is, the impossibility of forming supersymmetric black holes dynamically in finite time with non-supercharged matter, remains true when a negative cosmological constant is introduced \cite{McSharry_Reall_2025}.

\vspace{0.1em}

The case involving a positive cosmological constant $\Lambda>0$, on the other hand, has remained largely unexplored in the context of extremal black hole formation, despite representing the most natural scenario for the study of extremality \myuline{without charge or angular momentum}, due to the fundamental fact that there exist spherically symmetric, uncharged, extremal solutions to the vacuum Einstein equations with $\Lambda>0$. These are the \textit{extremal Schwarzschild--de Sitter spacetime}, which can be endowed with two different time orientations, and the constant area-radius solution known as the \textit{Nariai spacetime}.

\vspace{0.1em}

The aim of this work is to explore rigorously the dynamical formation in spherical symmetry of all these three extremal solutions by gravitational collapse in finite time, thus producing \textbf{counterexamples to the third law with $\Lambda>0$ without charge}.

\vspace{0.1em}

In order to have non-trivial dynamics in spherical symmetry, given the rigidity imposed by the analogue of Birkhoff's theorem with $\Lambda>0$ \cite{Tolman_1934_book, Bonnor_1962,Katona_Lucietti_2024}, we will consider the Einstein equations coupled to an uncharged scalar field $\phi:\mathcal{M}\rightarrow\mathbb R$ of mass $\mathfrak{m}\geq0$, where $\big(\mathcal{M},g\big)$ is a 4-dimensional Lorentzian manifold. The coupled system reads: 

\vspace{-1em}

\begin{equation}
    \begin{aligned}
        R_{\mu\nu}-\tfrac12 R g_{\mu\nu}+\Lambda g_{\mu\nu}&=2 T_{\mu\nu}[\phi],\\
        \square_g\phi\doteq g^{\mu\nu}\nabla_{\mu}\nabla_\nu \phi&=\mathfrak{m}^2\phi,
    \end{aligned}
    \label{eq:Einstein-scalar field system}
\end{equation}
with the energy-momentum tensor components $T_{\mu\nu}[\phi]$ given by
\begin{equation}
    T_{\mu\nu}[\phi]\doteq\partial_{\mu}\phi\partial_{\nu}\phi-\tfrac12g_{\mu\nu}\big(\partial^{\sigma}\phi\partial_{\sigma}\phi+\mathfrak{m}^2\phi^2\big). 
\end{equation}
Note that the value for the conformal mass for the (3+1)-dimensional spacetime with $\Lambda>0$ corresponds to $\mathfrak{m}^2=2\Lambda/3$, but we will not restrict our study to solely the conformal mass value, and we will assume instead that the scalar field mass $\mathfrak{m}$ is non-negative, unless otherwise stated. 

\begin{figure}[H] 
        \centering
        \includegraphics[width=0.55\textwidth]{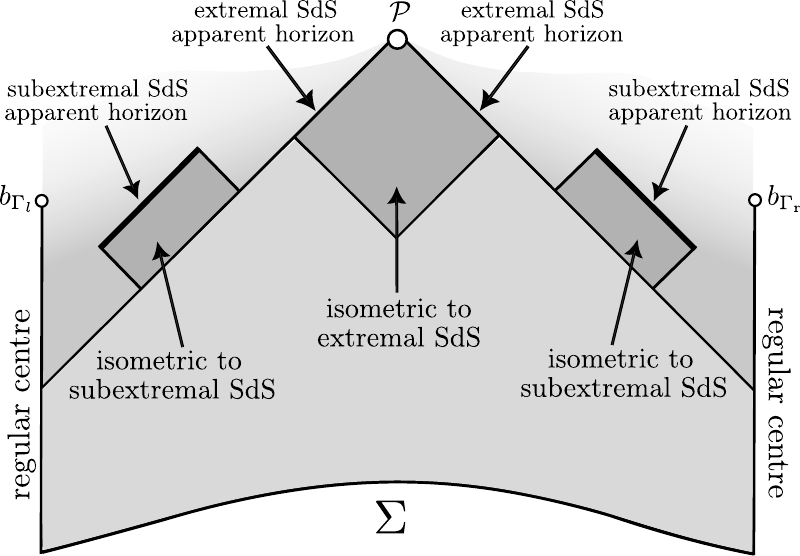}
        \caption{Penrose diagram illustrating the construction in Theorem \ref{thm:counterexample third law rough} involving the dynamical formation of a region of the exactly extremal Schwarzschild--de Sitter spacetime. Similar diagrams hold for the other time orientation of extremal Schwarzschild--de Sitter spacetime, as well as with the extremal Schwarzschild--de Sitter region replaced by a region of the Nariai spacetime.} 
        \label{fig:third law intro}
\end{figure}

Our main results can be unified, in rough terms, by the following statement:

\begin{customthm}{1}[Rough statement]
    \label{thm:counterexample third law rough}
    For any scalar field mass $\mathfrak{m}\geq0$, there exist spherically symmetric, regular initial data for the Einstein-scalar field system \eqref{eq:Einstein-scalar field system} on a Cauchy hypersurface $\Sigma\cong\mathbb S^3$, whose spacetime evolution exhibits first formation of subextremal Schwarzschild--de Sitter, and subsequently formation of a region of exactly extremal Schwarzschild--de Sitter in finite advanced time, as depicted in Figure \ref{fig:third law intro}. Moreover, the spacetime contains no other event horizons beyond those shown in Figure \ref{fig:third law intro}. If $\mathfrak{m}=0$, the resulting spacetimes have connected apparent horizons. 
    
    Additionally, if the scalar field mass satisfies $\mathfrak{m}>\sqrt{\Lambda}/2$ (including thus the conformally coupled case $\mathfrak{m}=\sqrt{{2\Lambda}/{3}}$), there exist spherically symmetric, regular initial data on $\Sigma\cong\mathbb S^3$, for which the same picture holds for the other time orientation of extremal Schwarzschild--de Sitter, as well as replacing extremal Schwarzschild--de Sitter by Nariai, although without the exclusion of additional horizons. Moreover, such initial data cannot exist if $\mathfrak{m}=0$. 
\end{customthm}

We defer a more precise statement of Theorem \ref{thm:counterexample third law rough} until after having presented in detail the family of spherically symmetric solutions to the vacuum Einstein equations with $\Lambda>0$ in Section \ref{subsec:SdS family of solutions introduction}.

Nevertheless, it is worth stressing that the spacetimes in Theorem \ref{thm:counterexample third law rough} involving the time orientation of extremal Schwarzschild--de Sitter that can be interpreted as a black hole spacetime constitute genuine counterexamples to the third law, and it is precisely the absence of additional event horizons that makes this interpretation possible: without this, one could not rule out that the extremal region belongs to a different black hole than the one that was subextremal at earlier times, and the solutions would not unambiguously describe a subextremal black hole apparent horizon that becomes extremal in finite time. For further discussion of the third law in this setting, see Section~\ref{subsec:intro third law}.

Our strategy to construct the Cauchy data relies on characteristic gluing techniques for the Einstein-scalar field system with $\Lambda>0$ in spherical symmetry, along the lines of \cite{Kehle_Unger_2025}. We describe this strategy in Section~\ref{subsec:strategy intro} and provide more precise statements of Theorem~\ref{thm:counterexample third law rough} in Section~\ref{subsec:main theorems intro}.

We emphasise from the outset that even these precise versions of Theorem~\ref{thm:counterexample third law rough} do not provide a complete global picture of the constructed solutions. Even for the solutions involving the extremal Schwarzschild--de Sitter black hole, for which we have shown the uniqueness of the two event horizons, we still lack a complete description near the boundary points $\mathcal{P}$, $b_{\Gamma_{l}}$, $b_{\Gamma_{\mathrm{r}}}$. The global structure is even less clear for the solutions involving the other time orientation of the extremal Schwarzschild--de Sitter spacetime and the Nariai spacetime, given that for these solutions Theorem \ref{thm:counterexample third law rough} does not rule out the presence of additional event horizons. We close the introduction with a discussion of some unresolved aspects of the geometry of the constructed solutions and their conjectured global structure in Section~\ref{subsec:global structure}.

\subsection{Spherically symmetric vacuum solutions with positive cosmological constant and the third law}

\subsubsection{The Schwarzschild--de Sitter family of solutions and Nariai}
\label{subsec:SdS family of solutions introduction}

The metric for the analogue of the Schwarzschild solution with $\Lambda>0$ was discovered by Kottler \cite{Kottler_1918}, Weyl \cite{Weyl_1919} and Trefftz \cite{Trefftz_1922} independently. For fixed $\Lambda>0$ and $M$ a real parameter identified as the mass of the solution, they found
\begin{equation}
    g^{_\mathrm{{SdS}}}_{M}=-\bigg(1-\frac{2M}{r}-\frac13\Lambda r^2\bigg)dt^2+\bigg(1-\frac{2M}{r}-\frac13\Lambda  r^2\bigg)^{-1}dr^2+r^2 \slashed{g}_{\mathbb S^2} , 
    \label{eq:line element}
\end{equation}
where $\slashed{g}_{\mathbb S^2}$ is the round metric on $\mathbb S^2$, $t\in \mathbb R$, and $r$ was originally considered to vary within the range where
\begin{equation}
    f_{M}(r)\doteq 1-\frac{2M}{r}-\frac13\Lambda r^2
\end{equation}
is positive. For $0<M<\tfrac{1}{3\sqrt{\Lambda}}$, $f_M$ has two distinct positive roots that define black hole and cosmological event horizons, $\mathcal{H}$ and $\overline{\mathcal{H}}$ respectively. We call $r_{\mathcal{H}}$ the root corresponding to the radius of the \textit{black hole event horizon}, and $r_{\overline{\mathcal{H}}}$ the one defining the radius of the \textit{cosmological event horizon}. We will often see $r_{\mathcal{H}}$, $r_{\overline{\mathcal{H}}}$ as functions of the spacetime mass. For $0<M<\frac{1}{3\sqrt{\Lambda}}$, they satisfy
\begin{equation}
    0<r_{\mathcal{H}}(M)<\frac{1}{\sqrt{\Lambda}}<r_{\overline{\mathcal{H}}}(M)<\sqrt{\frac{3}{\Lambda}}.
\end{equation}
It was later understood \cite{Geyer_1980,Bazanski_Ferrari_1986}  that the metric \eqref{eq:line element} can be extended beyond those horizons, giving rise to a globally hyperbolic spacetime with $\mathbb R \times \mathbb R \times \mathbb S^2$ topology, that we will refer to as \textit{Schwarzschild--de Sitter} spacetime. By performing suitable identifications, one can also construct a spatially compact version of Schwarzschild--de Sitter spacetime with global $\mathbb R \times \mathbb S^1 \times \mathbb S^2$ topology. 

As $M$ increases, the value of $r_{\mathcal{H}}(M)$ increases and the value of $r_{\overline{\mathcal{H}}}(M)$ decreases. The case of \eqref{eq:line element} when $M=\tfrac{1}{3\sqrt{\Lambda}}$ and $f_M$ has one double positive root
\begin{equation}
    r_{\mathcal{H}}\bigg(\frac{1}{3\sqrt{\Lambda}}\bigg)=\frac{1}{\sqrt{\Lambda}}=r_{\overline{\mathcal{H}}}\bigg(\frac{1}{3\sqrt{\Lambda}}\bigg)
\end{equation}
was not addressed in detail until later. The static region that in the subextremal case lay between the black hole horizon and the cosmological horizon is not present in this case. The lack of a static region results in the solution not having a discrete time-reversal symmetry, and two different time orientations can be considered for this solution. Depending on the time orientation chosen, we will observe:

\vspace{0.5em}

\begin{itemize}
    \item a solution with a past cosmological region and a black hole region, separated by an extremal black hole event horizon, which we will refer to as the \textit{extremal Schwarzschild--de Sitter ``black hole'' spacetime}; 
    \vspace{0.5em}
    \item a solution with a white hole region and a future cosmological region, separated by an extremal cosmological event horizon, which we will call the \textit{extremal Schwarzschild--de Sitter ``white hole'' spacetime}. 
\end{itemize}

\vspace{0.5em}

\begin{figure}[H]
        \centering
        \hspace{-2em}
        \begin{subfigure}[c]{0.43\textwidth}
            \centering
            \includegraphics[width=1.2\textwidth]{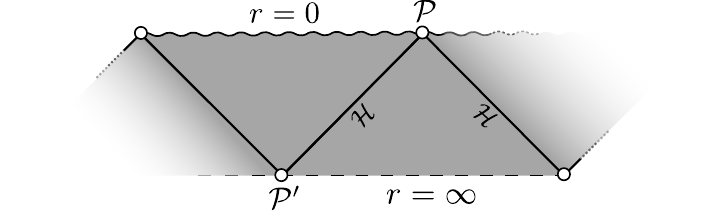}
            \caption{}
            \label{subfig:extremalSdS black hole intro}
        \end{subfigure}
        \hspace{2.7em}
        \begin{subfigure}[c]{0.43\textwidth}
            \centering
            \includegraphics[width=1.2\textwidth]{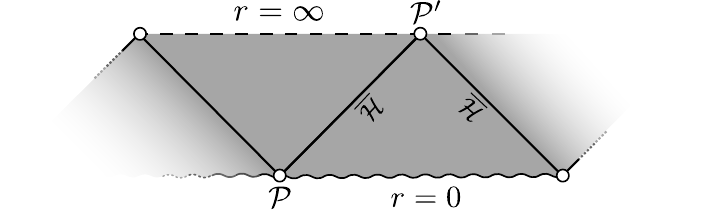}
            \caption{}
            \label{subfig:extremalSdS white hole intro}
        \end{subfigure}
        \caption{Penrose diagrams of: (a) the extremal Schwarzschild--de Sitter black hole spacetime; (b) the extremal Schwarzschild--de Sitter white hole spacetime. In (a), the boundary point $\mathcal{P}$ represents the future endpoint of all future complete timelike geodesics, whereas the boundary point $\mathcal{P}'$ represents the past endpoint of all past complete timelike geodesics that have $r\nrightarrow\infty$ in their infinite past. In (b), we have the opposite time orientation and the roles of $\mathcal{P}$ and $\mathcal{P}'$ are interchanged.} 
        \label{fig:extremalSdS intro}
\end{figure}

In either case, the spacetime is globally hyperbolic and has global $\mathbb R \times \mathbb R \times \mathbb S^2$ topology. As in the subextremal case, spatially compact versions of the extremal Schwarzschild--de Sitter black hole and white hole spacetimes can be constructed by performing suitable identifications. The Penrose diagrams for the extremal Schwarzschild--de Sitter black hole and white hole spacetimes are shown in Figure \ref{fig:extremalSdS intro}. For a more detailed explanation of the geometric features of these spacetimes see Section \ref{subsec:vacuum solutions}. 

\begin{figure}[H]
        \centering
        \includegraphics[width=0.27\textwidth]{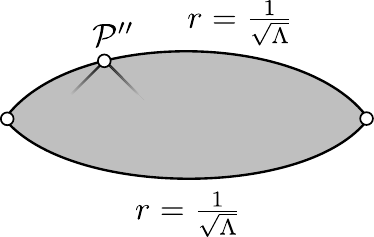}
        \caption{Penrose diagram of the Nariai spacetime. Causal geodesics are future and past complete, and \protect\myuline{every} point $\mathcal{P}''$ on the future (respectively, past) boundary of the spacetime is the future (respectively, past) endpoint of a causal geodesic.}
        \label{fig:Nariai intro}
\end{figure}

However, the spacetimes depicted in Figure \ref{fig:extremalSdS intro} do not constitute a complete account of the extremal case. Several decades before, Nariai \cite{Nariai_1950,Nariai_1951} found a spherically symmetric solution to the vacuum Einstein equations with $\Lambda>0$ whose line element was different from \eqref{eq:line element} and could not be transformed into such a form. This solution, now known as the \textit{Nariai spacetime}, is a constant area-radius spacetime that has $r=\frac1{\sqrt{\Lambda}}$ everywhere, and whose metric, in static coordinates, reads
\begin{equation}
    g^{\mathrm{Nar}}=-\big(1-\Lambda x^2\big)dt^2+\big(1-\Lambda  x^2\big)^{-1}dx^2+\frac1{\Lambda} \slashed{g}_{\mathbb S^2}, 
\end{equation}
with $t\in\mathbb R$ and $x\in\Big(-\frac1{\sqrt{\Lambda}}, \frac1{\sqrt{\Lambda}}\Big)$. This metric can be analytically extended to a globally hyperbolic and geodesically complete spacetime with $\mathbb R \times \mathbb R \times \mathbb S^2$ topology. See the corresponding Penrose diagram in Figure \ref{fig:Nariai intro}.  

Ginsparg and Perry \cite{Ginsparg_Perry_1983} explicitly linked the Nariai spacetime to the extremal Schwarzschild--de Sitter spacetime: they showed that the line element one obtains by zooming in on the region between the black hole horizon and the cosmological horizon in the subextremal Schwarzschild--de Sitter spacetime and taking the limit to the extremal case coincides with that of the Nariai spacetime. That is, they established that what is now called the ``near-horizon geometry'' of the extremal Schwarzschild--de Sitter solution is described by the Nariai metric, in the same way that the near-horizon geometry of the extremal Reissner--Nordström solution is described by the Bertotti--Robinson metric \cite{Bertotti_1959,Robinson_1959}. 

The Nariai spacetime, together with the extremal Schwarzschild--de Sitter ``black hole'' and ``white hole'' spacetimes, exhausts the set of spherically symmetric maximally extended solutions (up to local isometries) to the vacuum Einstein equations with $\Lambda>0$ that contain a static degenerate Killing horizon \cite{Katona_Lucietti_2024}. For this reason, we shall refer to these spacetimes collectively as \textit{extremal solutions}.

If $M>\tfrac{1}{3\sqrt{\Lambda}}$, then $f_M$ has no positive roots, and the spacetime has no horizons. In this case there is no discrete time-reversal symmetry again, and thus the solution can be endowed with two different time orientations, giving rise to a spacetime with a global singularity to the future, the \textit{future-singularity superextremal Schwarzschild--de Sitter solution}, or a spacetime with a global singularity to the past, the \textit{past-singularity superextremal Schwarzschild--de Sitter solution}. These solutions were first studied by Geyer \cite{Geyer_1980}, and have gained certain prominence recently, especially the future-singularity superextremal solution,  under the name of ``ultra-massive spacetimes'' \cite{,Senovilla_2023}. 

For completeness, we note that when $M=0$ the function $f_M$ has a single positive root and \eqref{eq:line element} reduces to the de Sitter metric \cite{deSitter_1918} written in static coordinates. This spacetime has two regular centres and global $\mathbb R \times \mathbb S^3$ topology. Around the two regular centres there are two static regions, bounded by cosmological horizons separating them from the past and future cosmological regions. Finally, if $M<0$, the solution again exhibits static regions bounded by cosmological horizons, but in this case the two centres at $r=0$ are in fact naked singularities where the curvature blows up.

For a detailed discussion of the geometric features of the members of the family that are relevant to our study, see Section~\ref{subsec:vacuum solutions}.

\subsubsection{The third law of black hole thermodynamics with $\Lambda>0$}

\label{subsec:intro third law}

Shortly after the laws of black hole horizons were first presented \cite{Bardeen_Carter_Hawking_1973}, Gibbons and Hawking \cite{Gibbons_Hawking_1977} showed that the zeroth, first and second laws could be extended not only to black hole event horizons in spacetimes with $\Lambda>0$, but also to cosmological event horizons, renaming the resulting statements more generally as \textit{laws of event horizons}. Since then, there have been numerous works dealing with horizon thermodynamics in the case where $\Lambda>0$. See \cite{Mann_2026} for a recent account on the subject and references therein. 

Although it was not originally conjectured by Gibbons and Hawking, it does not seem unreasonable to consider a version for the third law in the case with $\Lambda>0$, given that all the other laws extend naturally to this case: 

\begin{conjecture}[Third law of horizon thermodynamics with $\Lambda>0$]
    A subextremal black hole which is part of a solution to the Einstein equations with $\Lambda>0$  cannot become extremal in finite time by any continuous process, no matter how idealised, in which the spacetime and matter fields remain regular and obey the weak energy condition.
\end{conjecture}

Unlike in the asymptotically flat setting, the question of what it means for a subextremal black hole to ``become extremal'' requires careful formulation when $\Lambda>0$, given the existence of multi-black hole configurations. Any reasonable notion must account for this feature to ensure that one is following the evolution of the same black hole. 

One strategy to achieve this is to establish that the spacetime contains no black hole regions other than the one under consideration. A second possible strategy is to introduce a suitable notion of an apparent horizon, prove that it is connected (thereby ensuring it tracks a single black hole), and follow its evolution to show that it is initially isometric to a subextremal horizon and matches an extremal horizon at later times. Both approaches are pursued to ensure that the spacetimes constructed in Theorem~\ref{thm:counterexample third law rough} violate the third law.

Remarkably, the second alternative cannot work with $\Lambda=0$: Israel  showed \cite{Israel_1986} that a subextremal apparent horizon that becomes extremal in finite time in the asymptotically flat case must necessarily be disconnected, as observed in the counterexamples constructed in \cite{Kehle_Unger_2025}. Israel's result, however, exploits the fact that asymptotically flat extremal solutions, such as extremal Reissner--Nordström and extremal Kerr, have no trapped surfaces. Because the extremal Schwarzschild--de Sitter black hole spacetime is foliated by trapped spheres away from the horizon, Israel's argument does not apply in our case. 

The possibility of incorporating $\Lambda>0$  has already been explored in some of the recent works concerning the third law. Several of them have studied it in the presence of charge \cite{Marin_2024,Gadioux_Reall_Santos_2026}, analysing then the extremal case associated with the coincidence of the inner and outer horizons of the Reissner--Nordström--de Sitter black hole. The question of the formation of uncharged, non-rotating extremal solutions with $\Lambda>0$ has been addressed for the first time in the numerical study \cite{Dunajski_Szybka_2026}, which deals with the formation in vacuum of near-extremal Schwarzschild--de Sitter black holes in 4+1 dimensions. None of these studies, however, rules out the emergence of additional black hole regions, leaving open whether their constructions track the evolution of a single black hole. 

\subsection{Formation of extremal solutions in gravitational collapse}
\label{subsec:dynamical formation}

We are interested in the study of gravitational collapse as an evolutionary problem. More precisely, our aim is to construct Cauchy initial data for the Einstein-scalar field system whose associated future maximal globally hyperbolic development \cite{Foures-Bruhat_1952, Choquet-Bruhat_Geroch_1969,Ringstrom_2009} exhibits dynamical formation of Schwarzschild--de Sitter or Nariai spacetime. The initial data in this case will be a quintuple $\big(\Sigma, g_0, k_0, \phi_0, \phi_1\big)$, where $\big(\Sigma,g_0\big)$ is a 3-dimensional Riemannian manifold, $k_0$ is a symmetric 2-tensor on $\Sigma$, and $\phi_0$, $\phi_1$ are scalar fields on $\Sigma$, to be interpreted as the initial value and initial ``time derivative'' of the scalar field. 

In the asymptotically flat setting, the initial data hypersurface is taken to be $\Sigma\cong\mathbb R^3$, motivated by the topology of Cauchy hypersurfaces in Minkowski spacetime. In our case, given that the trivial solution to the vacuum Einstein equations with $\Lambda>0$ is not Minkowski but de Sitter spacetime, we want $\Sigma$ to have the same topology as Cauchy hypersurfaces in de Sitter spacetime, hence we will take $\Sigma \cong \mathbb S^3$. 

We now outline the strategy for constructing the desired Cauchy data, provide the statement of our results on gravitational collapse (including a more detailed version of Theorem \ref{thm:counterexample third law rough}), and comment on the properties and conjectured global structure of the constructed spacetimes. 

\subsubsection{Overview of the general strategy to construct the Cauchy data}
\label{subsec:strategy intro}

Our strategy will not be to construct the Cauchy data first and obtain the desired spacetimes by evolving such data. Instead, we first address the characteristic gluing problem for the Einstein-scalar field system with $\Lambda>0$ in spherical symmetry, inspired by the construction in \cite{Kehle_Unger_2025}. We will obtain results concerning the gluing to the black hole and cosmological event horizons. See already Section \ref{sec:characteristic gluing} and Section \ref{subsec:vacuum solutions} for the formulation of the gluing problem and Section \ref{sec:gluing results} for the statements of the gluing results.

Once we have obtained the characteristic gluing data, we can make use of the local existence results for the characteristic initial value problem for the Einstein-scalar field system in spherical symmetry to obtain a local solution that has the desired features. Since we require the constructed spacetime to have Cauchy hypersurfaces with $\mathbb S^3$ topology, we will have to carry out, within the same spacetime construction, gluings along two intersecting null hypersurfaces. This differs from the situation in \cite{Kehle_Unger_2025}, where studying the characteristic gluing problem along a single null hypersurface was enough to locally construct the desired spacetime. 

\begin{figure}[H]
        \centering
        \includegraphics[width=0.55\textwidth]{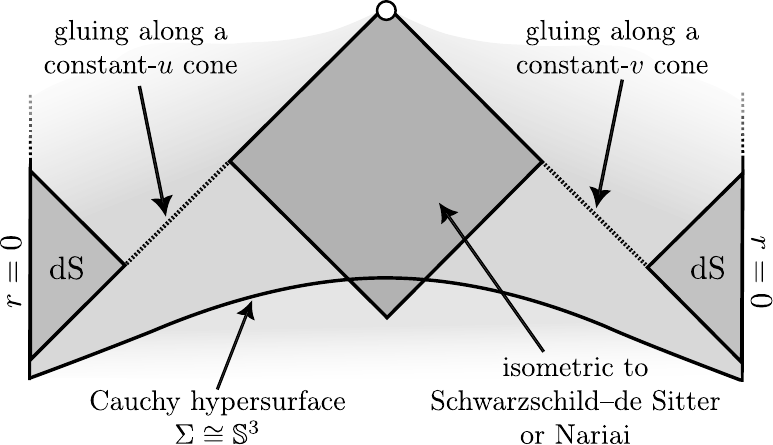}
        \caption{Diagram summarising the strategy for constructing the Cauchy data. Characteristic gluing along two intersecting null hypersurfaces, together with local existence for the characteristic initial value problem, yields the spacetime in a neighbourhood of the null hypersurfaces along which we glue. Cauchy data are obtained a posteriori as the data induced on a suitable Cauchy hypersurface $\Sigma\cong\mathbb S^3$.} 
        \label{fig:general strategy intro}
\end{figure}

In the spacetime obtained following the procedure we just described, we will identify \textit{a posteriori} a suitable Cauchy hypersurface with its corresponding induced initial data, whose  future maximal development includes the physically relevant regions we previously glued. This strategy is summarised in Figure \ref{fig:general strategy intro}. In practice, several of our constructions will require performing multiple gluings along each of the null directions rather than the single step shown.

\vspace{0.5em}

\subsubsection{Statements of the main results}
\label{subsec:main theorems intro}

We are now in a position to state a more detailed version of Theorem \ref{thm:counterexample third law rough}, distinguishing between the formation of the extremal black hole spacetime, the extremal white hole spacetime, and the Nariai spacetime. We start with the statement concerning the formation of the extremal Schwarzschild--de Sitter black hole:  

\begin{customthm}{1.1}
    \label{thm:grav collapse to ext SdS BH intro}
    For any regularity index $k\in \mathbb N$ and scalar field mass $\mathfrak{m}\geq0$, there exist spherically symmetric Cauchy data for the Einstein-scalar field system \eqref{eq:Einstein-scalar field system}, $\big(\Sigma, g_0, k_0, \phi_0, \phi_1\big)$, with $\Sigma \cong \mathbb S^3$ and two regular centres, such that the future maximal globally hyperbolic development $\big(\mathcal{M},g,\phi\big)$ has the following properties: 
    \begin{itemize}
            \item All dynamical quantities are at least $C^k$-regular. 
            \item The spacetime contains: 
            \begin{enumerate}[(i)]
                \item a region isometric to a double null rectangle in the extremal Schwarzschild--de Sitter black hole spacetime up to and including the boundary point $\mathcal{P}$, whose northwestern and northeastern edges can be identified with appropriate portions of the extremal black hole horizons; 
                \item a region, lying to the southwest of the extremal region, isometric to a double null rectangle in a subextremal Schwarzschild--de Sitter spacetime, whose northwestern edge can be identified with a portion of the subextremal black hole horizon; 
                \item a region, lying to the southeast of the extremal region, isometric to a double null rectangle in a subextremal Schwarzschild--de Sitter spacetime, whose northeastern edge can be identified with a portion of the subextremal black hole horizon;
            \end{enumerate}
            \item The spacetime contains no event horizons other than the two bounding the extremal region, and these have respective portions that are exactly isometric to the extremal Schwarzschild--de Sitter black hole horizons, as described in (i). 
            \item The initial data hypersurface $\Sigma$ does not intersect the black hole event horizons. 
            \item Additionally, if $\mathfrak{m}=0$, the apparent horizons present beyond the black hole event horizons are connected, achronal curves in the (1+1)-dimensional reduction of the spherically symmetric spacetime.
    \end{itemize} 
\end{customthm}

We emphasise that ruling out the presence of additional event horizons (and hence of additional black hole regions) in the spacetime is key to conclude that the spacetimes described by Theorem \ref{thm:grav collapse to ext SdS BH intro} constitute genuine uncharged, non-rotating counterexamples to the third law, arising from regular, physical initial data.  

Our gluing results also allow us to show that constructions similar to the one yielding Theorem~\ref{thm:grav collapse to ext SdS BH intro} can be carried out when the region in the extremal Schwarzschild--de Sitter black hole spacetime is replaced by a region in either the extremal Schwarzschild--de Sitter white hole spacetime (Theorem~\ref{thm:grav collapse to ext SdS WH intro}) or the Nariai spacetime (Theorem~\ref{thm:grav collapse to Nariai intro}): 

\begin{customthm}{1.2}
    \label{thm:grav collapse to ext SdS WH intro}
    For any regularity index $k\in \mathbb N$ and any positive scalar field mass $\mathfrak{m}>\sqrt{\Lambda}/2$, there exist spherically symmetric Cauchy data for the Einstein-scalar field system \eqref{eq:Einstein-scalar field system}, $\big(\Sigma, g_0, k_0, \phi_0, \phi_1\big)$, with $\Sigma \cong \mathbb S^3$ and two regular centres, such that the future maximal globally hyperbolic development $\big(\mathcal{M},g,\phi\big)$ is at least $C^k$-regular and contains:
    \begin{itemize}
            \item A region isometric to a double null rectangle in the extremal Schwarzschild--de Sitter white hole spacetime up to and including the boundary point $\mathcal{P}'$, whose northwestern and northeastern edges can be identified with appropriate portions of the extremal cosmological horizons. 
            \item Regions like the ones described in parts (ii) and (iii) of Theorem \ref{thm:grav collapse to ext SdS BH intro}.
    \end{itemize} 
    Moreover, if $\mathfrak{m}=0$, no such Cauchy data exist.
\end{customthm}

\begin{customthm}{1.3}
    \label{thm:grav collapse to Nariai intro}
    For any regularity index $k\in \mathbb N$ and any positive scalar field mass $\mathfrak{m}>\sqrt{\Lambda}/2$, there exist spherically symmetric Cauchy data for the Einstein-scalar field system \eqref{eq:Einstein-scalar field system}, $\big(\Sigma, g_0, k_0, \phi_0, \phi_1\big)$, with $\Sigma \cong \mathbb S^3$ and two regular centres, such that the future maximal globally hyperbolic development $\big(\mathcal{M},g,\phi\big)$ is at least $C^k$-regular and contains:
    \begin{itemize}
            \item A region isometric to a double null rectangle in the Nariai spacetime. 
            \item Regions like the ones described in parts (ii) and (iii) of Theorem \ref{thm:grav collapse to ext SdS BH intro}.
    \end{itemize} 
    Moreover, if $\mathfrak{m}=0$, no such Cauchy data exist.
\end{customthm}

The corresponding diagrams would look like the one shown in Figure \ref{fig:third law intro}, with the central double null rectangle being isometric to a patch of the extremal Schwarzschild--de Sitter white hole  spacetime in the case of Theorem \ref{thm:grav collapse to ext SdS WH intro}, or to a patch of the Nariai spacetime for Theorem \ref{thm:grav collapse to Nariai intro}. 

Crucially, both Theorem~\ref{thm:grav collapse to ext SdS WH intro} and Theorem~\ref{thm:grav collapse to Nariai intro} require $\mathfrak{m}>\sqrt{\Lambda}/2$. Moreover, we show that $\mathfrak{m}>0$ is a necessary condition for the dynamical formation of the extremal Schwarzschild--de Sitter white hole spacetime and the Nariai spacetime from physical initial data (see the discussion in Section~\ref{sec:gluing results} and the statement of Theorem~\ref{thm:impossible gluing}). Whether it is possible to form these solutions dynamically from data on $\mathbb S^3$ in the complementary regime $0<\mathfrak{m}\leq \sqrt{\Lambda}/2$ remains to be understood.

Finally, our gluing results also allow us to establish the existence of Cauchy data on $\mathbb S^3$ leading to the formation of a subextremal Schwarzschild--de Sitter black hole of prescribed mass. Although not directly related to the question of forming extremal solutions dynamically, this result is included here for its independent relevance in the study of black hole formation with $\Lambda>0$: 

\begin{customthm}{2}
    For any regularity index $k\in \mathbb N$, scalar field mass $\mathfrak{m}\geq0$, and spacetime mass $M$ satisfying $0<M<\frac{1}{3\sqrt{\Lambda}}$, there exist spherically symmetric Cauchy data for the Einstein-scalar field system \eqref{eq:Einstein-scalar field system}, $\big(\Sigma, g_0, k_0, \phi_0, \phi_1\big)$, with $\Sigma \cong \mathbb S^3$ and two regular centres, such that the future maximal globally hyperbolic development $\big(\mathcal{M},g,\phi\big)$ has the following properties: 
    \begin{itemize}
            \item All dynamical quantities are at least $C^k$-regular. 
            \item The spacetime contains a region isometric to a double null rectangle in the subextremal Schwarzschild--de Sitter spacetime of mass $M$, whose northwestern and northeastern edges can be identified with appropriate portions of the subextremal black hole and cosmological horizons respectively. 
            \item There are no other event horizons beyond the black hole horizon we construct.
            \item The initial data hypersurface $\Sigma$ does not intersect the subextremal black hole and cosmological horizons we construct.
    \end{itemize}
    \label{thm:grav collapse to subext SdS intro}
\end{customthm}

For detailed diagrams of the constructions underlying all the theorems stated here, as well as for their proofs, see Section~\ref{sec:applications}.

We note that Theorem~\ref{thm:grav collapse to ext SdS BH intro} and Theorem~\ref{thm:grav collapse to subext SdS intro} constitute, to the best of our knowledge, the first rigorous results ensuring the formation of black holes with $\Lambda>0$ in gravitational collapse from Cauchy data on $\mathbb S^3$ for the Einstein-scalar field system. For previous results on gravitational collapse, black hole formation and related questions with $\Lambda>0$, see \cite{Lemaitre_1933,Omer_1965,Garfinkle_Vuille_1991,Nakao_1992,Yamanaka_Nakao_Sato_1992,Markovic_Shapiro_2000,Lake_2000, Deshingkar_Chamorro_Jhingan_Joshi_2001,Debnath_Nath_Chakraborty_2006,Dafermos_Rendall_2016,Costa_2020,Yokoyama_2022} and references therein. For more general considerations regarding the study of black holes with $\Lambda>0$, see \cite{Hintz_Vasy_2018, Hintz_2021, Fournodavlos_Schlue_2026}.

\subsubsection{Global structure of the constructed spacetimes} 
\label{subsec:global structure}

Our results establish the existence of regular Cauchy data on a hypersurface $\Sigma\cong \mathbb S^3$ whose future maximal development contains regions isometric to regions of the extremal Schwarzschild--de Sitter spacetime (in any of its time orientations) and of the Nariai spacetime. However, they do not provide a complete picture of the constructed solutions. Even for the solutions described by Theorem \ref{thm:grav collapse to ext SdS BH intro} and Theorem \ref{thm:grav collapse to subext SdS intro}, for which we show that there are no additional event horizons beyond the black hole horizons we construct, several aspects of their global structure remain to be understood. 

\begin{figure}[H]
        \centering
        \includegraphics[width=0.36\textwidth]{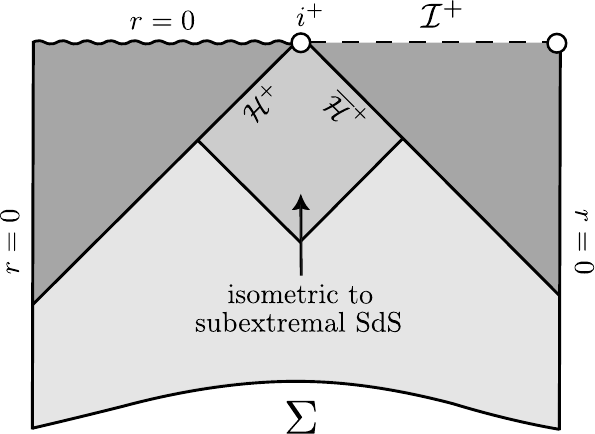}
        \caption{Expected Penrose diagram for some of the solutions described by Theorem \ref{thm:grav collapse to subext SdS intro}.} 
        \label{fig:global subextremal}
\end{figure}

For at least some of the Cauchy data sets described by Theorem \ref{thm:grav collapse to subext SdS intro}, one would expect the future maximal globally hyperbolic development to have the Penrose diagram shown in Figure \ref{fig:global subextremal}. The absence of event horizons beyond the subextremal black hole horizon we construct gives hope of establishing this picture. In contrast, an analogous result beyond the cosmological horizon is missing; the situation in this region is much less clear, and a priori further black hole regions could be present. Ruling these out is a necessary first step in confirming this global structure. 

\begin{figure}[H]
        \centering
        \includegraphics[width=0.36\textwidth]{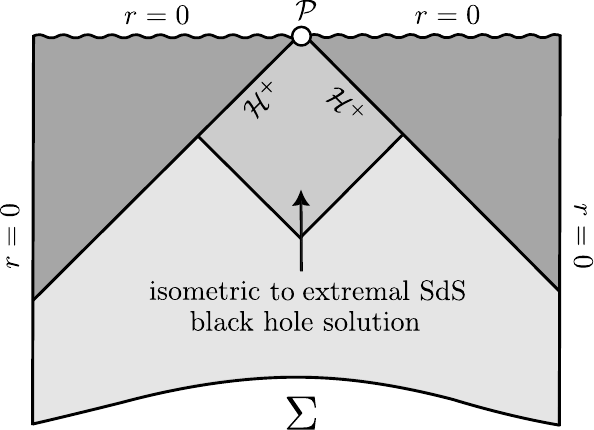}
        \caption{Expected Penrose diagram for some of the solutions described by Theorem \ref{thm:grav collapse to ext SdS BH intro}.} 
        \label{fig:global extremal BH}
\end{figure}

One potential obstacle in establishing the picture in Figure \ref{fig:global subextremal} is the presence of a Cauchy horizon emanating from $i^+$. In light of the results of \cite{Dafermos_Rendall_2016} on the global structure of solutions to the Einstein-massless Vlasov system with $\Lambda\geq0$, one expects to be able to rule out such Cauchy horizons beyond the subextremal black hole horizon. However, \cite{Dafermos_Rendall_2016} leaves open the possibility of Cauchy horizons when an extremal black hole forms dynamically, so the analysis is likely to be more difficult for the solutions involving the extremal Schwarzschild--de Sitter black hole spacetime. Nevertheless, it is expected that at least some of the solutions described by Theorem \ref{thm:grav collapse to ext SdS BH intro} look globally as shown in Figure \ref{fig:global extremal BH}.

\begin{figure}[H]
        \centering
        \includegraphics[width=0.36\textwidth]{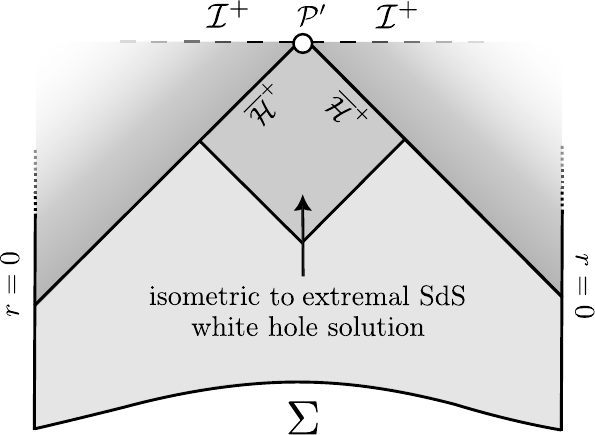}
        \caption{Expected Penrose diagram for the solutions described by Theorem \ref{thm:grav collapse to ext SdS WH intro} near $\mathcal{P}'$. The diagram is agnostic to the behaviour far from $\mathcal{P}'$.} 
        \label{fig:global extremal WH}
\end{figure}

\begin{figure}[H]
        \centering
        \includegraphics[width=0.36\textwidth]{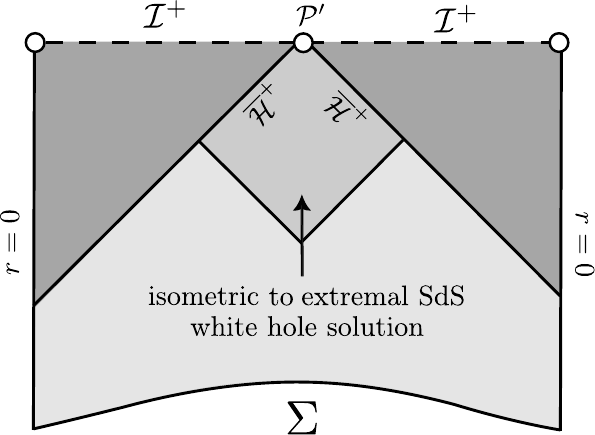}
        \caption{Expected Penrose diagram for some of the modified solutions involving the extremal Schwarzschild--de Sitter white hole spacetime, where the entire future boundary is specified.} 
        \label{fig:global extremal WH complete}
\end{figure}

The global structure of the solutions described by Theorem \ref{thm:grav collapse to ext SdS WH intro} is considerably more uncertain. Indeed, these solutions contain truncated null cones isometric to the subextremal black hole horizons, with trapped spheres beyond them, which makes it doubtful whether additional black hole regions can be ruled out in this case. Nevertheless, at least locally near the boundary point $\mathcal{P}'$, one would expect the Penrose diagram of these solutions to be as shown in Figure \ref{fig:global extremal WH}. We note that the construction in Theorem \ref{thm:grav collapse to ext SdS WH intro} can be modified to avoid the portion of the apparent horizon isometric to the subextremal Schwarzschild--de Sitter black hole horizon (see Remark \ref{rmk:possible gluign dS to ext cosm horizon or Nariai}). It would be interesting to analyse whether some of these modified solutions can be described by the Penrose diagram in Figure \ref{fig:global extremal WH complete}.

The solutions of Theorem \ref{thm:grav collapse to Nariai intro} are of a different nature, since the Nariai solution is highly unstable: arbitrarily small spherically symmetric perturbations of it develop multiple black hole regions \cite{Ginsparg_Perry_1983,Dafermos_Rendall_2016}. It is therefore natural to expect that their global structure depends sensitively on the choice of Cauchy data.

\paragraph{Outline of the paper.} This paper is organised as follows. In Section~\ref{sec:Einstein-scalar field system} we set up the characteristic initial value problem for the Einstein-scalar field system with $\Lambda>0$ in spherical symmetry. Building on this, Section~\ref{sec:characteristic gluing} gives a precise formulation of the characteristic gluing problem, including the definition of sphere data and how gluing of sphere data leads to gluing of spacetimes. In Section~\ref{subsec:vacuum solutions} we return to the spherically symmetric vacuum solutions that we wish to glue, analysing their geometry in more detail and characterising their sphere data. Section~\ref{sec:gluing results} contains the statements of our main gluing results, together with a more general result (Theorem~\ref{thm:impossible gluing}) showing that certain gluings are impossible, and Section~\ref{sec:proofs} is devoted to their proofs. Finally, in Section~\ref{sec:applications} we present the statements and proofs of the theorems on the dynamical formation of the various (sub)extremal solutions in gravitational collapse. 

\paragraph{Acknowledgements.} The author would like to thank Claude Warnick and Mihalis Dafermos for their guidance through many valuable discussions that helped shape this project, and for comments on earlier versions of this manuscript. The author also thanks Christoph Kehle, Jonathan Luk, Georgios Moschidis, Matt Smith and Ryan Unger for helpful conversations, and acknowledges support from the Gates Cambridge Trust (grant number OPP1144), as well as the hospitality of Princeton University through its VSRC program during April 2025, when this work began.

\paragraph{Declaration of AI use.} AI tools (Claude, Gemini) were used only during the final writing stages of this manuscript to identify typographical and grammatical errors and repeated terms, to improve the clarity and style of the text, and to identify relevant references. The mathematical content (statements, proofs and results) and all figures are the author's own, and no AI tool was used to produce them. The author takes full responsibility for the content of this manuscript.

\section{The Einstein-scalar field system with a positive cosmological constant in spherical symmetry}
\label{sec:Einstein-scalar field system}

In this section we introduce the Einstein-scalar field system with a positive cosmological constant $\Lambda>0$ in spherical symmetry, and give an overview of its characteristic initial value problem, highlighting the aspects that will be relevant for the characteristic gluing problem. The study of this system has a long tradition in the $\Lambda=0$ case; for previous work including $\Lambda>0$, see \cite{Costa_Alho_Natario_2013,Costa_2020,Costa_Mena_2021,Costa_Duarte_Mena_2023} and references therein.

\subsection{Spherically symmetric spacetimes in double null gauge}
\label{subsec:double null gauge}

We say that a spacetime, described by a connected and time-oriented, smooth (3+1)-dimensional Lorentzian manifold $(\mathcal{M}, g)$, is \textit{spherically symmetric} if there exists an isometric action of $SO(3)$ on $(\mathcal{M}, g)$. The orbit of a point $p\in\mathcal{M}$ under the action is either $\mathrm{Orb}(p)=\{p\}$ or $\mathrm{Orb}(p)\cong \mathbb S^2$. We then define the centre of symmetry $\Gamma$ to be the set of points fixed by the action:
\begin{equation}
    \Gamma\doteq\big\{p\in\mathcal{M}:\mathrm{Orb}(p)=\{p\}\big\}\subsetneq\mathcal{M}.
\end{equation}
In general, $\Gamma$ is a disjoint union of timelike geodesics. In particular, for the spacetimes constructed here, $\Gamma$ will consist of a pair of disjoint timelike geodesics, and thus there will be \textit{two centres of symmetry}, in contrast to the asymptotically flat case, where spherically symmetric solutions exhibiting gravitational collapse typically have a single centre of symmetry. Hence we will write $\Gamma$ as the union of the left and right centres of symmetry: 
\begin{equation}
    \Gamma=\Gamma_{l}\,\dot{\cup}\,\Gamma_{\mathrm{r}}.
\end{equation}

Under the $SO(3)$ action, $\mathcal{M}\backslash\Gamma$ splits diffeomorphically as 
\begin{equation}
    \mathcal{M}\backslash\Gamma\cong\mathring{\mathcal{Q}}\times\mathbb S^2,
\end{equation}
and the spacetime metric can be written as
\begin{equation}
    g=g_{\mathcal{Q}}+r^2g_{\mathbb S^2}, 
\end{equation}
where $(\mathcal{Q},g_{\mathcal{Q}})$, with $\mathcal{Q}\doteq\mathring{\mathcal{Q}}\cup\Gamma$, is a (1+1)-dimensional Lorentzian manifold with boundary $\Gamma$\footnote{Note that we are denoting by $\Gamma$ both the centres of symmetry in $\mathcal{M}$ and their projections onto $\mathcal{Q}$.}, $g_{\mathbb S^2}$ is the round metric on the unit sphere, and $r:\mathring{\mathcal{Q}}\rightarrow\mathbb R_{>0}$ is a function that can be geometrically interpreted as the \textit{area-radius} of the orbits of the $SO(3)$ action on $(\mathcal{M},g)$, and, as such, the function $r$ extends to 0 on $\Gamma$.

For the spacetimes that will be constructed here, the quotient $(\mathcal{Q},g_{\mathcal{Q}})$ admits global double null coordinates $(u,v)$, with $\partial_u$, $\partial_v$ future-directed, such that the metric can be written in the form 
\begin{equation}
    g=-\Omega^2(u,v)dudv+r^2(u,v)g_{\mathbb S^2},
    \label{eq:spherically symmetric metric}
\end{equation}
with $\Omega^2$ a positive function often referred to as the \textit{null lapse}. Note that the double null pair $(u,v)$ is not unique. In particular, given any two strictly increasing diffeomorphisms $f,h$ defined on the ranges of $u$ and $v$ respectively, we can define another pair of future-directed double null coordinates $(\bar{u},\bar{v})$ by
\begin{equation}
    \bar{u}=f(u), \quad \bar{v}=h(v),
    \label{eq:new double null}
\end{equation}
and the metric then takes the form 
\begin{equation}
    g=-\overline{\Omega}^2(\bar{u},\bar{v})d\bar{u}d\bar{v}+\overline{r}^2(\bar{u},\bar{v})g_{\mathbb S^2},
\end{equation}
with 
\begin{align}
    \overline{\Omega}^2(\bar{u},\bar{v})&=\Omega^2\big(f^{-1}(\bar{u}),h^{-1}(\bar{v})\big)(f^{-1})'(\bar{u})(h^{-1})'(\bar{v}),\\
    \overline{r}(\bar{u},\bar{v})&=r\big(f^{-1}(\bar{u}),h^{-1}(\bar{v})\big).
\end{align}

One can also compute how higher order derivatives of the metric functions $r$, $\Omega^2$ transform under a reparametrisation of the double null coordinates. This \textit{gauge freedom} will be important for the study of the characteristic initial value problem and crucial to address the question of characteristic gluing.

\begin{remark}
    In double null coordinates it is straightforward to identify the trapping character of the symmetry spheres: the symmetry sphere with double null coordinates $(u,v)$ is 
    \begin{itemize}
        \item \textit{trapped} if $\partial_ur(u,v)<0$, $\partial_vr(u,v)<0$,
        \vspace{0.7em}
        \item \textit{marginally trapped} if $\partial_ur(u,v)<0$, $\partial_vr(u,v)=0$, or $\partial_ur(u,v)=0$, $\partial_vr(u,v)<0$,
        \vspace{0.5em}
        \item \textit{antitrapped} if $\partial_ur(u,v)>0$, $\partial_vr(u,v)>0$,
        \vspace{0.5em}
        \item \textit{marginally antitrapped} if $\partial_ur(u,v)>0$, $\partial_vr(u,v)=0$, or $\partial_ur(u,v)=0$, $\partial_vr(u,v)>0$,
        \vspace{0.5em}
        \item \textit{untrapped} if $\partial_ur(u,v)<0$, $\partial_vr(u,v)>0$, or $\partial_ur(u,v)>0$, $\partial_vr(u,v)<0$.
        \vspace{0.5em}
    \end{itemize}
    Note that, since gauge transformations of the form \eqref{eq:new double null} do not change the signs of $\partial_ur$, $\partial_vr$, these notions are independent of the double null coordinate pair considered, as one would expect. 
\end{remark}

\vspace{0.5em}

Regarding the trapping character of the spheres along the centre(s) of symmetry, we will assume that 
\begin{equation}
    \partial_ur|_{\Gamma_l}<0,\, \partial_vr|_{\Gamma_l}>0 \quad \text{ and } \quad \partial_ur|_{\Gamma_{\mathrm{r}}}>0, \, \partial_vr|_{\Gamma_{\mathrm{r}}}<0.
\end{equation}

Finally, we recall the definition of the \textit{Hawking mass}, which can be seen as a function on $\mathcal{Q}$, given by 
\begin{equation}
    m\doteq \frac{r}{2}\left(1-g\left(\nabla r, \nabla r\right)\right)=\frac{r}{2}\left(1+\frac{4\partial_ur\partial_vr}{\Omega^2}\right),
    \label{eq:definition Hawking mass}
\end{equation}
though in the presence of $\Lambda>0$, it will be more useful to consider the \textit{renormalised Hawking mass}: 
\begin{equation}
        \widetilde{m}\doteq m-\frac16\Lambda r^3=\frac{r}{2}\left(1+\frac{4\partial_ur\partial_vr}{\Omega^2}-\frac13\Lambda r^2\right). 
\end{equation}

Although we have introduced the notions above in the smooth category for simplicity and clarity, in what follows all the structures involved will be assumed to have finite regularity, as appropriate for the characteristic initial value problem for the Einstein-scalar field system (see Section~\ref{subsec:characteristic initial value problem}). In particular, the spacetimes constructed here by characteristic gluing will be $C^k$ for any (arbitrarily large) $k\in\mathbb N$, but not necessarily smooth.

\subsection{The spherically symmetric Einstein equations coupled to a scalar field}
\label{subsec:eqns in double null gauge}
We focus our study on spacetimes that are spherically symmetric solutions to the Einstein-scalar field system with a non-negative scalar field mass $\mathfrak{m}\geq0$. We then assume the existence of double null coordinates $(u,v)$ such that the metric takes the form \eqref{eq:spherically symmetric metric}, and that $\phi$ is spherically symmetric, hence $\phi=\phi(u,v)$. In this setting, the system \eqref{eq:Einstein-scalar field system} reduces to a set of wave equations for the metric functions $r$, $\Omega^2$, and the scalar field $\phi$, namely
\begin{equation}
    \partial_u\partial_v\log\big(\Omega^2\big)=\frac{\Omega^2}{2r^2}+\frac{2\partial_ur\partial_vr}{r^2}-2\partial_u\phi\partial_v\phi,
    \label{eq:EE massive scalar field - eqn for Omega}
\end{equation}
\begin{equation}
    \partial_u\partial_vr=-\frac{\Omega^2}{4r}\Big(1-\Lambda r^2\Big)-\frac{\partial_ur\partial_vr}{r}+\frac{\Omega^2r}{4}\mathfrak{m}^2\phi^2,
    \label{eq:EE massive scalar field - eqn for r}
\end{equation}
\begin{equation}
    \partial_u\partial_v\phi=-\frac{\partial_ur\partial_v\phi}{r}-\frac{\partial_u\phi\partial_vr}{r}-\frac{\Omega^2}{4}\mathfrak{m}^2\phi,
    \label{eq:EE massive scalar field - eqn for phi}
\end{equation}
together with the constraint equations

\vspace{-0.5em}

\begin{equation}
    \partial_v\left(\frac{\partial_vr}{\Omega^2}\right)=-\frac{r}{\Omega^2}\big(\partial_v\phi\big)^2,
    \label{eq:EE massive scalar field - Raychaudhuri's v}
\end{equation}
\begin{equation}
    \partial_u\left(\frac{\partial_ur}{\Omega^2}\right)=-\frac{r}{\Omega^2}\big(\partial_u\phi\big)^2,
    \label{eq:EE massive scalar field - Raychaudhuri's u}
\end{equation}

\vspace{0.2em}

\noindent known as \textit{Raychaudhuri's equations}. Note that, if we perform a gauge transformation of the form \eqref{eq:new double null}, the resulting $\big(\overline{r}, \overline{\Omega}^2, \overline{\phi} \big)$, with $\overline{\phi}(\bar{u}, \bar{v})=\phi\big(f^{-1}(\bar{u}), h^{-1}(\bar{v})\big)$, will still satisfy the spherically symmetric Einstein-scalar field system, written in terms of the new double null coordinates $(\bar{u},\bar{v})$. This is simply the manifestation of the diffeomorphism invariance that characterises the Einstein equations, and which here translates into the invariance of the equations \eqref{eq:EE massive scalar field - eqn for Omega}--\eqref{eq:EE massive scalar field - Raychaudhuri's u} under transformations that preserve the double null gauge. 

\newpage

It will be useful to rewrite \eqref{eq:EE massive scalar field - eqn for r} as
\begin{equation}
    \partial_v(r\partial_u r)=-\frac{\Omega^2}{4}\big(1-\Lambda r^2\big)+\frac{\Omega^2r^2}{4}\mathfrak{m}^2\phi^2=\frac{\Omega^2}{4}\Big(r^2\big(\Lambda + \mathfrak{m}^2\phi^2\big)-1\Big),
    \label{eq:EE-massive scalar field - eqn for r partial_ur}
\end{equation}
and also to rewrite \eqref{eq:EE massive scalar field - eqn for phi} as
\begin{equation}
    \partial_v(r\partial_u\phi)=-\partial_ur\partial_v\phi-\frac{\Omega^2}{4}\mathfrak{m}^2r\phi.
    \label{eq:wave equation scalar field}
\end{equation}

Finally, using \eqref{eq:EE massive scalar field - eqn for r} and \eqref{eq:EE massive scalar field - Raychaudhuri's v}, we obtain 
\begin{equation}
    \partial_v\widetilde{m}=\frac{2r^2}{\Omega^2}(-\partial_ur)\big(\partial_v\phi\big)^2+\frac12 r^2 \partial_vr\,\mathfrak{m}^2\phi^2,
     \label{eq:massive scalar field v eqn for varpi}
\end{equation}
which will be useful when we consider the evolution of the renormalised Hawking mass along the gluing region. 

\subsection{The characteristic initial value problem}
\label{subsec:characteristic initial value problem}

We describe now the characteristic initial value problem for the Einstein-scalar field system in spherical symmetry, and summarise some of its features that will later be useful to study the characteristic gluing problem. This summary follows closely the discussion and conventions in \cite{Kehle_Unger_2025}.

In order to set up the characteristic initial value problem in spherical symmetry, we start considering (1+1)-dimensional Minkowski spacetime endowed with double null coordinates $(u,v)$, which we will denote $\mathbb R^{1+1}_{u,v}$, and a \textit{bifurcate null hypersurface}, ${C}\cup\underline{{C}}\subset\mathbb R^{1+1}_{u,v}$, with
\begin{align}
    {C}&\doteq \{u_0\}\times[v_0,v_1],\\
    \underline{{C}}&\doteq[u_0,u_1]\times\{v_0\}.
\end{align} 
The sphere where these two null hypersurfaces intersect, ${C}\cap\underline{{C}}=\{(u_0,v_0)\}$, is often referred to as the \textit{bifurcation sphere}. 

On ${C}\cup\underline{{C}}$ we prescribe \textit{bifurcate characteristic initial data}, which will give rise to a unique solution for system \eqref{eq:EE massive scalar field - eqn for Omega}--\eqref{eq:EE massive scalar field - Raychaudhuri's u} in a double null neighbourhood to the future of ${C}\cup\underline{{C}}$. The iteration argument that allows to prove the analogous result in the $\Lambda=0$ case can easily be adapted to conclude the local existence and uniqueness of solutions to the characteristic initial value problem for the Einstein-scalar field system with positive cosmological constant in spherical symmetry:

\begin{proposition}[Local existence and uniqueness]
    \label{prop:local existence and uniqueness characteristic IVP}
    Let $k\in \mathbb N$ and consider a $C^k$ bifurcate characteristic initial data set on ${C}\cup\underline{{C}}$, consisting of a real scalar field $\phi \in C^k({C}\cup\underline{{C}})$ and positive functions $r\in C^{k+1}({C}\cup\underline{{C}})$ and $\Omega^2\in C^k({C}\cup\underline{{C}})$, satisfying \eqref{eq:EE massive scalar field - Raychaudhuri's v}--\eqref{eq:EE massive scalar field - Raychaudhuri's u}. Then there exists a real number $\delta>0$, an associated double null neighbourhood of ${C}\cup\underline{{C}}$
    \begin{equation}
        \mathcal{Q}=\Big([u_0,u_0+\delta]\times[v_0, v_1]\Big)\cup\Big([u_0,u_1]\times[v_0,v_0 +\delta]\Big),
    \end{equation}
    and unique functions $r\in C^{k+1}(\mathcal{Q})$, $\Omega^2\in C^k(\mathcal{Q})$, $\phi \in C^k(\mathcal{Q})$, with $r>0$, $\Omega^2>0$, satisfying \eqref{eq:EE massive scalar field - eqn for Omega}--\eqref{eq:EE massive scalar field - Raychaudhuri's u} and extending the bifurcate characteristic initial data on ${C}\cup\underline{{C}}$. 
\end{proposition}

\begin{remark}
    We will sometimes talk about spherically symmetric solutions to the Einstein-scalar field system of the form $(\mathcal{M}, g, \phi)$ instead of $( \mathcal{Q},r, \Omega^2, \phi)$, understanding that the full spacetime manifold and metric are recovered as 
     \begin{equation}
        \mathcal{M}=\mathcal{Q}\times \mathbb S^2, \quad  g=-\Omega^2dudv+r^2g_{\mathbb S^2}.
    \end{equation}
\end{remark}

Note that, along ${C}$, the Einstein-scalar field system does not only constrain $r$, $\Omega^2$, and $\phi$ and their $v$-derivatives along ${C}$ (according to Raychaudhuri's equation \eqref{eq:EE massive scalar field - Raychaudhuri's v}). In fact, the $u$-derivatives of both metric functions and the scalar field along ${C}$ are likewise constrained: evaluated along ${C}$, the wave equations \eqref{eq:EE massive scalar field - eqn for Omega}--\eqref{eq:EE massive scalar field - eqn for phi} reduce to transport equations for the $u$-derivatives of $r$, $\Omega^2$, and $\phi$. This yields the following observation:

\begin{proposition}[\protect{\cite[Proposition 2.2]{Kehle_Unger_2025}}]
    \label{prop:compute u-derivatives from initial data}
    Given a $C^k$ bifurcate initial data set $(r,\Omega^2,\phi)$ on ${C}\cup\underline{{C}}$, it is possible to compute as many $u$-derivatives of $r, \Omega^2,$ and  $\phi$ along ${C}$ as the regularity allows, simply using the initial data on ${C}\cup\underline{{C}}$ and the Einstein-scalar field system. Similarly, it is possible to compute as many $v$-derivatives of $r, \Omega^2,$ and $\phi$ along $\underline{{C}}$ as the regularity allows. 
\end{proposition}
\begin{proof}
    Since $r\in C^{k+1}(\underline{{C}})$, $\Omega^2\in C^k(\underline{{C}})$ and $\phi\in C^{k}(\underline{{C}})$ are given as part of the bifurcate initial data, we can compute as many $u$-derivatives of $r$, $\Omega^2$, and $\phi$ at the bifurcation sphere ${C}\cap\underline{{C}}$ as the regularity of the initial data allows, that is, $k+1$ derivatives of $r$, and $k$ derivatives of $\Omega^2$ and $\phi$. Similarly, given that $r \in C^{k+1}({C})$, $\Omega^2\in C^k({C})$ and $\phi\in C^{k}({C})$ are also given as part of the bifurcate initial data, we can compute as many $v$-derivatives of $r$, $\Omega^2$, $\phi$ along ${C}$ as is consistent with the regularity of the initial data. 
    
    Now, we note that $\partial_ur$ satisfies the following transport equation along ${C}$:
    \begin{equation}
        \partial_v (\partial_ur)+\frac{\partial_vr}{r}\partial_u r=\frac{\Omega^2}{4r}\Big(r^2\big(\Lambda +\mathfrak{m}^2\phi^2\big)-1\Big), 
        \label{eq:transport equation partial_ur}
    \end{equation}
    and hence we can find $\partial_ur(u_0,v)$ by solving this ODE with initial value $\partial_ur(u_0,v_0)$. Once $\partial_ur$ along ${C}$ is known, we can find $\partial_u\phi(u_0,v)$ by solving
    \begin{equation}
        \partial_v (\partial_u\phi)+\frac{\partial_vr}{r}\partial_u \phi=-\frac{\partial_ur\partial_v\phi}{r}-\frac{\Omega^2}{4}\mathfrak{m}^2\phi,
        \label{eq:transport equation partial_uphi}
    \end{equation}
    with initial value $\partial_u\phi(u_0,v_0)$. Finally, considering $\partial_u \log\Omega^2(u_0,v_0)$ and the equation \eqref{eq:EE massive scalar field - eqn for Omega}, we can find $\partial_u \log\Omega^2(u_0,v)$.  

    By commuting equations \eqref{eq:transport equation partial_ur}, \eqref{eq:transport equation partial_uphi} and \eqref{eq:EE massive scalar field - eqn for Omega} with $\partial_u^i$, with $i=1, \dots, k-1$, we can iterate this sequential procedure to compute  the $u$-derivatives of $r$, $\Omega^2$, $\phi$ up to $k$-th order. In order to compute $\partial_u^{k+1}r(u_0, v)$, we commute equation \eqref{eq:transport equation partial_ur} with $\partial_u^{k}$ and observe that the expression we obtain only involves derivatives of $\Omega^2$ and $\phi$ up to order $k$, and that we can therefore solve, together with the initial value $\partial_u^{k+1}r(u_0, v_0)$, to find $\partial_u^{k+1}r$ along ${C}$ (alternatively, one can use \eqref{eq:EE massive scalar field - Raychaudhuri's u} differentiated an appropriate number of times). 

    The same procedure can be applied to compute $v$-derivatives of $r$, $\Omega^2$ and $\phi$ along $\underline{{C}}$, given that they satisfy analogous transport equations. 
\end{proof}

\begin{remark}
    \label{rmk:lapse-normalised characteristic initial data}
    As already discussed in the previous sections, we can always redefine the double null coordinates $(u,v)$ according to a reparametrisation of the form \eqref{eq:new double null}. In the set-up considered here, this gauge freedom can be used to make $\Omega^2$ constant on the bifurcate null hypersurface ${C}\cup\underline{{C}}$. In order to do it, we start rescaling one of the double null coordinates so that $\Omega^2(u_0,v_0)=1$. After that we just need to take $f,h \in \mathrm{Diff}_+(\mathbb R)$ given by 
    \begin{equation}
        \begin{aligned}
            f'(u)&=\Omega^2(u,v_0), \quad f(u_0)=u_0,\\
            h'(v)&=\Omega^2(u_0,v), \quad h(v_0)=v_0,
        \end{aligned}
    \end{equation}
    so that, in the new coordinates, $\Omega^2(u,v_0)=\Omega^2(u_0,v_0)=\Omega^2(u_0,v)$, and hence $\Omega^2\equiv1$ on ${C}\cup\underline{{C}}$. If a bifurcate initial data set $(r, \Omega^2, \phi)$ on ${C}\cup\underline{{C}}$ has $\Omega^2\equiv1$ on ${C}\cup\underline{{C}}$, we will say it is \textit{lapse-normalised}. 
\end{remark}
\begin{remark}
    \label{rmk:residual gauge freedom}
    Fixing $\Omega^2\equiv1$ on ${C}\cup\underline{{C}}$ does not exhaust all gauge freedom. In fact, the lapse-normalised condition is invariant under gauge transformations of the form $u\mapsto\gamma u$, $v\mapsto\gamma^{-1} v$, for any $\gamma \in \mathbb R_{>0}$. 
\end{remark}

We now note that, for lapse-normalised bifurcate initial data sets, not only the transversal derivatives of $r$, $\Omega^2$ and $\phi$ can be computed from their value at the bifurcation sphere using the Einstein-scalar field system, but rather the full initial data set can be recovered if one knows $\phi$ and the values of the first $u$- and $v$-derivatives of $r$ at the bifurcation sphere, as the following Proposition shows:

\begin{proposition}
    \label{prop:compute from phi}
    Given $\mathring{\phi}\in C^{k}({C}\cup\underline{{C}})$, and real numbers $\mathring{\nu}$, $\mathring{\lambda}$, and $\mathring{r}>0$, there exists a unique $C^k$ lapse-normalised bifurcate initial data set $(r, \Omega^2, \phi)$ on ${C}\cup\underline{{C}}$ with $\phi\equiv\mathring{\phi}$ and 
    \begin{equation}
        r({C}\cap\underline{{C}})=\mathring{r}, \quad \partial_ur({C}\cap\underline{{C}})=\mathring{\nu}, \quad \partial_vr({C}\cap\underline{{C}})=\mathring{\lambda}. 
    \end{equation}
\end{proposition}
\begin{proof}
    We set $\Omega^2\equiv1$ on ${C}\cup\underline{{C}}$ (cf. Remark \ref{rmk:lapse-normalised characteristic initial data}). Abusing the notation, we will call the new double null coordinates $(u,v)$, and assume that the bifurcation sphere is located at $(u_0,v_0)$. 
    
    Now, taking $\phi=\mathring{\phi}$, Raychaudhuri's equation along ${C}$ reads
    \begin{equation}
        \partial_v^2r=-r\big(\partial_v\mathring{\phi}\big)^2. 
    \end{equation}
    This is a second order ODE for $r(v)$, which has a unique $C^{k+1}$ solution with initial data 
    \begin{equation}
        r(v_0)=\mathring{r}, \quad \partial_vr(v_0)=\mathring{\lambda}. 
    \end{equation}
    Similarly, we obtain $r(u)$ solving Raychaudhuri's equation along $\underline{{C}}$ with initial data  
    \begin{equation}
        r(u_0)=\mathring{r}, \quad \partial_ur(u_0)=\mathring{\nu}. 
    \end{equation}
\end{proof}
\begin{remark}
    \label{rmk:everything from phi}
    Combining the results of Proposition \ref{prop:compute u-derivatives from initial data} and Proposition \ref{prop:compute from phi}, we conclude that, as long as we are willing to work with a lapse-normalised initial data set, once we know $\phi$ on ${C}\cup\underline{{C}}$ and $r$, $\partial_ur$, $\partial_vr$ at the bifurcation sphere ${C}\cap\underline{{C}}$, we obtain $r$, $\Omega^2$ and $\phi$ and all their $u$- and $v$-derivatives everywhere on ${C}\cup\underline{{C}}$. This fact will become very useful when we address the characteristic gluing problem. 
\end{remark}

\section{Characteristic gluing for the Einstein-scalar field system}
\label{sec:characteristic gluing}

Our approach to constructing solutions exhibiting dynamical formation of the Schwarzschild--de Sitter and Nariai spacetimes begins with a study of the characteristic gluing problem for the Einstein-scalar field system in spherical symmetry. This section is hence devoted to describing the set-up for such a problem and introducing the formalism needed to state and prove our characteristic gluing results in the following sections. 

Although the discussion here will focus on the gluing problem along a constant-$u$ cone, by the symmetry of the Einstein-scalar field system under transformations of the form $u\mapsto v$, $v\mapsto u$, everything we will say is immediately translatable to the gluing along a constant-$v$ cone. 

\subsection{Description of the characteristic gluing problem}
\label{subsec:description gluing}

The study of the characteristic gluing problem for the Einstein equations was initiated in $3+1$ dimensions in the asymptotically flat setting by Aretakis, Czimek and Rodnianski \cite{Aretakis_Czimek_Rodnianski_2023, Aretakis_Czimek_Rodnianski_2024, Aretakis_Czimek_Rodnianski_2025}. These results were later extended to the case where $\Lambda\neq0$ and to higher dimensions by Chru\'{s}ciel, Cong and Gray \cite{Chrusciel_Cong_2024,Chrusciel_Cong_Gray_2024,Chrusciel_Cong_Gray_2026}. All these works deal with the Einstein equations outside spherical symmetry, but are limited to the perturbative regime. We, however, will work in spherical symmetry, which makes it possible to tackle large-data problems that would be considerably harder to tackle in the general, non-spherically symmetric setting. In this framework, which is the analogue of that in \cite{Kehle_Unger_2025}, one sets up the characteristic gluing problem as follows.

Let us consider two spherically symmetric solutions to the Einstein-scalar field system, $\big(\mathcal{Q}_1,r_1, \Omega^2_1, \phi_1\big)$ and $\big(\mathcal{Q}_2,r_2, \Omega^2_2, \phi_2\big)$, endowed with double null coordinates $(u,v)$. We will moreover assume that these solutions are at least $C^k$-regular, with $k\in \mathbb N$. Rescaling the double null coordinate pairs on $\mathcal{Q}_1$ and $\mathcal{Q}_2$ if necessary, we assume that
\begin{align}
    \mathcal{Q}_1&=[u_0, u_1]\times [v_{-1},v_0],\\
    \mathcal{Q}_2&=[u_{-1},u_0]\times [v_1,v_2].
\end{align}
The $C^k$ gluing problem addresses the question of whether there exists a $C^k$ spherically symmetric solution to the Einstein-scalar field system $\big(\mathcal{Q}, r, \Omega^2, \phi\big)$, equipped with a double null pair extending the coordinate functions $u,v$ introduced before, such that $\mathcal{Q}_1,\mathcal{Q}_2\subset\mathcal{Q}$, and $r|_{\mathcal{Q}_1}=r_1$, $r|_{\mathcal{Q}_2}=r_2$, $\Omega^2|_{\mathcal{Q}_1}=\Omega^2_1$, $\Omega^2|_{\mathcal{Q}_2}=\Omega^2_2$, $\phi|_{\mathcal{Q}_1}=\phi_1$, $\phi|_{\mathcal{Q}_2}=\phi_2$.

\begin{figure}[H]
        \centering
        \includegraphics[width=0.57\textwidth]{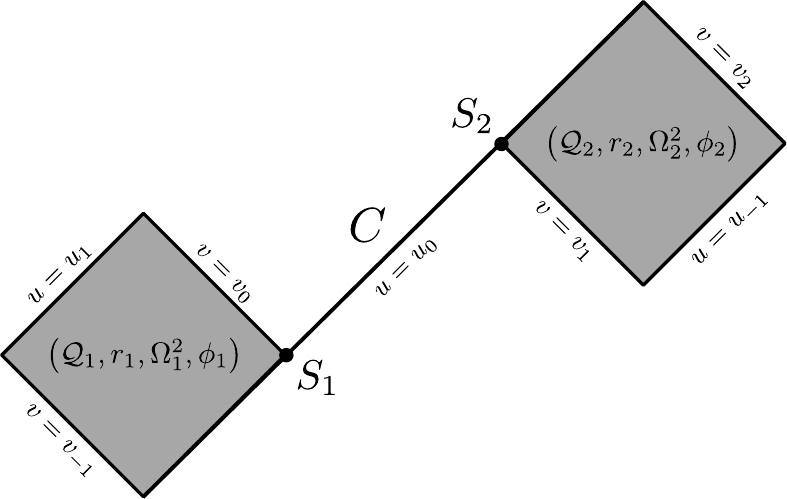}
        \caption{Diagram illustrating the set-up for the characteristic gluing of two spherically symmetric solutions to the Einstein-scalar field system $\big(\mathcal{Q}_1,r_1, \Omega^2_1, \phi_1\big)$ and $\big(\mathcal{Q}_2,r_2, \Omega^2_2, \phi_2\big)$ along a null cone ${C}=\{u=u_0\}$.} 
        \label{fig:gluing setup intro}
\end{figure}

In view of the local existence results for the characteristic initial value problem for the Einstein-scalar field system (see Proposition \ref{prop:local existence and uniqueness characteristic IVP}), the characteristic problem ultimately reduces to a question about characteristic data on the null cone ${C}\doteq~\{u=u_0\}$, namely, whether there exist characteristic data along ${C}$ inducing, on the spheres $S_1$ and $S_2$ (see Figure \ref{fig:gluing setup intro}), values for $r$, $\Omega^2$, $\phi$ and as many $u$- and $v$-derivatives of them as are consistent with $C^k$-regularity, that agree with the values induced by the solutions $\big(\mathcal{Q}_1,r_1, \Omega^2_1, \phi_1\big)$ and $\big(\mathcal{Q}_2,r_2, \Omega^2_2, \phi_2\big)$, respectively. 

\subsection{Formulation of the characteristic gluing problem in spherical symmetry}

In order to make precise the ideas described in the previous section, we introduce here the notions of \textit{sphere data, cone data} and \textit{seed data}. These will allow us to provide, at the end of this section, a concrete formulation of the characteristic gluing problem in spherical symmetry.   

\subsubsection{Sphere data and gauge freedom}

An idea that has already been mentioned in our description of the characteristic gluing problem is that any $C^k$ solution to the Einstein-scalar field system $(\mathcal{Q}, r, \Omega^2, \phi)$ induces, on each sphere of symmetry $(u,v)\in\mathcal{Q}$, a collection of real numbers corresponding to the values of the metric functions, the scalar field, and their $u$- and $v$- derivatives on that sphere: 
\begin{align}
    r(u,v)>0, \, \, \, \partial_u^ir(u,v), \, \, \,\partial_v^ir(u,v) \, \,  &\text{ for } \, i=1, \dots, k+1,\\[0.5em]
    \Omega^2(u,v)>0, \, \,  \partial_u^i\Omega^2(u,v), \, \,   \partial_v^i\Omega^2(u,v) \, \,  &\text{ for } \, i=1, \dots, k,\\[0.5em]
    \phi(u,v), \, \, \,  \partial_u^i\phi(u,v), \, \, \, \partial_v^i\phi(u,v) \, \,  &\text{ for } \,  i=1, \dots, k. 
\end{align}
These numbers are subject to the constraint equations \eqref{eq:EE massive scalar field - Raychaudhuri's v}, \eqref{eq:EE massive scalar field - Raychaudhuri's u} and to the identities for higher order derivatives that one can derive by formally differentiating them. The notion of a sphere data set precisely formalises this idea:  
\begin{definition}[\protect{Sphere data, cf. \cite[Definition 2.3]{Kehle_Unger_2025}}]
    Let $k\in \mathbb N$. A \textit{sphere data set with regularity index} $k$ for the Einstein-scalar field system is a list of real numbers:
    \begin{itemize}
        \item $\varrho>0$, \, $\varrho^i_u$, \,$\varrho^i_v$ \, for $i=1, \dots, k+1$,
        \item $\omega>0$, \, $\omega^i_u$, \, $\omega^i_v$  \, for $i=1, \dots, k$,
        \item $\varphi$, \, $\varphi^i_u$, \, $\varphi^i_v$, \, for $i=1, \dots, k$. 
    \end{itemize}
    which moreover satisfy: 
    \begin{enumerate}[(i)]
        \item $\varrho^{i+2}_u$ can be expressed as a function of $\varrho^{j}_u$, $\omega^{j}_u$ and  $\varphi^{j}_u$ for $0 \leq j \leq i+1$ by formally differentiating \eqref{eq:EE massive scalar field - Raychaudhuri's u}, 
        \item $\varrho^{i+2}_v$ can be expressed as a function of $\varrho^{j}_v$, $\omega^{j}_v$ and  $\varphi^{j}_v$, for $0 \leq j \leq i+1$ by formally differentiating \eqref{eq:EE massive scalar field - Raychaudhuri's v}, 
    \end{enumerate}
    where we are assuming $\varrho^0_u=\varrho^0_v=\varrho$, $\omega^0_u=\omega^0_v=\omega$, and $\varphi^0_u=\varphi^0_v=\varphi$. We will denote by $\mathcal{D}_k$ the set of all sphere data sets for the Einstein-scalar field system with regularity index $k$. 
\end{definition}

\begin{remark}
    We will normally refer to the components of a sphere data set $D\in\mathcal{D}_k$ by $\varrho(D)$, $\omega(D)$, $\varphi^1_u(D)$, etc., although we will sometimes drop the reference to $D$ when there is no risk of ambiguity. 
\end{remark}

From our discussion in Section \ref{sec:Einstein-scalar field system}, it becomes clear the importance of gauge freedom in the study of the Einstein-scalar field system, and this will also be the case at the level of sphere data. Let us then analyse how gauge transformations act on sphere data. Given $D\in\mathcal{D}_k$, we start looking for functions $r\in C^{k+1}(\mathbb R^2)$, $\Omega^2 \in C^{k}(\mathbb R^2)$, $\phi\in C^{k}(\mathbb R^2)$, such that 
\begin{align}
    r(0,0)=\varrho, \, \, \, \partial^i_ur(0,0)&=\varrho^i_u, \, \, \, \partial^i_vr(0,0)=\varrho^i_v,\\[0.5em]
    \Omega^2(0,0)=\omega, \, \, \, \partial^i_u\Omega^2(0,0)&=\omega^i_u, \, \, \, \partial^i_v\Omega^2(0,0)=\omega^i_v,\\[0.5em]
    \phi(0,0)=\varphi, \, \, \, \partial^i_u\phi(0,0)&=\varphi^i_u, \, \, \, \partial^i_v\phi(0,0)=\varphi^i_v.  
\end{align}
Now, we consider the \textit{group of gauge transformations} for the spherically symmetric Einstein-scalar field system that fix the origin, 
\begin{equation}
\mathcal{G}\doteq\big\{(f,h):f,h \in \mathrm{Diff}_+(\mathbb R) \textrm{ and } f(0)=h(0)=0 \big\}. 
\end{equation}
Thus $\sigma=(f,h)\in \mathcal{G}$ will act on $r$, $\Omega^2$, $\phi$, producing $\overline{r}, \overline{\Omega}^2, \overline{\phi}$, functions of the double null pair $(\bar{u},\bar{v})$, given by:
    \begin{align}
        \overline{r}(\bar{u},\bar{v})&=r\big(f^{-1}(\bar{u}),h^{-1}(\bar{v})\big), \label{eq:gauge transformation 1}\\
        \overline{\Omega}^2(\bar{u},\bar{v})&=\Omega^2\big(f^{-1}(\bar{u}),h^{-1}(\bar{v})\big)(f^{-1})'(\bar{u})(h^{-1})'(\bar{v}), \label{eq:gauge transformation 2}\\
        \overline{\phi}(\bar{u}, \bar{v})&=\phi\big(f^{-1}(\bar{u}), h^{-1}(\bar{v})\big). \label{eq:gauge transformation 3}
    \end{align}
In order to find the components of the transformed sphere data set, $\sigma D$, we evaluate the expressions for $\overline{r}, \overline{\Omega}^2, \overline{\phi}$ and their $\bar{u}$- and $\bar{v}$-derivatives, which we can obtain by formally differentiating \eqref{eq:gauge transformation 1}, \eqref{eq:gauge transformation 2}, \eqref{eq:gauge transformation 3}, at $\bar{u}=\bar{v}=0$, for example:
\begin{align}
    \varrho(\sigma D)=& \, \, \overline{r}(0,0)=\varrho(D), \\
    \omega( \sigma D)=& \, \, \overline{\Omega}^2(0,0)=\tfrac{1}{f'(0)h'(0)}\omega(D),\\
    \varrho^1_u(\sigma D)=& \, \, \partial_{\bar{u}}\overline{r} (0,0)=\tfrac{1}{f'(0)}\varrho^1_u(D),\\
    \varrho^1_v(\sigma D)=& \, \, \partial_{\bar{v}}\overline{r} (0,0)=\tfrac{1}{h'(0)}\varrho^1_v(D).
\end{align}

In relation to this, we introduce the notion of gauge equivalence for sphere data sets: 

\begin{definition}[Gauge equivalent sphere data sets]
    Let $D_1, D_2 \in \mathcal{D}_k$ be two sphere data sets. We will say that $D_1$ is \textit{gauge equivalent} to $D_2$, and denote it by $D_1\sim D_2$, if there exists $\sigma\in \mathcal{G}$ such that $D_1= \sigma D_2$.
\end{definition}

We have already discussed how gauge freedom allows us to fix the value of $\Omega^2$ to be constant on a given $C^k$ bifurcate initial data set, giving rise to the notion of lapse-normalised bifurcate characteristic initial data. Note that such data will have, at the bifurcation sphere, $\Omega^2=1$ and $\partial^i_u \Omega^2=0=\partial^i_v \Omega^2$ for $1\leq i \leq k$. Viewed in the context of sphere data sets, this fact motivates the following definition:

\begin{definition}[\protect{Lapse-normalised sphere data set, cf. \cite[Def. 2.5]{Kehle_Unger_2025}}]
    We will say that a sphere data set $D\in\mathcal{D}_k$ is \textit{lapse-normalised} if it has $\omega=1$ and $\omega^i_u=0=\omega^i_v$ for all $1\leq i \leq k$. 
\end{definition}

\begin{remark}
    Note that, in view of Remark \ref{rmk:lapse-normalised characteristic initial data}, and of how gauge transformations act on sphere data sets, any sphere data set is gauge equivalent to a lapse-normalised sphere data set. 
\end{remark}

Because it will be useful later, we also introduce now the definition of the renormalised Hawking mass for sphere data sets:

\begin{definition}[Renormalised Hawking mass of a sphere data set] 
    \label{def:renormalised Hawking mass sphere data set}
    We define the \textit{renormalised Hawking mass of a sphere data set} $D\in\mathcal{D}_k$ as
\begin{equation}
     \widetilde{m}(D)\doteq  \frac{\varrho(D)}{2}\bigg(1+\frac{4\varrho^1_u(D)\varrho^1_v(D)}{\omega(D)}-\frac13\Lambda \varrho(D)^2\bigg). 
\end{equation}
\end{definition}

\begin{remark}
    \label{rmk:mass invariance of sphere data}
    In view of our previous discussion about how the sphere data set components transform under the action of the group $\mathcal{G}$, we deduce that the renormalised Hawking mass of a sphere data set is invariant under gauge transformations. 
\end{remark}

\subsubsection{Cone data, seed data, and definition of characteristic gluing for sphere data sets}
\label{subsec:definition characteristic gluing}

We have already made precise how a spherically symmetric solution to the Einstein-scalar field system induces data on its symmetry spheres. Now, given two spheres along a null cone ${C}$, with data induced by two different solutions respectively, the question is whether there exists another spherically symmetric solution to the Einstein-scalar field system that induces these same data on the two spheres. Understanding this requires considering not only the data induced by a solution on spheres, but also the characteristic data that a solution would induce along the cone ${C}$, and whether this is compatible with the two prescribed sphere data sets. With this in mind, we introduce the notion of a \textit{cone data set}:

\begin{definition}[\protect{Cone data set, cf. \cite[Definition 2.6]{Kehle_Unger_2025}}]
    Let $k\in\mathbb N$ and $v_0, v_1 \in \mathbb R$ with $v_0<v_1$. A $C^k$ \textit{cone data set} for the spherically symmetric Einstein-scalar field system is a continuous map $D:[v_0,v_1]\rightarrow \mathcal{D}_k$ such that, upon identifying 
    \begin{align}
        r(u_0, v)&= \varrho\big(D(v)\big),\\
        \Omega^2(u_0, v)&=\omega\big(D(v)\big),\\
        \phi(u_0, v)&= \varphi\big(D(v)\big),
    \end{align}
    for all $v \in [v_0,v_1]$ and similarly for the $u$- and $v$-derivatives, with $u_0\in \mathbb R$, we have that $(r, \Omega^2, \phi)$ satisfies the Einstein-scalar field system along the cone ${C}=\{u_0\}\times[v_0,v_1]$. 
\end{definition}

Now we are ready to give a precise definition of characteristic gluing: 

\begin{definition}[\protect{Characteristic gluing, cf. \cite[Definition 3.1]{Kehle_Unger_2025}}]
    \label{def:characteristic gluing}
    Let $k\in\mathbb N$, and $D_0, D_1 \in \mathcal{D}_k$. We say that $D_0$ can be characteristically glued to $D_1$ to order $k$ within the Einstein-scalar field system in spherical symmetry if there exist $v_0<v_1$ and a $C^k$ cone data set $D:[v_0,v_1]\rightarrow \mathcal{D}_k$ such that $D(v_0)$ is gauge equivalent to $D_0$ and $D(v_1)$ is gauge equivalent to $D_1$.
\end{definition}

Therefore, the problem of characteristic gluing between two spheres in spherical symmetry boils down to being able to produce a suitable cone data set. In view of Proposition \ref{prop:compute u-derivatives from initial data}, we can produce a unique cone data set $D:[v_0,v_1]\rightarrow \mathcal{D}_k$ for the spherically symmetric Einstein-scalar field system given $D(v_0)$ and $r\in C^{k+1}\big([v_0,v_1]\big)$, $\Omega^2\in C^{k}\big([v_0,v_1]\big)$, $\phi\in C^{k}\big([v_0,v_1]\big)$. If we moreover impose that the cone data set is lapse-normalised, that is it has $\omega(D(v))=1$ for all $v\in[v_0,v_1]$, then it can be uniquely produced from $D(v_0)$ and $\phi\in C^{k}\big([v_0,v_1]\big)$ alone, in view of Proposition \ref{prop:compute from phi} and Remark \ref{rmk:everything from phi}. This motivates the definition of \textit{seed data}: 

\begin{definition}[\protect{Seed data,  cf. \cite[Proposition 2.4]{Kehle_Unger_2025}}] 
    Let $k\in\mathbb N$, $v_0, v_1 \in \mathbb R$ with $v_0<v_1$, and $D_0\in \mathcal{D}_k$ a lapse-normalised sphere data set. We say that $\phi\in C^{k}\big([v_0,v_1]\big)$ is \textit{seed data}, or simply a \textit{seed}, associated to the sphere data set $D_0$ if $\partial^i_v\phi(v_0)=\varphi^i_v(D_0)$, for $0\leq i \leq k$.  
\end{definition}

Hence, given a lapse-normalised $D_0\in \mathcal{D}_k$ and an associated seed $\phi\in C^{k}\big([v_0,v_1]\big)$, there exists a unique cone data set $D:[v_0,v_1]\rightarrow \mathcal{D}_k$ for the spherically symmetric Einstein-scalar field system such that $D(v_0)=D_0$, and $\omega\big(D(v)\big)=1$, $\varphi\big(D(v)\big)=\phi(v)$ for all $v\in[v_0,v_1]$. 

Going back to the characteristic gluing problem as stated in Definition \ref{def:characteristic gluing}, this tells us that, assuming the sphere $D_0$ to be lapse-normalised (which is always possible, since we can apply a gauge transformation to make it lapse-normalised), the characteristic gluing problem reduces to finding a seed $\phi\in C^{k}\big([v_0,v_1]\big)$ such that the unique lapse-normalised cone data set it generates has $D(v_1)$ gauge equivalent to $D_1$. This is precisely the view of the characteristic gluing problem that we shall keep in mind henceforth.  

\subsubsection{From sphere data gluing to spacetime gluing}

In Section \ref{subsec:description gluing}, we introduced the characteristic gluing problem at the level of spacetimes and claimed that, as a consequence of the local existence results for the characteristic initial value problem for the Einstein-scalar field system, it could be reduced to a question about gluing characteristic data along a null cone. We now justify this claim by showing that, if the sphere data sets induced by two spherically symmetric solutions can be glued in the sense of Definition \ref{def:characteristic gluing}, then the two solutions can be glued in the sense we gave in Section \ref{subsec:description gluing}.

\begin{proposition}[\protect{\cite[Proposition 3.1]{Kehle_Unger_2025}}]
    \label{prop:spacetime gluing}
    Let $\big(\mathcal{Q}_1,r_1, \Omega^2_1, \phi_1\big)$ and $\big(\mathcal{Q}_2,r_2, \Omega^2_2, \phi_2\big)$ be two spherically symmetric $C^k$ solutions to the Einstein-scalar field system, with 
    \begin{align}
        \mathcal{Q}_1&=[u_{0,1}, u_{1,1}]\times [v_{0,1},v_{1,1}],\\
        \mathcal{Q}_2&=[u_{0,2},u_{1,2}]\times [v_{0,2},v_{1,2}].
    \end{align} 
    Let $D_1$ be the sphere data set induced by $\big(\mathcal{Q}_1,r_1, \Omega^2_1, \phi_1\big)$ on $(u_{0,1}, v_{1,1})$ and $D_2$ be the sphere data set induced by $\big(\mathcal{Q}_2,r_2, \Omega^2_2, \phi_2\big)$ on $(u_{1,2}, v_{0,2})$. If $D_1$ can be characteristically glued to $D_2$ to order $k$ in the sense of Definition \ref{def:characteristic gluing}, then there exists a spherically symmetric $C^k$ solution of the Einstein-scalar field system, $\big(\mathcal{Q}, r, \Omega^2, \phi\big)$ that, upon being endowed with a global double null coordinate pair $(u,v)$, contains the double null rectangles 
    \begin{equation}
        \begin{aligned}
            \mathcal{R}_1&=[u_0, u_1]\times [v_{-1},v_0],\\
            \mathcal{R}_2&=[u_{-1},u_0]\times [v_1,v_2],
        \end{aligned} 
        \label{eq:regions spacetime gluing}
    \end{equation}
    with $\big(\mathcal{R}_i, r|_{\mathcal{R}_i}, \Omega^2|_{\mathcal{R}_i}, \phi|_{\mathcal{R}_i}\big)$ isometric to $\big(\mathcal{Q}_i, r_i, \Omega^2_i, \phi_i\big)$ for $i = 1, 2$, and moreover the sphere data set induced on $(u_0, v_0)$ is equal to $D_1$ to $k$-th order and the sphere data set induced on $(u_0, v_1)$ is gauge equivalent to $D_2$ to $k$-th order.
\end{proposition}

\begin{proof}
    The strategy for proving the proposition is the same as that given in the proof of \cite[Proposition 3.1]{Kehle_Unger_2025}, but we shall also outline it here for convenience. 
    If $D_1$ can be characteristically glued to $D_2$, then there exist $v_0<v_1$ and $D:[v_0,v_1]\rightarrow \mathcal{D}_k$ such that $D(v_0)\sim D_1$ and $D(v_1)\sim D_2$, i.e. there exist gauge transformations $\sigma_1,\sigma_2\in\mathcal{G}$ (possibly trivial) such that $\sigma_1 D_1$ is equal to $D(v_0)$ and $\sigma_2 D_2$ is equal to $D(v_1)$.
    
    We apply the gauge transformation $\sigma_1$ to the double null coordinates on $\mathcal{Q}_1$ and rescale them further so that finally $\mathcal{Q}_1=[u_0, u_1]\times [v_{-1},v_0]$. We also apply the gauge transformation $\sigma_2$ to the double null coordinates on $\mathcal{Q}_2$, and further rescale them so that finally $\mathcal{Q}_2=[u_{-1},u_0]\times [v_1,v_2]$. 

    As a consequence of the characteristic gluing, we also know there exist functions $r\in C^{k+1} \big([v_0,v_1]\big)$, $\Omega^2\in C^k \big([v_0,v_1]\big)$, $\phi\in C^k \big([v_0,v_1]\big)$, which can be thought of as satisfying the Einstein-scalar field system along the cone $\{u=u_0\}$. Now we solve the characteristic initial value problem on the bifurcate null hypersurface ${C}\cup\underline{{C}}$, with
    \begin{equation}
        \begin{aligned}
            {C}=\{u_0\}\times[v_0,v_2],\\
            \underline{{C}}=[u_0, u_1]\times\{v_0\}.
        \end{aligned}
    \end{equation}
    Along $\underline{{C}}$, we take as characteristic data the values of $r$, $\Omega^2$, $\phi$ induced by the solution $\big(\mathcal{Q}_1,r_1, \Omega^2_1, \phi_1\big)$. 
    Along ${C}$, we take $r\in C^{k+1} \big([v_0,v_1]\big)$, $\Omega^2\in C^k \big([v_0,v_1]\big)$, $\phi\in C^k \big([v_0,v_1]\big)$ as provided by the characteristic gluing procedure, and extend them by the values induced by the solution $\big(\mathcal{Q}_2,r_2, \Omega^2_2, \phi_2\big)$ on $\{u_0\}\times[v_1,v_2]$. In view of Proposition \ref{prop:local existence and uniqueness characteristic IVP}, we will obtain a spherically symmetric solution to the Einstein-scalar field system on a thin slab $\mathcal{U}$ to the future of ${C}\cup\underline{{C}}$ of the form $[u_0,u_0+\delta]\times[v_0,v_2]$, for some $\delta>0$. 
   
    We can repeat the procedure solving the characteristic initial value problem to the past on $\widetilde{{C}}\cup\widetilde{\underline{{C}}}$, with
    \begin{equation}
        \begin{aligned}
            \widetilde{{C}}=\{u_0\}\times[v_{-1},v_1],\\
            \widetilde{\underline{{C}}}=[u_{-1}, u_0]\times\{v_1\},
        \end{aligned}
    \end{equation}
    hence obtaining a solution on a thin slab $\widetilde{\mathcal{U}}=[u_0-\tilde{\delta},u_0]\times[v_{-1},v_1]$, for some $\tilde{\delta}>0$. 

    Considering now the solution defined on $\mathcal{Q}=\mathcal{Q}_1\cup\mathcal{Q}_2\cup\mathcal{U}\cup\widetilde{\mathcal{U}}$, we observe that, by virtue of the gluing procedure, the characteristic initial value problem, and how we have chosen our characteristic data, we will have 
    $r\in C^{k+1}(\mathcal{Q})$, $\Omega^2\in C^k (\mathcal{Q})$, $\phi\in C^k (\mathcal{Q})$. It is clear now that for $\mathcal{R}_1$, $\mathcal{R}_2 \subset \mathcal{Q}$ given by \eqref{eq:regions spacetime gluing}, we will have that $\big(\mathcal{R}_i, r|_{\mathcal{R}_i}, \Omega^2|_{\mathcal{R}_i}, \phi|_{\mathcal{R}_i}\big)$ is isometric to $\big(\mathcal{Q}_i, r_i, \Omega^2_i, \phi_i\big)$ for $i = 1, 2$. Note that, by construction, the sphere data set induced by $\big(\mathcal{Q}, r, \Omega^2, \phi\big)$ on $(u_0, v_0)$ is equal to $D_1$ to $k$-th order, and the sphere data set induced on $(u_0, v_1)$ is gauge equivalent to $D_2$ (in fact it is equal to $\sigma_2 D_2$) to $k$-th order.
\end{proof}

\section{The spherically symmetric vacuum solutions and their sphere data}
\label{subsec:vacuum solutions}
In this section we give a more detailed overview of the geometry of the family of spherically symmetric solutions to the vacuum Einstein equations with $\Lambda>0$, with the aim of ultimately introducing the concrete definitions of the sphere data sets that will be relevant to our gluing results.

\subsection{The Schwarzschild--de Sitter family}

As seen in our discussion in Section \ref{subsec:SdS family of solutions introduction}, the mass $M$ of a member in the Schwarzschild--de Sitter family is the parameter that allows us to classify such spacetime as belonging to the subextremal, extremal or superextremal regimes. In what follows, we assume $M\geq0$, and for convenience we shall consider, instead of the mass, the renormalised parameter
\begin{equation}
    \mathfrak{c}\doteq3M\sqrt{\Lambda},
    \label{eq:def extremality index}
\end{equation} 
which we will refer to as \textit{extremality index}. We recall the line element \eqref{eq:line element}, which written in terms of $\mathfrak{c}$ reads 
\begin{equation}
    g^{_\mathrm{{SdS}}}_{\mathfrak{c}}=-f_{\mathfrak{c}}(r)dt^2+\big(f_{\mathfrak{c}}(r)\big)^{-1}dr^2+r^2 g_{\mathbb S^2} \, , 
    \label{eq:generalsolution}
\end{equation}
where the metric coefficients function $f_{\mathfrak{c}}$ (formerly $f_M$) is given by 
\begin{equation}
    f_{\mathfrak{c}}(r)=1-\frac2r\frac{\mathfrak{c}}{3\sqrt{\Lambda}}-\frac13\Lambda r^2. 
\end{equation}

Note that for any symmetry sphere in a Schwarzschild--de Sitter spacetime of extremality index $\mathfrak{c}$, we have 
\begin{equation}
    \widetilde{m}=\frac{\mathfrak{c}}{3\sqrt{\Lambda}}, 
\end{equation}
that is, the renormalised Hawking mass coincides with the mass of the spacetime everywhere. 

\subsubsection{de Sitter spacetime}

When $\mathfrak{c}=0$, $f_0(r)$ has one single positive root, $r=\sqrt{3/\Lambda}$, which determines the location of the cosmological horizon, and the spacetime has no black hole horizon in this case. This is de Sitter spacetime \cite{deSitter_1918}. More precisely, the line element \eqref{eq:generalsolution} with $\mathfrak{c}=0$ describes the \textit{static region} of de Sitter spacetime, where $0\leq~r~<~\sqrt{3/\Lambda}$, $f_0(r)$ is positive and hence the Killing vector field $K\doteq \partial_t$ is timelike. 

\begin{figure}[H]
        \centering
        \includegraphics[width=0.35\textwidth]{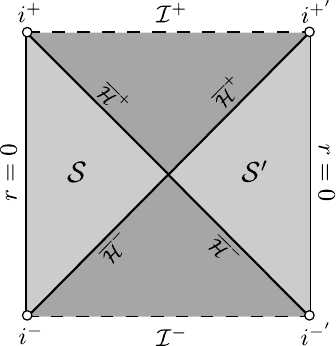}
        \caption{Penrose diagram of de Sitter spacetime.} 
        \label{fig:deSitter}
\end{figure}

Upon a change of coordinates, this line element can be analytically extended beyond the cosmological horizon to describe the \textit{past and future cosmological regions}, where $r>r_{\overline{\mathcal{H}}}(0)$ and $K$ is spacelike, and \textit{another static region}. The result is a globally hyperbolic and geodesically complete spacetime with global $\mathbb R \times \mathbb S^3$ topology, whose Penrose diagram looks as shown in Figure \ref{fig:deSitter}. The conformal boundary of the spacetime features \textit{future null infinity}, denoted $\mathcal{I}^+$, which is spacelike due to the presence of $\Lambda>0$, and constitutes the place where all future-directed causal geodesics whose radius grows without bound end up in the infinite future. Similarly with \textit{past null infinity}, $\mathcal{I}^-$. The points $i^+$, ${i^+}'$ are the future endpoints, in the conformal completion, of every orbit of the static timelike Killing field $K$ in the respective static region; similarly for $i^-$, ${i^-}'$ with the past endpoints of the static orbits\footnote{Note that $i^+$, ${i^+}'$(respectively $i^-$, ${i^-}'$) do not have here the same interpretation as in the asymptotically flat case, since timelike geodesics could, in general, terminate at $\mathcal{I}^+$ (resp. initiate at $\mathcal{I}^-$); hence we shall not refer to $i^+$, $i^-$, etc., as timelike infinities here.}. 

The static regions (shaded in light grey in Figure \ref{fig:deSitter}) are foliated by untrapped spheres. We shall assume global double null coordinates such that the static region on the left side of the diagram, $\mathcal{S}$, has $\partial_ur<0$, $\partial_vr>0$ everywhere up to and including the regular centre $r=0$. Thus the static region on the right, $\mathcal{S}'$, will have $\partial_ur>0$, $\partial_vr<0$. 

The \textit{future cosmological region} is foliated by antitrapped spheres with $\partial_ur>0$, $\partial_vr>0$, hence why sometimes we will refer to it as the \textit{expanding cosmological region}. This region is bounded by the constant-$v$ and constant-$u$ components of the future cosmological event horizon, denoted $\overline{\mathcal{H}}^+$, which are foliated by marginally antitrapped spheres (spheres on the constant-$v$ component have $\partial_vr>0$ whereas those on the constant-$u$ component have $\partial_ur>0$). 

The \textit{past cosmological region}, sometimes also called \textit{contracting cosmological region}, is foliated by trapped spheres with $\partial_ur<0$, $\partial_vr<0$, and bounded by the past cosmological event horizon, $\overline{\mathcal{H}}^-$, foliated by marginally trapped spheres. 

\subsubsection{The subextremal Schwarzschild--de Sitter spacetime}

For $0< \mathfrak{c}<1$, the metric $g^{_\mathrm{{SdS}}}_{\mathfrak{c}}$ describes the subextremal Schwarzschild--de Sitter spacetime of extremality index $\mathfrak{c}$ in static coordinates. In this range of $\mathfrak{c}$, $f_{\mathfrak{c}}$ has two positive roots, $r_{\mathcal{H}}$ and $r_{\overline{\mathcal{H}}}$, corresponding to the black hole and cosmological horizons, respectively. Seen as functions of $\mathfrak{c}$, these take the explicit form: 
\begin{equation}
    r_{\mathcal{H}}(\mathfrak{c})= \frac{2}{\sqrt{\Lambda}}\cos\bigg(\frac{\cos^{-1}(-\mathfrak{c})}{3}-\frac{2\pi}{3}\bigg),
    \label{eq:eventhorizon}
\end{equation}
\begin{equation}
    r_{\overline{\mathcal{H}}}(\mathfrak{c})=\frac{2}{\sqrt{\Lambda}}\cos\bigg(\frac{\cos^{-1}(-\mathfrak{c})}{3}\bigg).
    \label{eq:cosmologicalhorizon}
\end{equation}
The function $f_{\mathfrak{c}}(r)$ is positive for $r_{\mathcal{H}}(\mathfrak{c})<r<r_{\overline{\mathcal{H}}}(\mathfrak{c})$, and negative for $r<r_{\mathcal{H}}(\mathfrak{c})$ and $r>r_{\overline{\mathcal{H}}}(\mathfrak{c})$. We note that, for $\mathfrak{c}\in(0,1)$, $r_{\mathcal{H}}$ is a strictly increasing function of $\mathfrak{c}$, while $r_{\overline{\mathcal{H}}}$ is a strictly decreasing function of $\mathfrak{c}$. Moreover they satisfy: 
\begin{equation}
    0< r_{\mathcal{H}}(\mathfrak{c})<\frac{1}{\sqrt{\Lambda}}< r_{\overline{\mathcal{H}}}(\mathfrak{c})<\sqrt{\frac{3}{\Lambda}},
    \label{eq:inequality horizons radii subext SdS}
\end{equation}

\begin{figure}[H]
    \centering
    \includegraphics[width=0.65\textwidth]{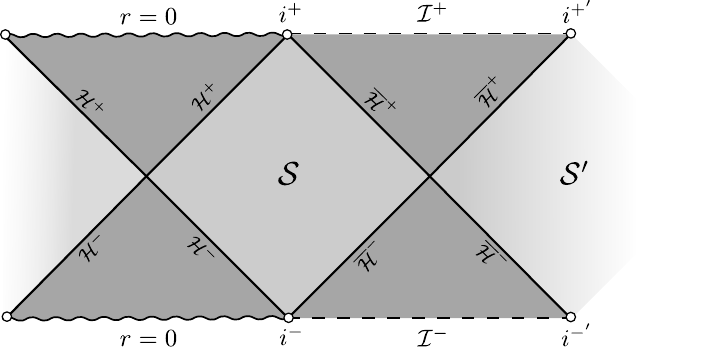}
    \caption{Penrose diagram for the subextremal Schwarzschild--de Sitter spacetime.} 
    \label{fig:subextremalSdS}
\end{figure}

The line element \eqref{eq:line element} can be analytically extended not only beyond the cosmological event horizon at $r_{\overline{\mathcal{H}}}(\mathfrak{c})$, but also beyond the black hole event horizon at $r_{\mathcal{H}}(\mathfrak{c})$. The resulting spacetime, which has global $\mathbb R\times\mathbb R\times\mathbb S^2$ topology, is globally hyperbolic but not geodesically complete, and consists of an infinite sequence of black hole, white hole, static and cosmological regions that repeats indefinitely in both horizontal directions. The repeating unit of this sequence is plotted in the Penrose diagram of Figure~\ref{fig:subextremalSdS}. We now describe the regions displayed in that diagram; the same description applies to every other copy.

The static regions are foliated by untrapped spheres. As in de Sitter spacetime, we choose double null coordinates such that the spheres in $\mathcal{S}$ have $\partial_ur<0$, $\partial_vr>0$, and the spheres in $\mathcal{S}'$ have $\partial_ur>0$, $\partial_vr<0$. The future (resp. past) cosmological region is foliated by antitrapped (resp. trapped) spheres, and extends to future null infinity $\mathcal{I}^+$ (resp. past null infinity $\mathcal{I}^-$). The boundary points $i^+$, ${i^+}'$, $i^-$, ${i^-}'$ are also present in this case, again with the same interpretation as in de Sitter spacetime.

The \textit{black hole region} is defined in the usual way as the region from which no future-directed causal curve reaches $\mathcal{I}^+$, that is, the region which is not in the causal past of $\mathcal{I}^+$. Its interior is foliated by trapped spheres, and it is bounded to the past by the black hole event horizon $\mathcal{H}^+$ (whose constant-$v$ component has $\partial_ur=0$, $\partial_vr<0$, and whose constant-$u$ component has $\partial_ur<0$, $\partial_vr=0$), and to the future by the spacelike singularity at $r=0$. The \textit{white hole region} is defined analogously, as the region which cannot be reached from $\mathcal{I}^-$ by a future-directed causal curve, that is, the region not in the causal future of $\mathcal{I}^-$. All spheres in its interior are antitrapped, and it is bounded to the future by the white hole event horizon $\mathcal{H}^-$, which is foliated by marginally antitrapped spheres, and to the past by the spacelike singularity at $r=0$.

As mentioned in Section \ref{subsec:SdS family of solutions introduction}, it is possible to construct a spatially compact version of the subextremal Schwarzschild--de Sitter spacetime with global $\mathbb R \times \mathbb S^1 \times \mathbb S^2$ topology. We highlight that our results do not depend on the global spatial topology assumed for the Schwarzschild--de Sitter spacetime. 

\subsubsection{The extremal Schwarzschild--de Sitter spacetime}

For $\mathfrak{c}=1$, $f_1(r)$ has just one double positive root at $r_{\mathcal{H}}(1)=r_{\overline{\mathcal{H}}}(1)=1/{\sqrt{\Lambda}}$, and it is negative everywhere else on $(0,\infty)$. In this case, the line element \eqref{eq:line element} describes a globally hyperbolic, geodesically incomplete spacetime that we refer to as \textit{extremal Schwarzschild--de Sitter} spacetime. As discussed in Section \ref{subsec:SdS family of solutions introduction}, this spacetime can be endowed with two different time orientations. We highlight here the main geometric features of each of them, and refer the reader to \cite{Podolsky_1999} for a detailed analysis of this solution. As in the subextremal case, the Penrose diagrams below depict the repeating unit of the spacetime, and the same description applies to every other copy.

\subsubsection*{The extremal Schwarzschild--de Sitter black hole spacetime}
This spacetime features black hole and contracting cosmological regions, both foliated by trapped spheres. These two alternating types of regions are separated by the \textit{extremal black hole event horizon}, foliated by marginally trapped spheres. See the resulting Penrose diagram in Figure \ref{subfig:extremalSdS black hole}. 

The future boundary of the spacetime includes the $r=0$ spacelike singularity and the point $\mathcal{P}$, which is the future endpoint of inextendible causal geodesics whose radius asymptotically approaches $r=1/\sqrt{\Lambda}$ in the infinite future. Note that, although there are future-directed causal geodesics in this spacetime that escape from the black hole and exist for infinite affine time, the geometry severely restricts their asymptotic radius (in particular, there are no future-directed causal geodesics whose radius grows unboundedly)\footnote{This allows us to adapt the definition of the black hole region to our setting: in the absence of future null infinity, it is the region of spacetime that is not in the causal past of the boundary point $\mathcal{P}$, which can be thought of as the future endpoint of observers who ``live'' forever.}.

\vfill

\newpage 

The past boundary of the spacetime includes past null infinity $\mathcal{I}^-$ and the point $\mathcal{P}'$, the past endpoint of inextendible causal geodesics whose radius asymptotically approaches $r=1/\sqrt{\Lambda}$ in the infinite past.

\subsubsection*{The extremal Schwarzschild--de Sitter white hole spacetime}

\begin{figure}[H]
        \centering
        \hspace{-4em}
        \begin{subfigure}[c]{0.4\textwidth}
            \centering
            \includegraphics[width=1.2\textwidth]{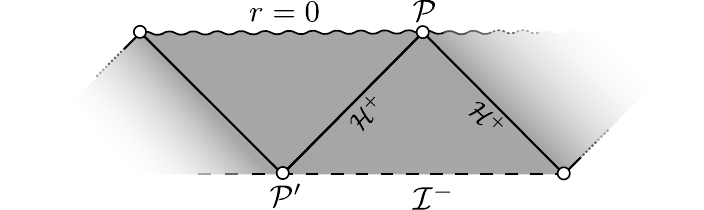}
            \caption{}
            \label{subfig:extremalSdS black hole}
        \end{subfigure}
        \hspace{4.0em}
        \begin{subfigure}[c]{0.4\textwidth}
            \centering
            \includegraphics[width=1.2\textwidth]{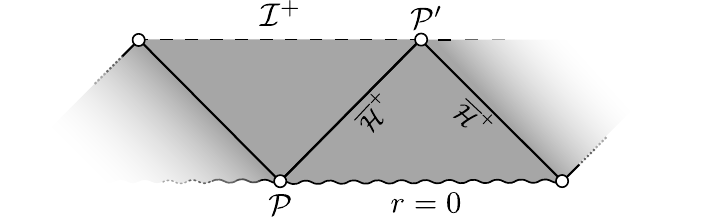}
            \caption{}
            \label{subfig:extremalSdS white hole}
        \end{subfigure}
        \caption{Penrose diagrams of: (a) the extremal Schwarzschild--de Sitter black hole spacetime; (b) the extremal Schwarzschild--de Sitter white hole spacetime.} 
        \label{fig:extremalSdS}
\end{figure}

This spacetime, which is the time-reverse of the extremal Schwarzschild--de Sitter black hole spacetime, includes white hole and expanding cosmological regions, both foliated by antitrapped spheres, and separated by a horizon foliated by marginally antitrapped spheres that we will refer to as the \textit{extremal cosmological event horizon}. In this case the boundary points $\mathcal{P}$ and $\mathcal{P}'$ are also present, and their interpretation is, as expected, the time reverse of the one they have in the black hole case. The Penrose diagram of this spacetime is depicted in Figure \ref{subfig:extremalSdS white hole}.

As mentioned already in the discussion about the subextremal Schwarzschild--de Sitter spacetime, our results will be independent of the choice of global topology for the extremal Schwarzschild--de Sitter spacetime, which in principle is $\mathbb R \times \mathbb R \times \mathbb S^2$, although a spatially compact alternative is also possible.

\subsubsection{The superextremal Schwarzschild--de Sitter spacetime} 

If $\mathfrak{c}>1$, $f_{\mathfrak{c}}(r)$ is negative for all $r\in(0,\infty)$, and $g_{\mathfrak{c}}^{\textrm{SdS}}$ describes the superextremal Schwarzschild--de Sitter spacetime, already introduced in Section \ref{subsec:SdS family of solutions introduction}. Although this spacetime is not directly relevant to our results, for the sake of completeness we briefly discuss the geometry of its two possible time orientations.

\begin{figure}[H]
    \centering
    \begin{subfigure}[c]{0.4\textwidth}
        \centering
        \includegraphics[width=0.6\textwidth]{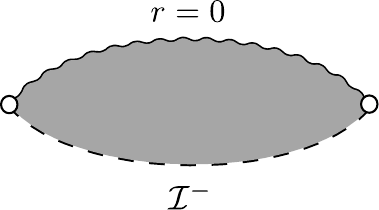}
        \caption{}
        \label{subfig:superextremalSdS black hole}
    \end{subfigure}
    \hspace{2em}
    \begin{subfigure}[c]{0.4\textwidth}
        \centering
        \includegraphics[width=0.6\textwidth]{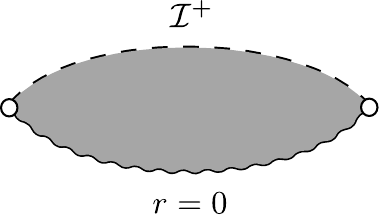}
        \caption{}
        \label{subfig:superextremalSdS white hole}
    \end{subfigure}
    \caption{Penrose diagrams of: (a) the future-singularity superextremal Schwarzschild--de Sitter spacetime; (b) the past-singularity superextremal Schwarzschild--de Sitter spacetime.} 
    \label{fig:superextremalSdS}
\end{figure}

\paragraph{The future-singularity superextremal
Schwarzschild--de Sitter spacetime.} This spacetime is foliated by trapped spheres and contracts from past null infinity $\mathcal{I}^-$ to a global spacelike singularity at $r=0$ (see the corresponding Penrose diagram in Figure \ref{subfig:superextremalSdS black hole}). 

\paragraph{The past-singularity superextremal
Schwarzschild--de Sitter spacetime.} This is the time reverse of the future-singularity spacetime. Accordingly, in this case all symmetry spheres are antitrapped, and the spacetime expands from a spacelike singularity at $r=0$ to future null infinity $\mathcal{I}^+$ (see Figure \ref{subfig:superextremalSdS white hole}). 

\vfill

\subsection{The Nariai spacetime}

As argued in Section~\ref{subsec:SdS family of solutions introduction}, the Schwarzschild--de Sitter metric~\eqref{eq:line element} does not describe all spherically symmetric solutions to the vacuum Einstein equations with $\Lambda>0$. In order to get the complete set of solutions with non-negative renormalised Hawking mass, we need to include the Nariai spacetime~\cite{Nariai_1950,Nariai_1951}. This is a globally hyperbolic, geodesically complete spacetime with global $\mathbb R \times \mathbb R \times \mathbb S^2$ topology (a version with global $\mathbb R\times\mathbb S^1\times\mathbb S^2$ topology is also possible). We refer back to Figure~\ref{fig:Nariai intro} for the Penrose diagram of this spacetime. 

The most distinctive feature of this spacetime, at least from the point of view of the results presented in this paper, is that it has constant area-radius, $r = 1/\sqrt{\Lambda}$, and hence $\partial_ur=0=\partial_vr$ everywhere. Despite having $r = 1/\sqrt{\Lambda}$, the future (resp. past) boundary of the spacetime lies at infinite affine distance: it is reached by future-directed (resp. past-directed) geodesics in infinite affine time.

To motivate why Nariai should be regarded as an extremal solution, let us note that, for any symmetry sphere in the Nariai spacetime, 
\begin{equation}
    \widetilde{m}=\frac{1}{2\sqrt{\Lambda}}\Bigg(1-\frac13\Lambda\bigg(\frac{1}{\sqrt{\Lambda}}\bigg)^2\Bigg)=\frac{1}{3\sqrt{\Lambda}},
\end{equation}
so the Nariai solution has the same renormalised Hawking mass as the extremal Schwarzschild--de Sitter solution, and, as mentioned at the beginning of this section, the mass is precisely the parameter that distinguishes the subextremal, extremal and superextremal regimes. Moreover, as pointed out in Section~\ref{subsec:SdS family of solutions introduction}, the connection between the extremal Schwarzschild--de Sitter spacetime and the Nariai spacetime was first made explicit by Ginsparg and Perry \cite{Ginsparg_Perry_1983}: they showed that zooming in on the region between the black hole and cosmological horizons of the subextremal Schwarzschild--de Sitter solution, and then taking the limit to the extremal case, yields the Nariai metric.

\subsection{Sphere data in de Sitter, (sub)extremal Schwarzschild--de Sitter, and Nariai}

Now that we have outlined the main geometric features of the different Schwarzschild–de Sitter spacetimes and of Nariai, we are in a position to introduce the precise definitions of the sphere data sets that our gluing theorems will be concerned with. We emphasise already that more definitions will be required than in \cite{Kehle_Unger_2025}, given the wider variety of spacetime regions and event horizons. Nonetheless, we aim to keep the list of definitions as short as possible, including only those that are directly relevant to our study. 

Following our discussion in Section \ref{subsec:definition characteristic gluing}, we will work with lapse-normalised sphere data sets (of regularity
index $k\in \mathbb N$). Moreover, since the sphere data sets we are interested in correspond to vacuum solutions, they will all have
\begin{equation}
    \varphi^i_u = \varphi^i_v = 0 \, \, \, \text{ for } \,  i=0, \dots, k,  
\end{equation}
which, in view of the definition of sphere data sets, also implies
\begin{equation}
    \hspace{1em} \varrho^i_u = \varrho^i_v = 0 \, \, \, \text{ for } \,  i=2, \dots, k+1.  
\end{equation}
Therefore, the only non-trivial components (besides $\omega=1$) will be $\varrho$, $\varrho^1_u$ and $\varrho^1_v$. The residual gauge freedom after setting the lapse-normalised condition (recall Remark \ref{rmk:residual gauge freedom}) allows us to make transformations of the form 
\begin{equation}
    u\mapsto\gamma u, \quad v\mapsto\gamma^{-1} v \quad \text{ for } \,  \gamma \in \mathbb R_{>0}.  
    \label{eq:residual gauge transf}
\end{equation}
In view of how gauge transformations act on sphere data sets, we deduce that transformations of the form \eqref{eq:residual gauge transf} will leave $\varrho$ invariant, and change the other non-trivial components according to
\begin{equation}
    \varrho^1_u\mapsto\gamma^{-1} \varrho^1_u, \quad \varrho^1_v\mapsto\gamma \varrho^1_v.
    \label{eq:residual gauge transf on derivatives}
\end{equation} 

Since our formulation of the characteristic gluing problem (Definition \ref{def:characteristic gluing}) cares about the sphere data sets only up to gauge transformations, our definitions of sphere data sets will also be up to gauge transformations, and hence will be concerned with the signs of $ \varrho^1_u$, $ \varrho^1_v$ rather than their specific values, given that, in light of \eqref{eq:residual gauge transf on derivatives}, their signs are what is truly gauge-invariant. We shall also include in our definitions the renormalised Hawking mass of the sphere data set, as introduced in Definition \ref{def:renormalised Hawking mass sphere data set}, which is also invariant under gauge transformations (see Remark \ref{rmk:mass invariance of sphere data}).

\begin{definition}[de Sitter sphere data] 
    \label{def:dS sphere data}
    Let $k \in \mathbb N$. 
    \vspace{0.5em}
    \begin{enumerate}[(a)]
        \item Given $0<R<\sqrt{3/\Lambda}$, we define a \textit{de Sitter static sphere data set of radius $R$}, denoted $_{\mathcal{S}}D^{\mathrm{dS}}_{R,k}$, to be a lapse-normalised sphere data set satisfying: 
        \begin{itemize}
            \item $\varrho=R$,
            \item $\varrho^1_u\varrho^1_v<0$,
            \item $1+4\varrho^1_u\varrho^1_v-\frac13\Lambda \varrho^2 =0$,
            \item all other components are zero.
        \end{itemize}
        \vspace{0.5em}
        \item Given $R>\sqrt{3/\Lambda}$, we define a \textit{de Sitter cosmological sphere data set of radius $R$}, denoted $_{\mathcal{C}}D^{\mathrm{dS}}_{R,k}$, to be a lapse-normalised sphere data set satisfying: 
        \begin{itemize}
            \item $\varrho=R$,
            \item $\varrho^1_u\varrho^1_v>0$,
            \item $1+4\varrho^1_u\varrho^1_v-\frac13\Lambda \varrho^2 =0$,
            \item all other components are zero.
        \end{itemize}
    \end{enumerate}
\end{definition}

\vspace{0.5em}

\begin{remark}
    The above definition identifies (up to a gauge transformation) a de Sitter static or cosmological sphere if moreover the sign of $\varrho^1_u$, or equivalently the sign of $\varrho^1_v$, is specified:
    \vspace{0.5em}
    \begin{itemize}
        \item If a sphere data set $_{\mathcal{S}}D^{\mathrm{dS}}_{R,k}$ has $\varrho^1_u<0$, it corresponds to a sphere in the region we denoted $\mathcal{S}$ in the Penrose diagram of de Sitter spacetime (we refer back to Figure \ref{fig:deSitter}). If, on the contrary, it has $\varrho^1_u>0$, then the sphere data set corresponds to a sphere in region $\mathcal{S}'$. 
        \vspace{0.5em}
        \item Similarly, if $_{\mathcal{C}}D^{\mathrm{dS}}_{R,k}$ has $\varrho^1_u>0$, then it corresponds to a sphere in the future cosmological region, and if it has $\varrho^1_u<0$, to a sphere in the past cosmological region. 
    \end{itemize}
\end{remark}   

\vspace{0.5em}

\begin{definition}[Schwarzschild--de Sitter sphere data]
    \label{def:SdS sphere data} 
    Let $k \in \mathbb N$ and $\mathfrak{c}\in (0,1]$.
    \vspace{0.5em}
    \begin{enumerate}[(a)]
         \item For $r_{\mathcal{H}}(\mathfrak{c})<R<r_{\overline{\mathcal{H}}}(\mathfrak{c})$, we define a \textit{Schwarzschild--de Sitter static sphere data set of extremality index $\mathfrak{c}$ and radius $R$}, denoted $_{\mathcal{S}}D^{\mathrm{SdS}}_{\mathfrak{c},R,k}$, to be a lapse-normalised sphere data set satisfying: 
        \begin{itemize}
            \item $\varrho=R$
            \item $\varrho^1_u\varrho^1_v<0$,
            \item $\frac12\varrho\Big(1+4\varrho^1_u\varrho^1_v-\frac13\Lambda \varrho^2\Big)=\frac{\mathfrak{c}}{3\sqrt{\Lambda}}$,
            \item all other components are zero.
        \end{itemize} 
        \vspace{0.5em}
        \item Given $R>r_{\overline{\mathcal{H}}}(\mathfrak{c})$, we define a \textit{Schwarzschild--de Sitter cosmological sphere data set of extremality index $\mathfrak{c}$ and radius $R$}, denoted $_{\mathcal{C}}D^{\mathrm{SdS}}_{\mathfrak{c},R,k}$, to be a lapse-normalised sphere data set satisfying: 
        \begin{itemize}
            \item $\varrho=R$,
            \item $\varrho^1_u\varrho^1_v>0$,
            \item $\frac12\varrho\Big(1+4\varrho^1_u\varrho^1_v-\frac13\Lambda \varrho^2\Big)=\frac{\mathfrak{c}}{3\sqrt{\Lambda}}$,
            \item all other components are zero.
        \end{itemize}
        \vspace{0.5em}
        \item We define a \textit{Schwarzschild--de Sitter black hole horizon sphere data set of extremality index $\mathfrak{c}$}, denoted $_{\mathcal{H}^+}D^{\mathrm{SdS}}_{\mathfrak{c},k}$, to be a lapse-normalised sphere data set satisfying: 
        \begin{itemize}
            \item $\varrho=r_{\mathcal{H}}(\mathfrak{c})=\frac{2}{\sqrt{\Lambda}}\cos\Big(\frac{\cos^{-1}(-\mathfrak{c})}{3}-\frac{2\pi}{3}\Big)$,
            \item $\big\{\text{sgn}(\varrho^1_u),\text{sgn}(\varrho^1_v)\big\}=\{0,-\}$, or $\varrho^1_u=\varrho^1_v=0$,
            \item all other components are zero.
        \end{itemize} 
        \vspace{0.5em}
        \item We define a \textit{Schwarzschild--de Sitter cosmological horizon sphere data set of extremality index $\mathfrak{c}$}, denoted $_{\overline{\mathcal{H}}^+}D^{\mathrm{SdS}}_{\mathfrak{c},k}$, to be a lapse-normalised sphere data set satisfying: 
        \begin{itemize}
            \item $\varrho =r_{\overline{\mathcal{H}}}(\mathfrak{c}) =\frac{2}{\sqrt{\Lambda}}\cos\Big(\frac{\cos^{-1}(-\mathfrak{c})}{3}\Big)$,
            \item $\big\{\text{sgn}(\varrho^1_u),\text{sgn}(\varrho^1_v)\big\}=\{0,+\}$, or $\varrho^1_u=\varrho^1_v=0$,
            \item all other components are zero.
        \end{itemize} 
    \end{enumerate}
\end{definition}

\vspace{0.2em}

\begin{remark}
    Note that parts (a), (b) and (d) of this definition can be extended to the case $\mathfrak{c}=0$, then coinciding with the definitions of the de Sitter sphere data sets introduced in Definition \ref{def:dS sphere data}, that is, 
    \begin{equation}
        _{\mathcal{S}}D^{\mathrm{SdS}}_{0,R,k}=\, _{\mathcal{S}}D^{\mathrm{dS}}_{R,k}
        \quad \quad \quad 
        _{\mathcal{C}}D^{\mathrm{SdS}}_{0,R,k}=\, _{\mathcal{C}}D^{\mathrm{dS}}_{R, k} \quad \quad \quad
        _{\overline{\mathcal{H}}^+}D^{\mathrm{SdS}}_{0,k}=\, _{\overline{\mathcal{H}}^+}D^{\mathrm{dS}}_{k}.
    \end{equation}
\end{remark}

\begin{remark}
    If $\mathfrak{c}<1$, the sphere data sets introduced in Definition \ref{def:SdS sphere data} correspond to subextremal Schwarzschild--de Sitter spheres, and moreover:  
    \begin{enumerate}[i)]
        \item Part (c) of Definition \ref{def:SdS sphere data} identifies a Schwarzschild--de Sitter black hole horizon sphere if either $\varrho^1_u=\varrho^1_v=0$, and hence the sphere is the bifurcating sphere of the black hole horizon, or, in the case that $\varrho^1_u,\varrho^1_v$ are not both zero, we additionally specify which one is negative: if $_{\mathcal{H}^+}D^{\mathrm{SdS}}_{\mathfrak{c},k}$ has $\varrho^1_u<0$, then it corresponds to a sphere on the constant-$u$ component of $\mathcal{H}^+$, whereas if it has $\varrho^1_v<0$, then it identifies a sphere on the constant-$v$ component of $\mathcal{H}^+$ (we refer back to the Penrose diagram in Figure \ref{fig:subextremalSdS}). 
        \item Similarly, part (d) of Definition \ref{def:SdS sphere data} corresponds to a unique Schwarzschild--de Sitter cosmological horizon sphere if either $\varrho^1_u=\varrho^1_v=0$ or, if $\varrho^1_u,\varrho^1_v$ are not both zero, we additionally specify which one is positive. 
        \item Parts (a) and (b) of Definition \ref{def:SdS sphere data} point to Schwarzschild--de Sitter static and cosmological spheres respectively if additionally the sign of $\varrho^1_u$ or $\varrho^1_v$ is specified, just as for de Sitter sphere data sets in Definition \ref{def:dS sphere data}.
    \end{enumerate}
\end{remark}

\begin{remark}
    When $\mathfrak{c}=1$, the sphere data sets introduced in Definition \ref{def:SdS sphere data} correspond to extremal Schwarzschild--de Sitter (or Nariai) spheres. In this case: 
    \begin{enumerate}[i)]
        \item There will be no sphere data sets satisfying Part (a) of Definition \ref{def:SdS sphere data}, given that $r_{\mathcal{H}}(1)=r_{\overline{\mathcal{H}}}(1)=1/{\sqrt{\Lambda}}$. 
        \item The sphere data sets in part (b) of Definition \ref{def:SdS sphere data} will correspond to spheres in the extremal black hole or white hole spacetimes depending on the sign of $\varrho^1_u$  (equivalently $\varrho^1_v$): if $_{\mathcal{C}}D^{\mathrm{SdS}}_{1,R,k}$ has $\varrho^1_u<0$, it corresponds to a sphere in the past cosmological region of the extremal Schwarzschild--de Sitter black hole spacetime, whilst if it has $\varrho^1_u>0$, it corresponds to a sphere in the future cosmological region of the extremal Schwarzschild--de Sitter white hole spacetime. 
        \item The sphere data sets $_{\mathcal{H}^+}D^{\mathrm{SdS}}_{1,k}$ with $\big\{\text{sgn}(\varrho^1_v),\text{sgn}(\varrho^1_u)\big\}=\{0,-\}$ correspond to spheres on the extremal black hole horizon present in the extremal Schwarzschild--de Sitter black hole spacetime. 
        \item The sphere data sets $_{\overline{\mathcal{H}}^+}D^{\mathrm{SdS}}_{1,k}$ with $\big\{\text{sgn}(\varrho^1_v),\text{sgn}(\varrho^1_u)\big\}=\{0,+\}$ correspond to spheres on the extremal cosmological horizon present in the extremal Schwarzschild--de Sitter white hole spacetime.
        \item The sphere data sets $_{\overline{\mathcal{H}}^+}D^{\mathrm{SdS}}_{1,k}$ or $_{\mathcal{H}^+}D^{\mathrm{SdS}}_{1,k}$ with $\varrho^1_u=\varrho^1_v=0$ correspond to spheres in the Nariai spacetime\footnote{This agrees with the well-known fact that extremal horizons of solutions to the Einstein equations cannot have bifurcating spheres, as $\partial_u r=\partial_v r =0$ leads to constant area-radius $r$ in the solution.}. We will normally refer to these spheres as $D^{\mathrm{Nar}}_k$ instead.
    \end{enumerate}
\end{remark}

\begin{remark}
    Note that
    \begin{align}
            \widetilde{m}\Big(\, _{\overline{\mathcal{H}}^+}D^{\mathrm{SdS}}_{\mathfrak{c},k}\Big)&=\frac{r_{\overline{\mathcal{H}}}(\mathfrak{c})}{2}\bigg(1-\frac13\Lambda \big(r_{\overline{\mathcal{H}}}(\mathfrak{c})\big)^2\bigg)=\\[0.5em]&=\frac{1}{\sqrt{\Lambda}}\bigg(\cos\bigg(\frac{\cos^{-1}(-\mathfrak{c})}{3}\bigg)-\frac43\cos^3\bigg(\frac{\cos^{-1}(-\mathfrak{c})}{3}\bigg)\bigg)=\\[0.5em]&=\frac{1}{\sqrt{\Lambda}}\bigg(\cos\bigg(\frac{\cos^{-1}(-\mathfrak{c})}{3}\bigg)-\frac13\cos\big(\cos^{-1}(-\mathfrak{c})\big) -\cos\bigg(\frac{\cos^{-1}(-\mathfrak{c})}{3}\bigg)\bigg)=\frac{\mathfrak{c}}{3\sqrt{\Lambda}}
    \end{align}

    \vspace{0.5em}

    \noindent where we have used a rearrangement of the triple angle identity:
    \begin{equation}
        \cos^3(\vartheta)=\frac{\cos(3\vartheta)+3\cos({\vartheta})}{4}.
    \end{equation}

    An analogous computation with the same result holds for $\widetilde{m}\Big(\, _{\mathcal{H}^+}D^{\mathrm{SdS}}_{\mathfrak{c},k}\Big)$. This is why there is no need to prescribe an extra condition related to the renormalised Hawking mass for these sphere data sets.
    \label{rmk:mass horizon spheres}
\end{remark} 

\section{The gluing theorems}
\label{sec:gluing results}
Having introduced the notion of characteristic gluing and the definitions of the relevant sphere data sets, we are ready to state our gluing results for the Einstein-scalar field system in spherical symmetry. For the sake of clarity, in this section we will only include the statements of the various gluing theorems, and their proofs will be presented separately in Section \ref{sec:proofs}. 

Although all the results are stated here for the gluing of two spheres along a constant-$u$ cone, we note that they can be translated into results along a constant-$v$ cone by simply mapping $v\mapsto u$, $u\mapsto v$. The analogues of some of the results presented here along a constant-$v$ cone will be essential when, in Section \ref{sec:applications}, we construct Cauchy data on a hypersurface $\Sigma \cong \mathbb S^3$. 

\subsection{Black hole horizon gluing}
\label{subsec:black hole horizon gluing}

The first result we establish regarding the gluing to the Schwarzschild--de Sitter black hole event horizon is the following:

\begin{customthm}{B0}
    \label{thm:dS to BHhorizon}
    For any $k\in \mathbb N$, $\mathfrak{m}\geq0$, $\mathfrak{c}\in (0,1]$ and $0< R_i <r_{\mathcal{H}}(\mathfrak{c})$, the de Sitter static sphere of radius $R_i$, $_{\mathcal{S}}D^{\mathrm{dS}}_{R_i,k}$, with $\varrho^1_u<0$, can be characteristically glued to the Schwarzschild--de Sitter black hole horizon sphere of extremality index $\mathfrak{c}$,\,  $_{\mathcal{H}^+}D^{\mathrm{SdS}}_{\mathfrak{c}, k}$, with $\varrho^1_u<0$, to order $k$ within the Einstein-scalar field model with scalar field mass $\mathfrak{m}$. 
\end{customthm}

This result would show that it is possible to glue a de Sitter static sphere to the black hole horizon of a Schwarzschild--de Sitter spacetime of any mass, up to and including the extremal case. However, we shall not prove the statement in Theorem \ref{thm:dS to BHhorizon} as it is, but rather the following extended version of it: 

\begin{customthm}{B1}
    \label{thm:subext SdS to BHhorizon}
    For any $k\in \mathbb N$, $\mathfrak{m}\geq0$, $\mathfrak{c}_1,\mathfrak{c}_2\in[0,1]$, with $\mathfrak{c}_1<\mathfrak{c}_2$, and $r_{\mathcal{H}}(\mathfrak{c}_1)<R_i <r_{\mathcal{H}}(\mathfrak{c}_2)$, the Schwarzschild--de Sitter static sphere of extremality index $\mathfrak{c}_1$ and radius $R_i$,\, $_{\mathcal{S}}D^{\mathrm{SdS}}_{\mathfrak{c}_1, R_i,k}$, with $\varrho^1_u<0$, can be characteristically glued to the Schwarzschild--de Sitter black hole horizon sphere of extremality index $\mathfrak{c}_2$, $_{\mathcal{H}^+}D^{\mathrm{SdS}}_{\mathfrak{c}_2, k}$, with $\varrho^1_u<0$, to order $k$ within the Einstein-scalar field model with scalar field mass $\mathfrak{m}$.
\end{customthm}

The result covers, in particular, the gluing from a de Sitter static sphere to a subextremal Schwarzschild--de Sitter black hole horizon sphere, which can be seen as the analogue of \cite[Thm. 2A]{Kehle_Unger_2025}, and the gluing from a subextremal Schwarzschild--de Sitter static sphere to an extremal Schwarzschild--de Sitter black hole horizon sphere (cf. \cite[Thm. 2B]{Kehle_Unger_2025}). This will be crucial to construct third-law-violating solutions, for which we want to ensure that an extremal black hole forms from an already existing subextremal one, which, in turn, was formed previously by gravitational collapse.    

The proof of Theorem \ref{thm:subext SdS to BHhorizon}, which is a straightforward generalisation of the proof of \cite[Thm. 2A]{Kehle_Unger_2025} to include a positive cosmological constant and to allow for a positive scalar field mass, is given in Section \ref{subsec:proof subext SdS to BHhorizon}. 

\subsection{Cosmological horizon gluing}
\label{subsec:cosmological horizon gluing}

The first result we prove regarding the gluing to the cosmological event horizon solely concerns the subextremal case, and it reads as follows: 

\begin{customthm}{C1}
   \label{thm:dS to subext COSMhorizon} 
    For any $k\in \mathbb N$, $\mathfrak{m}\geq0$, $\mathfrak{c}\in [0,1)$ and $0<R_i <r_{\overline{\mathcal{H}}}(\mathfrak{c})$, the de Sitter static sphere of radius $R_i$, $_{\mathcal{S}}D^{\mathrm{dS}}_{R_i, k}$, with $\varrho^1_u<0$, can be characteristically glued to the Schwarzschild--de Sitter cosmological horizon sphere of extremality index $\mathfrak{c}$,\,  $_{\overline{\mathcal{H}}^+}D^{\mathrm{SdS}}_{\mathfrak{c}, k}$, with $\varrho^1_u>0$, to order $k$ within the Einstein-scalar field model with scalar field mass $\mathfrak{m}$.
\end{customthm}

The fundamental difference with respect to the problem of gluing to the black hole event horizon is that, while spheres on the black hole horizon are marginally trapped, those on the cosmological horizon are marginally antitrapped. Therefore, although both in Theorem \ref{thm:subext SdS to BHhorizon} and Theorem \ref{thm:dS to subext COSMhorizon} we glue from static spheres with $\varrho^1_v>0$, $\varrho^1_u<0$ to horizon spheres with $\varrho^1_v=0$, in the black hole case the horizon sphere will have $\varrho^1_u<0$, while the cosmological horizon sphere will have $\varrho^1_u >0$. Ensuring this change of sign in $\varrho^1_u$ at the final sphere makes the gluing to cosmological horizon spheres significantly more intricate than the one to black hole horizon spheres.

In fact, as we will see in Section \ref{subsec:proof dS to subext COSMhorizon}, the strategy to prove Theorem \ref{thm:dS to subext COSMhorizon} is exclusive to the subextremal case and does not generalise to the extremal case. Nevertheless, it is reasonable to wonder if it is possible to extend the statement in Theorem \ref{thm:dS to subext COSMhorizon} to allow for $\mathfrak{c}=1$,  even if the proof would require a different approach. However, the analogue of Theorem \ref{thm:dS to BHhorizon} for the cosmological horizon could never be true, at least not for any scalar field mass $\mathfrak{m}\geq0$, as our next result illustrates:

\begin{customthm}{I}
    \label{thm:impossible gluing}
    Let $\big(\mathcal{Q},r,\Omega^2,\phi\big)$ be a spherically symmetric solution to the Einstein-\underline{massless} scalar field system with $\Lambda>0$ endowed with double null coordinates $(u,v)$. Consider ${C}\doteq\{u=u_0\}\cap\mathcal{Q}$ a null cone in $\mathcal{Q}$. Suppose that the symmetry sphere $(u_0,v_1)\in{C}$ has $\varrho=1/{\sqrt{\Lambda}}$, $\varrho^1_v=0$ and $\varrho^1_u\geq0$. Then, it is not possible for any of the spheres along ${C}\cap\{v<v_1\}$ to be gauge equivalent to a subextremal Schwarzschild--de Sitter or de Sitter static sphere with $\varrho^1_u<0$. 
\end{customthm}

The proof of Theorem \ref{thm:impossible gluing}, which relies solely on monotonicity properties imposed by Raychaudhuri's equation and the wave equation for $r$, is presented in Section \ref{subsec:proof impossible gluing}.

Theorem \ref{thm:impossible gluing} in itself is not a characteristic gluing result, but it is connected to the characteristic gluing problem in the sense that it rules out the possibility of gluing, already at the $C^0$ level, a de Sitter or subextremal Schwarzschild--de Sitter static sphere to an extremal cosmological horizon sphere within the spherically symmetric Einstein-massless scalar field model. Nevertheless, we show that gluing to the extremal cosmological horizon spheres is possible, provided that the scalar field mass is sufficiently large: 

\begin{customthm}{C2} 
    \label{thm:subext SdS to ext COSMhorizon}
    For any $k\in \mathbb N$ and $\mathfrak{m}>\sqrt{\Lambda}/2$, there exist $\tilde{\mathfrak{c}}\in(0,1)$ and $r_{\mathcal{H}}(\tilde{\mathfrak{c}})< R <1/{\sqrt{\Lambda}}$ such that the Schwarzschild--de Sitter static sphere of radius $R$ and extremality index $\tilde{\mathfrak{c}}$,\, $_{\mathcal{S}}D^{\mathrm{SdS}}_{\tilde{\mathfrak{c}},R, k}$, with $\varrho^1_u<0$, can be characteristically glued to the extremal Schwarzschild--de Sitter cosmological horizon sphere, $_{\overline{\mathcal{H}}^+}D^{\mathrm{SdS}}_{1, k}$, with $\varrho^1_u>0$, to order $k$ within the Einstein--massive scalar field model with scalar field mass $\mathfrak{m}$.
\end{customthm}

\subsection{Gluing to Nariai spheres}
\label{subsec:Nariai gluing}

Given that Nariai spheres have $\varrho=1/{\sqrt{\Lambda}}$ and $\varrho^1_u=0=\varrho^1_v$, Theorem \ref{thm:impossible gluing} also forbids the gluing to a Nariai sphere from a de Sitter or subextremal Schwarzschild--de Sitter static sphere with a massless scalar field. However, it can be shown that the gluing to Nariai spheres is possible, provided that the scalar field mass is sufficiently large, as was the case with the extremal cosmological horizon:
\begin{customthm}{N}
    \label{thm:subext SdS to Nariai}
    For any $k\in \mathbb N$ and $\mathfrak{m}>\sqrt{\Lambda}/2$, there exist $\tilde{\mathfrak{c}}\in(0,1)$ and $r_{\mathcal{H}}(\tilde{\mathfrak{c}})< R <1/{\sqrt{\Lambda}}$ such that the Schwarzschild--de Sitter static sphere of radius $R$ and extremality index $\tilde{\mathfrak{c}}$,\, $_{\mathcal{S}}D^{\mathrm{SdS}}_{\tilde{\mathfrak{c}},R, k}$, with $\varrho^1_u<0$, can be characteristically glued to the Nariai sphere, $D^{\mathrm{Nar}}_k$, to order $k$ within the Einstein--massive scalar field model with scalar field mass $\mathfrak{m}$.
\end{customthm}

\begin{remark}
    \label{rmk:possible gluign dS to ext cosm horizon or Nariai}
    The gluings to the extremal cosmological horizon and to Nariai, as presented in Theorem \ref{thm:subext SdS to ext COSMhorizon} and Theorem \ref{thm:subext SdS to Nariai}, are performed from a subextremal Schwarzschild--de Sitter sphere with $\tilde{\mathfrak{c}}>0$, so that, in particular, the gluing from a de Sitter static sphere is not covered by these results. Nevertheless, one could combine these gluings with a previous gluing from the static region of de Sitter to the static region of subextremal Schwarzschild--de Sitter to complete the results and make them more similar to Theorem \ref{thm:subext SdS to BHhorizon}. 
\end{remark}

\section{Proofs of the gluing theorems}
\label{sec:proofs}

We provide in this section the proofs for the results introduced in Section \ref{sec:gluing results}. We start describing the basic set-up and outlining the general features that all the proofs will have in common, to then go on to study the details of each of the constructions separately. 

\subsection{Set-up and general strategy}
\label{subsec:proofs general strategy}

The set-up for all the results proved in this section will be the same, and the same that we already introduced when we first presented the characteristic gluing problem in Section \ref{subsec:definition characteristic gluing}: we assume that the gluing is performed along a portion of a constant-$u$ null cone where $v\in[0,1]$. We then consider a dissection 
\begin{equation}
    0=v_0<v_1<\dots<v_{k+1}=1, 
\end{equation}
and associated smooth bump functions of disjoint, compact support $\chi_j\in C^{\infty}_c\big((v_{j-1},v_j)\big)$. We assume the functions $\chi_1,\dots,\chi_{k+1}$ to be fixed, and all the gluing constructions will depend on the choice of bump functions made. 

Following our discussion in Section \ref{subsec:definition characteristic gluing} about how the $C^k$ characteristic gluing problem can be rephrased as the problem of finding a suitable seed $\phi$ along the gluing region, we will make the following ansatz for the scalar field: 
\begin{equation}
    \phi_{\alpha}(v)=\phi(v;\alpha)\doteq\sum_{j=1}^{k+1} \alpha_j\chi_j(v), 
    \label{eq:gluing - ansatz scalar field 1}
\end{equation}
where $\alpha=(\alpha_1,\dots,\alpha_{k+1})\in\mathbb R^{k+1}$. This ansatz is analogous to the one made for the uncharged scalar field in \cite{Kehle_Unger_2025}. Note that the profile choice for the scalar field guarantees:
\begin{equation}
    \phi_{\alpha}(0)=\partial_v\phi_{\alpha}(0)=\dots=\partial_v^{k}\phi_{\alpha}(0)=0,
    \label{eq:gluing - v-derivatives scalar field at 0}
\end{equation}
\begin{equation}
    \phi_{\alpha}(1)=\partial_v\phi_{\alpha}(1)=\dots=\partial_v^{k}\phi_{\alpha}(1)=0,
    \label{eq:gluing - v-derivatives scalar field at 1}
\end{equation}
and we will moreover prescribe in all constructions
\begin{equation}
    \partial_u\phi_{\alpha}(0)=\dots=\partial_u^{k}\phi_{\alpha}(0)=0.
    \label{eq:gluing - u-derivatives scalar field at 0}
\end{equation}

Motivated by the fact that we want to glue lapse-normalised sphere data sets, we set $\Omega^2\equiv1$ along the gluing region, and hence the quantities whose gluing is non-trivial\footnote{The gluing of $\phi$ and its $v$-derivatives is trivially achieved by the ansatz \eqref{eq:gluing - ansatz scalar field 1}, and so is the gluing of higher $v$-derivatives of $r$. Once the gluing of the $u$-derivatives of $\phi$ is achieved, we automatically achieve the gluing of the $u$-derivatives of $r$ too. The gluing of $\Omega^2$ and its derivatives is guaranteed since we always glue up to gauge transformations, and we can apply a gauge transformation to the final sphere to make it lapse-normalised while respecting the gluing of the other quantities.} and that we have to deal with are $r$, $\partial_vr$, $\partial_ur$, and $\partial^i_u\phi$ for $1\leq i \leq k$. Their evolution along the gluing region will be determined by Raychaudhuri's equation
\begin{equation}
    \partial_v^2r(v; \alpha)=-r(v;\alpha)\big(\partial_v\phi_{\alpha}(v)\big)^2,
    \label{eq:gluing - Raychaudhuri's v}
\end{equation}
the wave equation for $r$ (rewritten as in \eqref{eq:EE-massive scalar field - eqn for r partial_ur})
\begin{equation}
    \partial_v(r\partial_u r)(v;\alpha)=\frac14\Big(r^2(v;\alpha)\big(\Lambda + \mathfrak{m}^2\phi^2_{\alpha}(v)\big)-1\Big),
    \label{eq:gluing - eqn for r partial_ur}
\end{equation}
and the wave equation for $\phi$, that we can rewrite as
\begin{equation}
    \partial_v(r\partial_u\phi_{\alpha})(v;\alpha)=-\partial_ur(v;\alpha)\partial_v\phi_{\alpha}(v)-\frac14\mathfrak{m}^2r(v;\alpha)\phi_{\alpha}(v).
    \label{eq:gluing - eqn phi}
\end{equation}

Note that we write $r(v;\alpha)$, $\partial_vr(v;\alpha)$,  $\partial_u\phi(v;\alpha)$, etc., to emphasise that these quantities also depend on the choice of $\alpha$. 

Given $\phi_{\alpha}$, we prescribe appropriate initial values for $r$ and $\partial_vr$ to obtain $r(v;\alpha)$ using Raychaudhuri's equation \eqref{eq:gluing - Raychaudhuri's v}. Depending on the specific gluing construction, these appropriate initial values will be $r(0;\alpha)$, $\partial_vr(0;\alpha)$ or $r(1;\alpha)$, $\partial_vr(1;\alpha)$. 

Once $r(v;\alpha)$ is known, $\partial_ur(v;\alpha)$ can be determined as the unique solution to equation \eqref{eq:gluing - eqn for r partial_ur} upon prescription of an appropriate initial value for $\partial_ur$; as before, whether we prescribe $\partial_ur(0;\alpha)$ or $\partial_ur(1;\alpha)$ will depend on the particular gluing construction. Sometimes, we will also consider the renormalised Hawking mass $\widetilde{m}(v;\alpha)$, whose evolution along the gluing region can be determined once $r(v;\alpha)$ and $\partial_ur(v;\alpha)$ are known, using
\begin{equation}
    \partial_v\widetilde{m}(v;\alpha)=-2r^2(v;\alpha)\partial_ur(v;\alpha)\big(\partial_v\phi_{\alpha}(v)\big)^2+\frac12\mathfrak{m}^2r^2(v;\alpha)\partial_vr(v;\alpha)\phi^2_{\alpha}(v).
    \label{eq:gluing - v eqn for renorm Hawking mass}
\end{equation}

Finally, regarding the $u$-derivatives of $\phi$, one could use \eqref{eq:gluing - eqn phi} (and the equations obtained when differentiating \eqref{eq:gluing - eqn phi} with respect to $u$ repeatedly) to obtain the value of the first $k$ $u$-derivatives of $\phi$ along the gluing region and try to show that indeed they all vanish at $v=1$. However, following the approach in \cite{Kehle_Unger_2025}, what we will do instead is to take advantage of the well-defined parity of $r(v;\alpha)$, $\phi(v;\alpha)$ and their $u$- and $v$-derivatives under the change $\alpha\rightarrow-\alpha$. This will allow us to use the Borsuk--Ulam theorem to conclude that there is a specific choice of coefficients, $\alpha_*$, that ensures: 
\begin{equation}
    \partial_u\phi_{\alpha_*}(1)=\dots=\partial_u^{k}\phi_{\alpha_*}(1)=0.
    \label{eq:gluing - u-derivatives scalar field at 1}
\end{equation}

We will assume in general $\alpha\neq0$, and it will then be convenient to rewrite \eqref{eq:gluing - ansatz scalar field 1} as
\begin{equation}
    \phi_{\alpha}(v)=|\alpha|\sum_{j=1}^{k+1}\frac{\alpha_j}{|\alpha|}\chi_j(v)=|\alpha|\sum_{j=1}^{k+1}\hat{\alpha}_j\chi_j(v)=|\alpha|\phi_{\hat{\alpha}}(v),
    \label{eq:gluing - ansatz scalar field 2}
\end{equation}
with $\hat{\alpha}=(\hat{\alpha}_1,\dots,\hat{\alpha}_{k+1})\in\mathbb S^k$. Then we will have
  \begin{equation}
    \phi_{\hat{\alpha}}^2=\sum_{j=1}^{k+1}\hat{\alpha}^2_j\chi^{2}_j\,, \quad \quad \quad \quad \big(\partial_v\phi_{\hat{\alpha}}\big)^2=\sum_{j=1}^{k+1}\hat{\alpha}^2_j\chi'^{2}_j,
 \end{equation}
since the functions $\chi_j$, $j=1,\dots,k+1$ have disjoint compact support, and therefore
 \begin{equation}
    C_{\mathrm{low}}\leq \int_{0}^{1}\phi_{\hat{\alpha}}^2 \, dv \leq C_{\mathrm{upp}},
    \label{eq:gluing - integral scalar field}
 \end{equation}
 \begin{equation}
    C'_{\mathrm{low}}\leq\int_{0}^{1}\big(\partial_v\phi_{\hat{\alpha}}\big)^2 dv\leq C'_{\mathrm{upp}},
    \label{eq:gluing - integral derivative scalar field}
 \end{equation}
for some positive constants $C_{\mathrm{upp}},C_{\mathrm{low}},C'_{\mathrm{upp}},C'_{\mathrm{low}}$ depending on the functions $\chi_1,\dots,\chi_{k+1}$, but \underline{not on $\hat{\alpha}$}. 

\begin{remark}
    Note that one can make $\frac{C_{\mathrm{low}}}{C_{\mathrm{upp}}}$ and $\frac{C'_{\mathrm{low}}}{C'_{\mathrm{upp}}}$ as close to 1 as desired. In particular, choosing the functions $\chi_1,\dots,\chi_{k+1}$ to be translations of the same profile of compact support, we get    
    \begin{align}
        \int_{0}^{1}\hspace{-0.2em}\chi_j^2\, dv&=\int_{v_{j-1}}^{v_j}\hspace{-0.7em}\chi_j^2\,dv=N_0,\\[0.5em]
        \int_{0}^{1}\hspace{-0.2em} \big(\chi'_j\big)^2 dv&=\int_{v_{j-1}}^{v_j}\hspace{-0.7em}\big(\chi'_j\big)^2 dv=N_1
    \end{align}
    with $N_0, N_1$ the same non-negative constants for all $j=1,\dots,k+1$. Therefore 
    \begin{align}
        \int_{0}^{1}\hspace{-0.2em}\phi_{\hat{\alpha}}^2 \, dv &=\sum_{j=1}^{k+1} \hat{\alpha}_j^2\int_{0}^{1}\hspace{-0.2em}\chi_j^2 \, dv =\sum_{j=1}^{k+1} \hat{\alpha}_j^2 \cdot \, N_0=N_0,
        \label{eq:gluing - integral scalar field special} \\[0.5em]
        \int_{0}^{1}\hspace{-0.2em}\big(\partial_v\phi_{\hat{\alpha}}\big)^2 dv &=\sum_{j=1}^{k+1} \hat{\alpha}_j^2\int_{0}^{1}\hspace{-0.2em}\big(\chi'_j\big)^2 dv =\sum_{j=1}^{k+1} \hat{\alpha}_j^2 \cdot \, N_1 =N_1,
        \label{eq:gluing - integral derivative scalar field special}
    \end{align}
    with $N_0,N_1$ depending solely on the base profile we consider to be translated, and not on $\hat{\alpha}$.
    \label{rmk:norm scalar field} 
\end{remark}

\subsection{Proof of gluing to the black hole horizon (Theorem \ref{thm:subext SdS to BHhorizon})}
\label{subsec:proof subext SdS to BHhorizon}
In Theorem \ref{thm:subext SdS to BHhorizon}, we want to glue to the Schwarzschild--de Sitter black hole horizon sphere of extremality index $\mathfrak{c}_2$ with $\varrho^1_u<0$ (and hence $\varrho^1_v=0$) from the Schwarzschild--de Sitter static sphere of extremality index $\mathfrak{c}_1<\mathfrak{c}_2$ and radius $r_{\mathcal{H}}(\mathfrak{c}_1)<R_i<r_{\mathcal{H}}(\mathfrak{c}_2)$ with $\varrho^1_u<0$ (and thus $\varrho^1_v>0$). In order to do it, let us prescribe 
\begin{equation}
    \begin{aligned}
        \hspace{3em}r(1;\alpha&)=  r_{\mathcal{H}}(\mathfrak{c}_2),\\
        \partial_vr(1;\alpha&)=  0.
        \label{eq: proof gluing 1 - initialisation}
    \end{aligned}
\end{equation}
In view of Raychaudhuri's equation \eqref{eq:gluing - Raychaudhuri's v}, and by continuity of $r$ and $\partial_vr$ in $v$ we have that
\begin{equation}
    \begin{aligned}
        r(v;\alpha&)\geq0, \\
        \partial_vr(v;\alpha&)\geq  0,
    \label{eq: proof gluing 1 - bootstrap assumption}
    \end{aligned}
\end{equation}
for $v\in[v_*,1]\subset[0,1]$, with $v_*$ close enough to 1. We will see now that by choosing $|\alpha|^2$ to be small enough, we can extend this estimate to $v\in[0,1]$ by a bootstrap argument. 

\begin{lemma}
    \label{lem:proof gluing 1 - positivity}
    There exists a constant $\delta>0$, depending on the baseline scalar field profile, $R_i$ and $\mathfrak{c}_2$, such that, if $0<|\alpha|^2\leq\delta$, then 
    \begin{equation}
        \begin{aligned}
            r(v;\alpha&)\geq R_i, \\
            \partial_vr(v;\alpha&)\geq  0,
        \label{eq: proof gluing 1 - r and derivative}
        \end{aligned}
    \end{equation}
    for all $v\in[0,1]$, and moreover
    \begin{equation}
         \partial_vr(0;\alpha)>0.
         \label{eq: proof gluing 1 - derivative at 0}
    \end{equation}
\end{lemma}
\begin{proof}
Taking \eqref{eq: proof gluing 1 - bootstrap assumption} as our bootstrap assumption, we deduce that $r$ is a non-negative monotone increasing function of $v$ on $[v_*,1]$, and from Raychaudhuri’s equation \eqref{eq:gluing - Raychaudhuri's v} we also get that $\partial_vr$ is a monotone decreasing function of $v$ on $[v_*,1]$. We can now use \eqref{eq:gluing - Raychaudhuri's v} to estimate
\begin{equation}
    \partial_vr(v_*;\alpha)=\int^1_{v_*}|\alpha|^2r\big(\partial_v\phi_{\hat{\alpha}})^2dv\leq |\alpha|^2r(1;\alpha)C'_{\mathrm{upp}},
\end{equation}
and therefore
\begin{equation}
    r(v_*;\alpha)\geq r(1;\alpha)-\partial_vr(v_*;\alpha)\geq r_{\mathcal{H}}(\mathfrak{c}_2)(1-|\alpha|^2C'_{\mathrm{upp}}).
\end{equation}

It is clear now that by choosing 
\begin{equation}
    0<|\alpha|^2\leq\delta \doteq \frac{1}{C'_{\mathrm{upp}}}\left(1-\frac{R_i}{r_{\mathcal{H}}(\mathfrak{c}_2)}\right),
\end{equation} 
we can make $r(v_*;\alpha)\geq R_i>0$, improving the bootstrap assumption and proving that \eqref{eq: proof gluing 1 - r and derivative} holds for $v\in [0,1]$. Note also that, for $0<|\alpha|^2\leq\delta$, $\partial_vr$ is not only monotonically decreasing in $v$, but also strictly decreasing whenever $\partial_v\phi_{\hat{\alpha}}(v)\neq0$, which proves \eqref{eq: proof gluing 1 - derivative at 0}.
\end{proof}

For $\alpha\in \mathbb S^k_{\sqrt{\delta}}=\{\alpha \in \mathbb R^{k+1}:|\alpha|=\sqrt{\delta}\}$, we define $D_{\alpha}(0)$ as the sphere data set satisfying: 
\begin{itemize}
    \item $\varrho=r(0;\alpha)$, with $R_i\leq r(0;\alpha)<r_{\mathcal{H}}(\mathfrak{c}_2)$,
    \item $\varrho^1_v=\partial_vr(0;\alpha)>0$,
    \item $\varrho^1_u=-\frac{1}{4\partial_vr(0;\alpha)}\Big(1-\frac{2\mathfrak{c}_1}{3\sqrt{\Lambda}r(0;\alpha)}-\frac13\Lambda r^2(0;\alpha)\Big)$,
    \item $\omega =1$,
    \item all other components are zero. 
\end{itemize}

We have chosen the value for $\varrho^1_u$ so that
\begin{equation}
    \widetilde{m}\left(D_{\alpha}(0)\right)=\frac{\varrho}{2}\left(1+4\varrho^1_v\varrho^1_u-\frac13\Lambda\varrho^2\right)=\frac{\mathfrak{c}_1}{3\sqrt{\Lambda}},
\end{equation}
that is, the renormalised Hawking mass of any Schwarzschild--de Sitter sphere of extremality index $\mathfrak{c}_1$. Note that, because \[r_{\mathcal{H}}(\mathfrak{c}_1)<R_i\leq r(0;\alpha)<r_{\mathcal{H}}(\mathfrak{c}_2)\leq r_{\overline{\mathcal{H}}}(\mathfrak{c}_2)<r_{\overline{\mathcal{H}}}(\mathfrak{c}_1),\] we have 
\begin{equation}
    1-\frac{2\mathfrak{c}_1}{3\sqrt{\Lambda}r(0;\alpha)}-\frac13\Lambda r^2(0;\alpha)=f_{\mathfrak{c}_1}\big(r(0;\alpha)\big)>0, 
\end{equation}
and thus $\varrho^1_u\left(D_{\alpha}(0)\right)<0$. Hence $D_{\alpha}(0)$ is gauge equivalent to the Schwarzschild--de Sitter static sphere $_{\mathcal{S}}D^{\mathrm{SdS}}_{\mathfrak{c}_1,r(0;\alpha), k}$ with $\varrho^1_u<0$. 

Now, with $D_{\alpha}(0)$ and the seed $\phi_{\alpha}$, we generate a cone data set $D_{\alpha}:[0,1]\longrightarrow \mathcal{D}_k$, continuous both in $\alpha$ and $v$ by standard ODE theory. Then, by construction
\begin{equation}
    \varrho\big(D_{\alpha}(v)\big)=r(v;\alpha)\,, \qquad \varrho^1_v\big(D_{\alpha}(v)\big)=\partial_vr(v;\alpha)\, , \qquad  \varrho^1_u\big(D_{\alpha}(v)\big)=\partial_ur(v;\alpha)\,,
\end{equation}
and also 
\begin{equation}
    \varphi\big(D_{\alpha}(v)\big)=\phi(v;\alpha)\,, \qquad \quad \varphi^i_v\big(D_{\alpha}(v)\big)=\partial_v^i\phi(v;\alpha) \quad i=1, \dots,k.
\end{equation}
We will moreover identify 
\begin{equation}
    \partial^i_u\phi(v;\alpha)\doteq\varphi^i_u\big(D_{\alpha}(v)\big) \qquad i=1, \dots,k
\end{equation}
for all $v\in[0,1]$. 
    
We will see now that, by possibly making the constant $\delta$ from Lemma \ref{lem:proof gluing 1 - positivity} even smaller, we can ensure $\partial_ur(1;\alpha)<0$ for every $\alpha$ with $0<|\alpha|^2\leq\delta$. 

\begin{lemma}
    By possibly making the constant $\delta>0$ from Lemma \ref{lem:proof gluing 1 - positivity} smaller (depending on $\mathfrak{c}_2$, $\mathfrak{m}$ and the baseline scalar field profile), we can ensure 
    \begin{equation}
        \partial_ur(1;\alpha)<0
        \label{eq:proof gluing 1 - estimate partial_ur final} 
    \end{equation}
    for every $\alpha$ with $0<|\alpha|^2\leq \delta$.
    \label{lem:proof gluing 1 - estimate partial_ur} 
\end{lemma}
\begin{proof}
    Let us consider $\alpha$ satisfying  $0<|\alpha|^2\leq \delta$, with $\delta$ given by Lemma \ref{lem:proof gluing 1 - positivity}. Hence $r$ is a positive monotone increasing function of $v$ on $[0,1]$, and moreover, because of Raychaudhuri's equation \eqref{eq:gluing - Raychaudhuri's v}, $\partial_vr$ is a non-negative, monotone decreasing function of $v$ on $[0,1]$, which is strictly decreasing whenever $\partial_v\phi_{\hat{\alpha}}(v)\neq0$. Given the positivity of $r$ along the gluing region, it suffices to show
    \begin{equation}
        (r\partial_ur)(1;\alpha)<0,
        \label{eq: proof gluing 1 - estimate partial_ur}
    \end{equation}
    in order to establish \eqref{eq:proof gluing 1 - estimate partial_ur final}.
    
    \vspace{0.5em}

    \noindent\underline{Case $\mathfrak{m}=0$}
    
    \noindent Since $r(v;\alpha)\leq r_{\mathcal{H}}(\mathfrak{c}_2)\leq\frac{1}{\sqrt{\Lambda}}$ for all $v\in[0,1]$, setting $\mathfrak{m}=0$ in \eqref{eq:gluing - eqn for r partial_ur} we obtain
    \begin{equation}
        \partial_v(r\partial_ur)(v;\alpha)=\frac14\left(\Lambda r^2(v;\alpha)-1\right)\leq0,
    \end{equation}
    and therefore $r\partial_ur$ is a decreasing function of $v$ and $(r\partial_ur)(1;\alpha)\leq(r\partial_ur)(0;\alpha)<0$ for any $\alpha$ satisfying $0<|\alpha|^2\leq \delta$, with $\delta$ given by Lemma \ref{lem:proof gluing 1 - positivity}. 

    \vspace{0.5em}
    
    \noindent \underline{Case $\mathfrak{m}>0$} 

    \noindent From equation \eqref{eq:gluing - eqn for r partial_ur} in this case we get 
    \begin{equation}
    \partial_v(r\partial_ur)(v;\alpha)=\frac14\Big(r^2(v;\alpha)\big(\Lambda +\mathfrak{m}^2|\alpha|^2\phi^2_{\hat{\alpha}}(v)\big)-1\Big)\leq\frac14\left(\frac{1}{\Lambda}\mathfrak{m}^2|\alpha|^2\phi^2_{\hat{\alpha}}(v)\right),
    \end{equation}
    and thus $r\partial_ur$ could a priori increase along the gluing region, which makes it less straightforward to conclude \eqref{eq: proof gluing 1 - estimate partial_ur}. Integrating Raychaudhuri's equation \eqref{eq:gluing - Raychaudhuri's v} along the gluing region and using the monotonicity of $r$ we get
    \begin{equation}
        \partial_vr(0;\alpha) \leq |\alpha|^2r(1;\alpha)C'_{\mathrm{upp}}.
        \label{eq: proof gluing 1 - upp bound partial_vr at 0}
    \end{equation}
    Now, integrating $\partial_vr$ in $v$ and using the monotonicity of $\partial_vr$ and \eqref{eq: proof gluing 1 - upp bound partial_vr at 0}, we find
    \begin{equation}
        r(1;\alpha)\big(1-|\alpha|^2C'_{\mathrm{upp}}\big)\leq r(0;\alpha)\leq r(1;\alpha).
        \label{eq: proof gluing 1 - upp and low bound r at 0}
    \end{equation}
    
    Finally, we integrate \eqref{eq:gluing - eqn for r partial_ur} to get 
    \begin{equation}
        (r\partial_ur)(1;\alpha)\leq \frac14\Big(r^2(1;\alpha)\big(\Lambda+\mathfrak{m}^2|\alpha|^2C_{\mathrm{upp}}\big)-1\Big) +(r\partial_ur)(0;\alpha),
    \end{equation}
    where we can moreover use \eqref{eq: proof gluing 1 - upp bound partial_vr at 0} and \eqref{eq: proof gluing 1 - upp and low bound r at 0} to get an upper bound for $(r\partial_ur)(0;\alpha)$:
    \begin{equation}
        (r\partial_ur)(0;\alpha)\leq-\frac{1-|\alpha|^2C'_{\mathrm{upp}}}{4|\alpha|^2C'_{\mathrm{upp}}}\bigg(1-\frac{2\mathfrak{c}_1}{3\sqrt{\Lambda}}\Big(r(1;\alpha)\big(1-|\alpha|^2C'_{\mathrm{upp}}\big)\Big)^{-1}-\frac13\Lambda r^2(1;\alpha)\bigg).
    \end{equation}
    
    Therefore, in order to conclude \eqref{eq: proof gluing 1 - estimate partial_ur}, it is sufficient to show that we can find $\delta$ such that, for $0<|\alpha|^2\leq \delta$, the following estimate holds:
    \begin{equation}
        r^2(1;\alpha)\big(\Lambda+\mathfrak{m}^2|\alpha|^2C_{\mathrm{upp}}\big)-1 -\frac{1-|\alpha|^2C'_{\mathrm{upp}}}{|\alpha|^2C'_{\mathrm{upp}}}\bigg(1-\frac{2\mathfrak{c}_1}{3\sqrt{\Lambda}}\Big(r(1;\alpha)\big(1-|\alpha|^2C'_{\mathrm{upp}}\big)\Big)^{-1}-\frac13\Lambda r^2(1;\alpha)\bigg)<0.
        \label{eq: proof gluing 1 - rewriting the problem 1}
    \end{equation}
    Rewriting $\mathfrak{m}^2=\beta \Lambda$ with $\beta \in \mathbb R_{>0}$, substituting the prescribed value for $r(1;\alpha)$ and rearranging, \eqref{eq: proof gluing 1 - rewriting the problem 1} is equivalent to 
    \begin{equation}
        \mathfrak{p}\big(|\alpha|^2\big)\doteq-1+ \frac{2\mathfrak{c}_1}{3\sqrt{\Lambda}}\frac{1}{r_{\mathcal{H}}(\mathfrak{c}_2)}+\frac13\Lambda\big(r_{\mathcal{H}}(\mathfrak{c}_2)\big)^2\Big(1+2C'_{\mathrm{upp}} |\alpha|^2+3\beta C_{\mathrm{upp}}C'_{\mathrm{upp}}\big(|\alpha|^2\big)^2\Big)<0, 
        \label{eq: proof gluing 1 - rewriting the problem 2}
    \end{equation}
    with $\mathfrak{p}\big(|\alpha|^2\big)$ a quadratic polynomial in $|\alpha|^2$. Moreover, note that 
    \begin{equation}
        \mathfrak{p}(0)=-\bigg(1-\frac{2\mathfrak{c}_1}{3\sqrt{\Lambda}}\frac{1}{r_{\mathcal{H}}(\mathfrak{c}_2)}-\frac13\Lambda\big(r_{\mathcal{H}}(\mathfrak{c}_2)\big)^2\bigg)=-f_{\mathfrak{c}_1}\big(r_{\mathcal{H}}(\mathfrak{c}_2)\big)<0,
        \label{eq: proof gluing 1 - p at 0}
    \end{equation}
    since $f_{\mathfrak{c}_1}(r)$ is positive for $r_{\mathcal{H}}(\mathfrak{c}_1)<r<r_{\overline{\mathcal{H}}}(\mathfrak{c}_1)$, and $r_{\mathcal{H}}(\mathfrak{c}_1)<r_{\mathcal{H}}(\mathfrak{c}_2)<r_{\overline{\mathcal{H}}}(\mathfrak{c}_1)$ because $0\leq\mathfrak{c}_1<\mathfrak{c}_2\leq1$. 

    As a consequence of \eqref{eq: proof gluing 1 - p at 0}, in a neighbourhood of $x=0$, we have $\mathfrak{p}(x)<0$. This implies that there exists a constant $\delta>0$, depending on $C_{\mathrm{upp}}, C'_{\mathrm{upp}}, \beta$ (equivalently $\mathfrak{m}$), $\mathfrak{c}_1$ and $\mathfrak{c}_2$, possibly smaller than the one obtained in Lemma \ref{lem:proof gluing 1 - positivity}, such that, for every $|\alpha|^2$ with $0<|\alpha|^2\leq \delta$, the constraint \eqref{eq: proof gluing 1 - rewriting the problem 2} is satisfied, hence ensuring \eqref{eq: proof gluing 1 - estimate partial_ur}.
\end{proof}

Therefore, if $\alpha \in \mathbb S^k_{\sqrt{\delta}}$, with $\delta>0$ given by Lemma \ref{lem:proof gluing 1 - estimate partial_ur}, $D_{\alpha}(1)$ will satisfy:
\begin{itemize}
    \item $\varrho=r(1;\alpha)=r_{\mathcal{H}}(\mathfrak{c}_2 )$ (see \eqref{eq: proof gluing 1 - initialisation}),
    \item $\varrho^1_v=\partial_vr(1;\alpha)=0$  (see \eqref{eq: proof gluing 1 - initialisation}),
    \item $\varrho^1_u=\partial_ur(1;\alpha)<0$ (consequence of Lemma \ref{lem:proof gluing 1 - estimate partial_ur}),
    \item $\omega =1$, 
    \item $\varphi^i_v=0$ for all $0\leq i \leq k$. 
\end{itemize}

It only remains to show that one can choose $\alpha\in\mathbb S^k_{\sqrt{\delta}}$ such that $\varphi^i_u=0$ (equivalently $\partial_u^{i}\phi(1;\alpha)=0$) for all $1\leq i \leq k$  to conclude that $D_{\alpha}(1)$ is gauge equivalent to the Schwarzschild--de Sitter black hole horizon sphere $_{\mathcal{H}^+}D^{\mathrm{SdS}}_{\mathfrak{c}_2, k}$, with $\varrho^1_u<0$. This will be achieved by the same procedure developed in \cite{Kehle_Unger_2025}, making use of the Borsuk--Ulam theorem, which we outline here now. We start noting that the metric functions and the scalar field have a well-defined parity as functions of $\alpha$:

\begin{lemma}
    As functions on $[0,1]\times \mathbb S_{\sqrt{\delta}}^k$, the scalar field $\phi(v;\alpha)$ and its $u$- and $v$-derivatives are odd functions of $\alpha$, and the metric functions $r(v;\alpha)$, $\Omega^2(v;\alpha)$ and their $u$- and $v$-derivatives are even functions of $\alpha$.
    \label{lem:parity lemma}
\end{lemma}
\begin{proof}
    The scalar field $\phi(v;\alpha)$ is an odd function of $\alpha$ by definition \eqref{eq:gluing - ansatz scalar field 1}, and so are its $v$-derivatives. The lapse is fixed to 1 along the gluing region irrespective of the value of $\alpha$, so $\Omega^2$ and all of its $v$-derivatives are even functions of $\alpha$. Given that $r(v;\alpha)$ is obtained using Raychaudhuri's equation \eqref{eq:gluing - Raychaudhuri's v}, and this only involves $\big(\partial_v\phi(v;\alpha)\big)^2$, we conclude that $r(v;\alpha)$ and all of its $v$-derivatives are even functions of $\alpha$.  Remember also that $\partial_ur(v;\alpha)$ was found by integrating \eqref{eq:gluing - eqn for r partial_ur}; since the right-hand side of \eqref{eq:gluing - eqn for r partial_ur} is an even function of $\alpha$, and so is $r(v;\alpha)$, $\partial_ur(v;\alpha)$ will be also an even function of $\alpha$. Now $\partial_u\phi(v;\alpha)$  is obtained using the wave equation for $\phi$ \eqref{eq:gluing - eqn phi}. The right-hand side of \eqref{eq:gluing - eqn phi} is odd in $\alpha$, so we deduce $(r\partial_u\phi) (v;\alpha)$ is odd and hence that $\partial_u\phi(v;\alpha)$ is an odd function of $\alpha$. 

    In order to compute higher derivatives, we proceed iteratively as in Proposition \ref{prop:compute u-derivatives from initial data}, noting that the transport equations satisfied by the $u$-derivatives of $r$ and $\Omega^2$ involve even powers of $\phi$ and its derivatives, and thus they are even functions of $\alpha$; the transport equations for the higher $u$-derivatives of $\phi$, however, involve only odd powers of $\phi$ and its derivatives.
\end{proof}
Let us then define a function $\mathfrak{F}$ given by
\begin{equation}
    \begin{aligned}
        \mathfrak{F}:\, &\mathbb S^k_{\sqrt{\delta}}\longrightarrow \mathbb R^k\\
        &\alpha\longmapsto\big(\partial_u\phi(1;\alpha),\dots,\partial_u^{k}\phi(1;\alpha)\big),
    \end{aligned} 
\end{equation}
which will be an odd function of $\alpha$ by the previous lemma. Now we can simply apply the Borsuk--Ulam theorem to conclude that there exists $\alpha_*\in\mathbb S^k_{\sqrt{\delta}}$ such that $\mathfrak{F}(\alpha_*)=0$, and therefore
\begin{equation}
    \partial_u\phi(1;\alpha_*)=\dots =\partial_u^{k}\phi(1;\alpha_*)=0.
\end{equation}
Hence  $D_{\alpha_*}(1)$ is gauge equivalent to the Schwarzschild--de Sitter horizon sphere $_{\mathcal{H}^+}D^{\mathrm{SdS}}_{\mathfrak{c}_2, k}$ with $\varrho^1_u<0$. 

Note that we have characteristically glued from the Schwarzschild--de Sitter static sphere $_{\mathcal{S}}D^{\mathrm{SdS}}_{\mathfrak{c}_1,r(0;\alpha_*), k}$. In the case that $r(0;\alpha_*)~>~R_i$, we can glue trivially, i.e. with vanishing scalar field, the sphere $_{\mathcal{S}}D^{\mathrm{SdS}}_{\mathfrak{c}_1,R_i,k}$ with $\varrho^1_u<0$ to the sphere $_{\mathcal{S}}D^{\mathrm{SdS}}_{\mathfrak{c}_1,r(0;\alpha_*), k}$ with $\varrho^1_u<0$, to complete the proof of Theorem \ref{thm:subext SdS to BHhorizon}.

\subsection{Proof of gluing to the subextremal cosmological horizon (Theorem \ref{thm:dS to subext COSMhorizon})}
\label{subsec:proof dS to subext COSMhorizon}

Remember that in Theorem \ref{thm:dS to subext COSMhorizon} we want to glue the de Sitter static sphere $_{\mathcal{S}}D^{\mathrm{dS}}_{R_i, k}$ with $\varrho^1_u<0$, to the subextremal Schwarzschild--de Sitter cosmological horizon sphere  $_{\overline{\mathcal{H}}^+}D^{\mathrm{SdS}}_{\mathfrak{c}, k}$ with $\varrho^1_u>0$. 

\begin{figure}[H]
        \centering
        \includegraphics[width=0.75\textwidth]{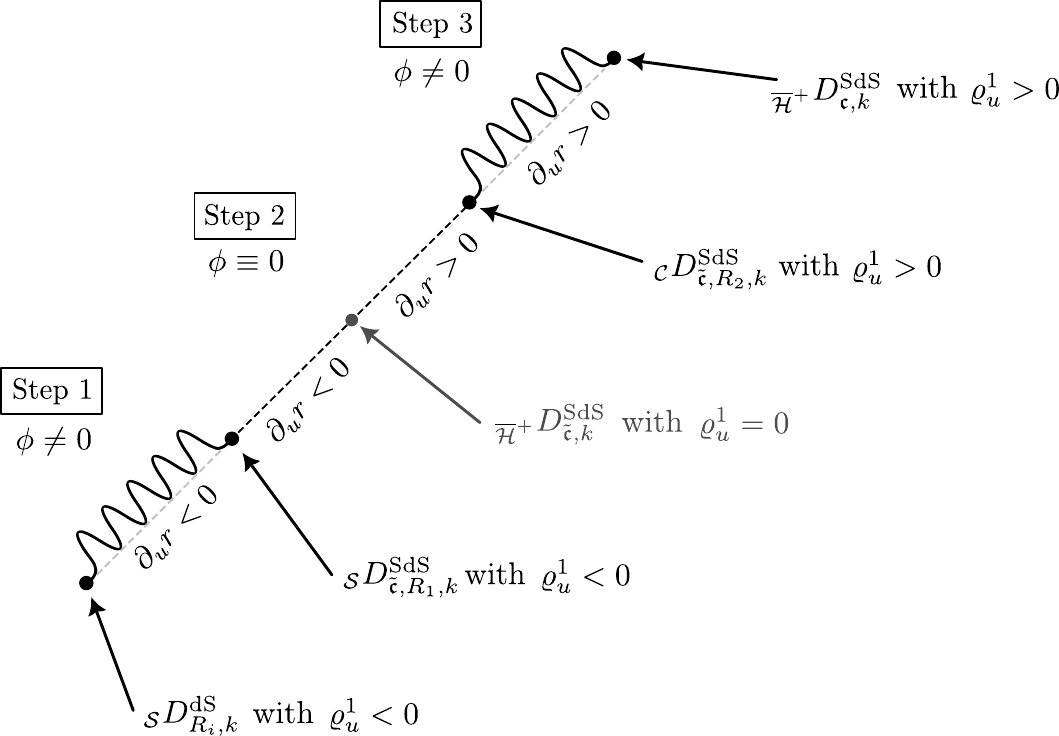}
        \caption{Diagram illustrating the three steps in the proof of Theorem \ref{thm:dS to subext COSMhorizon}.} 
        \label{fig:3 steps}
\end{figure}

We have already mentioned how one of the main difficulties in the gluing to the cosmological horizon is ensuring the change of sign in $\partial_ur$ along the gluing region. We will get around this difficulty and prove Theorem \ref{thm:dS to subext COSMhorizon} making use of a 3-step gluing scheme (see also Figure \ref{fig:3 steps}): 

\vspace{0.5em}

\begin{itemize}
    \item \textbf{Step 1:} the de Sitter static sphere $_{\mathcal{S}}D^{\mathrm{dS}}_{R_i, k}$ with $\varrho^1_u<0$ is glued to the subextremal Schwarzschild--de Sitter static sphere $_{\mathcal{S}}D^{\mathrm{SdS}}_{\tilde{\mathfrak{c}}, R_1, k}$, with $\varrho^1_u<0$, for some extremality index $\tilde{\mathfrak{c}}$ and radius $R_1$ satisfying $\mathfrak{c}<\tilde{\mathfrak{c}}<1$ and $R_i<R_1<r_{\overline{\mathcal{H}}}(\tilde{\mathfrak{c}})$, with a non-trivial scalar field of mass $\mathfrak{m}\geq0$. 
    \vspace{0.5em}
    \item \textbf{Step 2:}  $_{\mathcal{S}}D^{\mathrm{SdS}}_{\tilde{\mathfrak{c}}, R_1, k}$ is glued to a cosmological sphere in the same spacetime $_{\mathcal{C}}D^{\mathrm{SdS}}_{\tilde{\mathfrak{c}}, R_2, k}$ with $\varrho^1_u>0$, for some $R_2$ satisfying $r_{\overline{\mathcal{H}}}(\tilde{\mathfrak{c}})<R_2<r_{\overline{\mathcal{H}}}(\mathfrak{c})$, with a vanishing scalar field. 
    \vspace{0.5em}
    \item \textbf{Step 3:} $_{\mathcal{C}}D^{\mathrm{SdS}}_{\tilde{\mathfrak{c}}, R_2, k}$ is glued to the Schwarzschild--de Sitter cosmological horizon sphere of extremality index $\mathfrak{c}$,  $_{\overline{\mathcal{H}}^+}D^{\mathrm{SdS}}_{\mathfrak{c}, k}$ with $\varrho^1_u>0$, with a non-trivial scalar field of mass\footnote{In principle, one could take the scalar field masses in Step 1 and Step 3 to be different, since there is nothing in the proof that forces them to have the same value. Nevertheless, we make the physically motivated assumption that the scalar field mass is a fixed parameter of the matter model considered, and thus we take the scalar field mass in both steps to be the same.} $\mathfrak{m}\geq0$. 
    \vspace{2.5em}
\end{itemize}

With this construction, the change of sign in $\partial_ur$ is achieved by simply taking advantage of the geometry of the intermediate spacetime. Note that, precisely because $\partial_ur$ is positive and $\partial_vr$ is positive and small in Step 3, one would expect, in view of equation \eqref{eq:gluing - v eqn for renorm Hawking mass}, the renormalised Hawking mass $\widetilde{m}$ (or equivalently, the extremality index) to decrease in a neighbourhood of the final sphere. This is why we choose the intermediate spacetime to have extremality index $\tilde{\mathfrak{c}}$ greater than the desired extremality index $\mathfrak{c}$ for the final sphere in the global construction. Moreover, we recall that the radius of the cosmological horizon, $r_{\overline{\mathcal{H}}}$, seen as a function of the spacetime mass, satisfies
\begin{equation}
    r_{\overline{\mathcal{H}}}(\mathfrak{c}_1)<r_{\overline{\mathcal{H}}}(\mathfrak{c}_2) \quad \quad \text{ if } \, \mathfrak{c}_1>\mathfrak{c}_2.
\end{equation} 
Given that $\tilde{\mathfrak{c}}>\mathfrak{c}$, it seems perfectly reasonable to glue from a sphere with radius $r>r_{\overline{\mathcal{H}}}(\tilde{\mathfrak{c}})$ to a cosmological horizon sphere of radius $r_{\overline{\mathcal{H}}}(\mathfrak{c})$ with $\partial_vr$ non-negative and therefore $r$ increasing along the gluing, which is precisely the situation in Step 3. 

One could try to prove Theorem \ref{thm:dS to subext COSMhorizon} avoiding any intermediate gluings. However, this may result in an additional positive lower bound on the scalar field mass, which is sufficient but not necessary to make the gluing possible. With the three consecutive gluings described above we are able to prove the result \myuline{for any non-negative scalar field mass}. We prove now each of the three gluings separately.

\subsubsection*{Step 1: gluing $_{\mathcal{S}}D^{\mathrm{dS}}_{R_i, k}$ to $_{\mathcal{S}}D^{\mathrm{SdS}}_{\tilde{\mathfrak{c}}, R_1, k}$}

Note that, if  $R_i<1/{\sqrt{\Lambda}}$, it is always possible to glue trivially from $_{\mathcal{S}}D^{\mathrm{dS}}_{R_i, k}$ to $_{\mathcal{S}}D^{\mathrm{dS}}_{\frac{1}{\sqrt{\Lambda}}, k}$. Therefore, we will assume without loss of generality $R_i\geq1/\sqrt{\Lambda}$, and rewrite for convenience $$R_i=\frac{r_0}{\sqrt{\Lambda}},$$ for some $1\leq r_0<2\cos\Big(\frac{\cos^{-1}(-\mathfrak{c})}{3}\Big)$. Thus for this gluing we prescribe initially 
\begin{equation}
    \begin{aligned}
        \hspace{3em}r(0;\alpha&)= \frac{r_0}{\sqrt{\Lambda}},\\
        \partial_vr(0;\alpha&)=  \frac{\varepsilon}{\sqrt{\Lambda}},
        \label{eq: proof gluing 2.1 - initialisation}
    \end{aligned}
\end{equation}
 for some $\varepsilon>0$ possibly very small to be determined later\footnote{Remember that we always have the freedom to rescale $\partial_v r$ and $\partial_u r$ by a positive constant $\gamma$: $\partial_u r\mapsto \gamma\partial_ur, \partial_vr\mapsto\gamma^{-1}\partial_vr$.}. By continuity of $\partial_vr$, we have
\begin{equation}
    \begin{aligned}
        \partial_vr(v;\alpha&)\geq  0,
    \label{eq: proof gluing 2.1 - bootstrap assumption}
    \end{aligned}
\end{equation}
for all $v\in[0,v_*]$ for $v_*$ close enough to $0$. Choosing $|\alpha|^2$ to be small enough, we can extend this estimate to $v\in[0,1]$:

\begin{lemma}
    \label{lem:proof gluing 2.1 - positivity}
    There exists a constant $\delta>0$, depending on the baseline scalar field profile, $r_0$, and $\varepsilon$, such that if $0<|\alpha|^2<\delta$, then 
    \begin{equation}
        \begin{aligned}
            \partial_vr(v;\alpha&)> 0
        \label{eq: proof gluing 2.1 - r derivative}
        \end{aligned}
    \end{equation}
\end{lemma}
\begin{proof}
Taking \eqref{eq: proof gluing 2.1 - bootstrap assumption} as the bootstrap assumption, we deduce $r$ is an increasing function on $[0,v_*]$, and thus integrating Raychaudhuri’s equation \eqref{eq:gluing - Raychaudhuri's v} in $v$ we find
\begin{equation}
    \partial_vr(v_*;\alpha)=\partial_vr(0;\alpha) - \int^{v_*}_0|\alpha|^2r\big(\partial_v\phi_{\hat{\alpha}}\big)^2dv\geq \partial_vr(0;\alpha)-|\alpha|^2r(v_*;\alpha)C'_{\mathrm{upp}}.
    \label{eq: proof gluing 2.1 - lemma intermediate 1}
\end{equation}
On the other hand, we also have 
\begin{equation}
    r(v_*;\alpha)\leq r(0;\alpha)+\partial_vr(0;\alpha),
    \label{eq: proof gluing 2.1 - lemma intermediate 2}
\end{equation}
since $\partial_vr$ is monotone decreasing in $v$ (again, remember Raychaudhuri's equation \eqref{eq:gluing - Raychaudhuri's v} together with the fact that $r$ is positive).
Combining \eqref{eq: proof gluing 2.1 - lemma intermediate 1} and \eqref{eq: proof gluing 2.1 - lemma intermediate 2}, we get 
\begin{equation}
    \partial_vr(v_*;\alpha)\geq \partial_vr(0;\alpha)-\big(r(0;\alpha)+\partial_vr(0;\alpha)\big)|\alpha|^2C'_{\mathrm{upp}}
\end{equation}
It is clear now that by choosing 
\begin{equation}
    0<|\alpha|^2<\delta \doteq \frac{1}{C'_{\mathrm{upp}}}\frac{\partial_vr(0;\alpha)}{r(0;\alpha)+\partial_vr(0;\alpha)}=\frac{1}{C'_{\mathrm{upp}}}\frac{\varepsilon}{r_0+\varepsilon},
    \label{eq: proof gluing 2.1 upper bound alpha 1}
\end{equation} 
we can make $\partial_vr(v_*;\alpha)> 0$, improving the bootstrap assumption \eqref{eq: proof gluing 2.1 - bootstrap assumption} and proving the result. 
\end{proof}

\begin{remark}
    Note that, for any $\alpha$ satisfying $0<|\alpha|^2<\delta$, we can extend the estimates derived in the proof of Lemma \ref{lem:proof gluing 2.1 - positivity} to $v=1$, obtaining: 
    \begin{equation}
        r(1;\alpha)\leq r(0;\alpha)+\partial_vr(0;\alpha),
        \label{eq: proof gluing 2.1 - upper estimate r}
    \end{equation}
    \begin{equation}
        \partial_vr(1;\alpha)\geq \partial_vr(0;\alpha)-\big(r(0;\alpha)+\partial_vr(0;\alpha)\big)|\alpha|^2C'_{\mathrm{upp}}.
        \label{eq: proof gluing 2.1 - lower estimate partial_u r}
    \end{equation}
    \label{rmk: proof gluing 2.1 - estimates r partial_v r}
\end{remark}

We will also want to prescribe 
\begin{equation}
    \partial_ur(0;\alpha)= -\frac{1}{4\partial_vr(0;\alpha)}\left(1-\frac13\Lambda r(0;\alpha)^2\right),
    \label{eq: proof gluing 2.1 - initialisation 2}
\end{equation}
so that the initial renormalised Hawking mass is indeed 0 and the sphere at $v=0$ is gauge equivalent to a static sphere in de Sitter spacetime. For $\varepsilon$ sufficiently small, $0<|\alpha|^2<\delta$ also ensures $\partial_ur(1;\alpha)<0$, as shown in the following lemma. 
\begin{lemma}
    \label{lem:proof gluing 2.1 - negativity}
    We can choose $\varepsilon$ small enough so that 
    \begin{equation}
        \partial_ur(1;\alpha)< 0,
        \label{eq: proof gluing 2.1 - partial_u r at 1}
    \end{equation}
    for all $\alpha$ satisfying $0<|\alpha|^2<\delta$, with $\delta$ as in Lemma \ref{lem:proof gluing 2.1 - positivity}.
\end{lemma}
\begin{proof}
    Let us assume $0<|\alpha|^2<\delta$. By Lemma \ref{lem:proof gluing 2.1 - positivity}, $\partial_vr$ is a positive decreasing function of $v$, and $r$ is a strictly increasing function of $v$. Integrating equation \eqref{eq:gluing - eqn for r partial_ur}, and using the monotonicity properties of $r$ and $\partial_v r$, we find:

    \newpage 
    \begin{align}
        \hspace{-3em}(r\partial_ur)(1;\alpha)&\leq \frac14\bigg(r(1;\alpha)^2\big(\Lambda+\mathfrak{m}^2|\alpha|^2C_{\mathrm{upp}}\big)-1\bigg) + (r\partial_ur)(0;\alpha)\\[0.3em]
        &\leq \frac14\Bigg(\big(r(0;\alpha)+\partial_vr(0;\alpha)\big)^2\big(\Lambda+\mathfrak{m}^2|\alpha|^2C_{\mathrm{upp}}\big)-1 -\frac{r(0;\alpha)}{\partial_vr(0;\alpha)}\bigg(1-\frac13\Lambda r(0;\alpha)^2\bigg)\Bigg), 
        \label{eq: proof of gluing 2.1 rpartial_u r estimate}
    \end{align}
    where in the last step we used \eqref{eq: proof gluing 2.1 - upper estimate r}. 
    Thus, choosing $ |\alpha|^2$ such that
    \begin{align}
        \beta|\alpha|^2<\frac{1}{\Lambda C_{\mathrm{upp}}}&\left(\frac{1}{\big(r(0;\alpha)+\partial_vr(0;\alpha)\big)^2}\left(1+\frac{r(0;\alpha)}{\partial_vr(0;\alpha)}\left(1-\frac13\Lambda r(0;\alpha)^2\right)\right)-\Lambda\right)\\[0.3em]&\hspace{9em}=\frac{1}{C_{\mathrm{upp}}}\left(\frac{1}{\big(r_0+\varepsilon\big)^2}\left(1+\frac{r_0}{\varepsilon}\bigg(1-\frac{r_0^2}{3} \bigg)\right)-1\right),
        \label{eq: proof gluing 2.1 upper bound alpha 2}
    \end{align}
    where we have rewritten $\mathfrak{m}^2=\beta\Lambda$ with $\beta\in \mathbb R_{\geq0}$, we ensure $(r\partial_ur)(1;\alpha)<0$, and thus $\partial_ur(1;\alpha)<0$. 
    
    However, for \eqref{eq: proof gluing 2.1 upper bound alpha 2} to make sense (even when $\beta \rightarrow 0$), we need the right-hand side in \eqref{eq: proof gluing 2.1 upper bound alpha 2} to be positive. Note that this right-hand side can be rewritten as 
    \begin{equation}
        \frac{1}{C_{\mathrm{upp}}}\frac{1}{\varepsilon\big(r_0+\varepsilon\big)^2}\mathcal{E}  (\varepsilon)
    \end{equation}
    with
    \begin{equation}
        \mathcal{E}(\varepsilon)\doteq\varepsilon+r_0\bigg(1-\frac{r_0^2}{3} \bigg)-\varepsilon\big(r_0+\varepsilon\big)^2.
    \end{equation}
    Note that $\mathcal{E}(0)=r_0\Big(1-\frac{r_0^2}{3} \Big)>0$ for $1\leq r_0<2\cos\big(\frac{\cos^{-1}(-\mathfrak{c})}{3}\big)$ and hence, for $\varepsilon$ small enough\footnote{For $1\leq r_0<\sqrt{3}$, $\mathcal{E}$ has one single positive root,  $\varepsilon_*$, which can be explicitly computed in terms of $r_0$. Thus by ``$\varepsilon$ small enough'' we simply mean $\varepsilon<\varepsilon_*(r_0)$. By way of illustration, $\varepsilon_*(1)\simeq 0.515$ and $\varepsilon_*(\sqrt{3})=0$, with $\varepsilon_*$ decreasing monotonically as $r_0$ increases between these two values.}, we will have $\mathcal{E}(\varepsilon)>0$.
    
    In the case $\beta>0$, it still remains to show that 
    \begin{equation}
        \bar{\delta}\doteq\frac{1}{\beta C_{\mathrm{upp}}}\frac{1}{\varepsilon\big(r_0+\varepsilon\big)^2}\mathcal{E}  (\varepsilon),
        \label{eq: proof gluing 2.1 - definition delta tilde}
    \end{equation}
    
    satisfies $\delta\leq\bar{\delta}$, at least for $\varepsilon$ small enough. Remember
    \begin{equation}
        \delta=\frac{1}{C'_{\mathrm{upp}}}\frac{\varepsilon}{r_0+\varepsilon}.
    \end{equation}
    Comparing this to \eqref{eq: proof gluing 2.1 - definition delta tilde}, we see that it suffices to show that we can choose $\varepsilon$ so that 
    \begin{equation}
        \beta\frac{C_{\mathrm{upp}}}{C'_{\mathrm{upp}}}\frac{r_0+\varepsilon}{\mathcal{E}(\varepsilon)}\varepsilon^2\leq1, 
        \label{eq: proof gluing 2.1 - final cond negativity}
    \end{equation}
    In fact, we will always be able to choose $\varepsilon>0$ small enough such that \eqref{eq: proof gluing 2.1 - final cond negativity} holds, since all the quantities involved in the left-hand side of the inequality remain positive and finite as $\varepsilon\rightarrow0$. 
\end{proof}

Thereby, we assume in what follows that $0<|\alpha|^2<\delta$, and that $\varepsilon$ is small enough so that Lemma \ref{lem:proof gluing 2.1 - negativity} holds, and consequently $\partial_vr(1;\alpha)>0, \partial_ur(1;\alpha)<0$. This is already enough to give a satisfactory upper bound on the final renormalised Hawking mass: 
\begin{lemma}
    \label{lemma:proof gluing 2.1 - upper bound mass}
    For $\varepsilon$ small enough so that Lemma \ref{lem:proof gluing 2.1 - negativity} holds, and $0<|\alpha|^2<\delta$, we have  
    \begin{equation}
        \widetilde{m}(1;\alpha)<\frac{1}{3\sqrt{\Lambda}}.
    \end{equation}
\end{lemma}   
\begin{proof}   
    For $\varepsilon$ small enough so that Lemma \ref{lem:proof gluing 2.1 - negativity} holds, and $0<|\alpha|^2<\delta$, we have $\partial_vr(1;\alpha)>0, \partial_ur(1;\alpha)<0$, and hence
    \begin{equation}
        \widetilde{m}(1;\alpha)=\frac{r(1;\alpha)}2\bigg(1+4\partial_vr(1;\alpha)\partial_ur(1;\alpha)-\frac13\Lambda r(1;\alpha)^2\bigg)<\frac{r(1;\alpha)}2\bigg(1-\frac13\Lambda r(1;\alpha)^2\bigg).
        \label{eq: proof gluing 2.1 - estimate final mass 1}
    \end{equation}
    Let us consider 
    \begin{equation}
        \xi(r)\doteq r\Big(1-\frac13\Lambda r^2\Big).
    \end{equation}
    Note that $\xi(r)$ has a maximum at $r=\frac1{\sqrt{\Lambda}}$ and is strictly decreasing for $\frac1{\sqrt{\Lambda}}\leq r \leq \sqrt{\frac{3}{\Lambda}}$. In view of this, and remembering that $r(1;\alpha)>r(0;\alpha)\geq\frac{1}{\sqrt{\Lambda}}$, we conclude 
    \begin{equation}
        \widetilde{m}(1;\alpha)<\frac12\xi\big(r(1;\alpha)\big)<\frac12\xi\bigg(\frac{1}{\sqrt{\Lambda}}\bigg)=\frac{1}{3\sqrt{\Lambda}}.
    \end{equation}
\end{proof}

We will see now that, by possibly choosing $\varepsilon$ even smaller, we can also ensure the desired lower bound for the final renormalised Hawking mass. 
\begin{lemma}
    \label{lem:proof gluing 2.1 - lower bound mass}
    By possibly choosing $\varepsilon$ even smaller, we can find a constant $\underline{\delta}>0$  depending on $r_0$, $\varepsilon$, $\mathfrak{c}$ and the baseline scalar field profile, such that  if $\underline{\delta}<|\alpha|^2<\delta$, then
    \begin{equation}
        \widetilde{m}(1;\alpha)>\frac{\mathfrak{c}}{3\sqrt{\Lambda}}.
        \label{eq: proof gluing 2.1 - mass lower statement}
    \end{equation}
\end{lemma}
\begin{proof}
    Integrating \eqref{eq:gluing - v eqn for renorm Hawking mass}, we find
    \begin{equation}
        \widetilde{m}(1;\alpha)-\widetilde{m}(0;\alpha)=\int_0^1\bigg(2r\big(-r\partial_ur\big)|\alpha|^2\big(\partial_v\phi_{\hat{\alpha}}\big)^2+\frac12\mathfrak{m}^2r^2\partial_vr|\alpha|^2\phi^2_{\hat{\alpha}}\bigg)dv.
        \label{eq: proof gluing 2.1 - mass first}
    \end{equation}
    We are assuming  $0<|\alpha|^2<\delta$ and $\varepsilon$ small enough so that both Lemma \ref{lem:proof gluing 2.1 - positivity} and Lemma \ref{lem:proof gluing 2.1 - negativity} hold. Hence $\partial_vr$ is positive and decreasing, and $r$ is increasing and satisfies $r(v;\alpha)\geq\frac{1}{\sqrt{\Lambda}}$ for all $v\in[0,1]$. Thus
    \begin{equation}
        \partial_v(r\partial_ur)(v;\alpha)\geq\frac14\bigg(\frac{1}{\Lambda}\mathfrak{m}^2|\alpha|^2\phi^2_{\hat{\alpha}}(v)\bigg)\geq0,
    \end{equation}
    that is, $r\partial_ur$ is also monotone increasing. Therefore, from \eqref{eq: proof gluing 2.1 - mass first}, we can write
    \begin{align}
        \widetilde{m}(1;\alpha)\geq 2r(0;\alpha)\big((-r\partial_ur)(1;\alpha)\big)|\alpha|^2C'_{\mathrm{low}}+\frac12\mathfrak{m}^2r(0;\alpha)^2\partial_vr(1;\alpha)|\alpha|^2C_{\mathrm{low}}
        \label{eq: proof gluing 2.1 - m lower pre 1}, 
    \end{align}
        
    Now, introducing the estimates \eqref{eq: proof gluing 2.1 - upper estimate r} and \eqref{eq: proof gluing 2.1 - lower estimate partial_u r} from Remark \ref{rmk: proof gluing 2.1 - estimates r partial_v r}, and \eqref{eq: proof of gluing 2.1 rpartial_u r estimate} from Lemma \ref{lem:proof gluing 2.1 - negativity}, we obtain
    \begin{equation}
        \begin{aligned}
            \widetilde{m}(1;\alpha)\geq\frac12\Bigg(r(0;\alpha)&\bigg(1+\frac{r(0;\alpha)}{\partial_vr(0;\alpha)}\Big(1-\frac13\Lambda r(0;\alpha)^2\Big)-\big(r(0;\alpha)+\partial_vr(0;\alpha)\big)^2\big(\Lambda+\mathfrak{m}^2|\alpha|^2C_{\mathrm{upp}}\big)\bigg)|\alpha|^2C'_{\mathrm{low}}\\[-0.1em]&\hspace{8em}+\mathfrak{m}^2r(0;\alpha)^2\Big(\partial_vr(0;\alpha)-\big(r(0;\alpha)+\partial_vr(0;\alpha)\big)|\alpha|^2C'_{\mathrm{upp}}\Big)|\alpha|^2C_{\mathrm{low}}\Bigg)
        \end{aligned}
        \label{eq: proof gluing 2.1 - m lower}
    \end{equation}
    The right-hand side of this inequality can be rearranged as
    \begin{align}
        \frac{r(0;\alpha)}{2}\Bigg(\bigg(1+&\frac{r(0;\alpha)}{\partial_vr(0;\alpha)}\Big(1-\frac13\Lambda r(0;\alpha)^2\Big)-\big(r(0;\alpha)+\partial_vr(0;\alpha)\big)^2\Lambda\bigg)C'_{\mathrm{low}}+\mathfrak{m}^2r(0;\alpha) \partial_vr(0;\alpha)C_{\mathrm{low}}\Bigg)|\alpha|^2
        \\[0.5em]&-\frac{\mathfrak{m}^2}{2}\bigg(r(0;\alpha)\big(r(0;\alpha)+\partial_vr(0;\alpha)\big)^2C'_{\mathrm{low}}C_{\mathrm{upp}}+r(0;\alpha)^2\big(r(0;\alpha)+\partial_vr(0;\alpha)\big)C_{\mathrm{low}}C'_{\mathrm{upp}}\bigg)|\alpha|^4,
    \end{align}
    and hence, in order to ensure \eqref{eq: proof gluing 2.1 - mass lower statement} and complete the proof of the lemma, we want to show that we can find $0<|\alpha|^2<\delta$ satisfying
    \begin{align}
        \frac{r(0;\alpha)}{2}\Bigg(\bigg(&1+\frac{r(0;\alpha)}{\partial_vr(0;\alpha)}\Big(1-\frac13\Lambda r(0;\alpha)^2\Big)-\big(r(0;\alpha)+\partial_vr(0;\alpha)\big)^2\Lambda\bigg)C'_{\mathrm{low}}+\mathfrak{m}^2r(0;\alpha) \partial_vr(0;\alpha)C_{\mathrm{low}}\Bigg)|\alpha|^2
        \\[0.5em]&-\frac{\mathfrak{m}^2}{2}\bigg(r(0;\alpha)\big(r(0;\alpha)+\partial_vr(0;\alpha)\big)^2C'_{\mathrm{low}}C_{\mathrm{upp}}+r(0;\alpha)^2\big(r(0;\alpha)+\partial_vr(0;\alpha)\big)C_{\mathrm{low}}C'_{\mathrm{upp}}\bigg)|\alpha|^4 >\frac{\mathfrak{c}}{3\sqrt{\Lambda}}.
        \label{eq: proof gluing 2.1 - aim first statement}
    \end{align}
    We analyse now the massless and massive cases separately. 

    \vspace{0.5em}

    \noindent\underline{Case $\mathfrak{m}=0$} 

    \noindent Setting $\mathfrak{m}=0$ in \eqref{eq: proof gluing 2.1 - aim first statement} gives  
    \begin{equation}
        \frac{r(0;\alpha)}{2}\bigg(1+\frac{r(0;\alpha)}{\partial_vr(0;\alpha)}\Big(1-\frac13\Lambda r(0;\alpha)^2\Big)-\big(r(0;\alpha)+\partial_vr(0;\alpha)\big)^2\Lambda\bigg)C'_{\mathrm{low}}|\alpha|^2> \frac{\mathfrak{c}}{3\sqrt{\Lambda}},
    \end{equation}
    or, written in terms of the prescribed initial values,
    \begin{equation}
        \frac{r_0}{2}\bigg(1+\frac{r_0}{\varepsilon}\Big(1-\frac{r_0^2}{3}\Big)-\big(r_0+\varepsilon\big)^2\bigg)C'_{\mathrm{low}}|\alpha|^2=\frac12\frac{r_0}{\varepsilon}f(\varepsilon)C'_{\mathrm{low}}|\alpha|^2> \frac{\mathfrak{c}}{3}. 
        \label{eq: proof gluing 2.1 - pre condition massless 1}
    \end{equation}

    Since we are assuming that $\varepsilon$ is small enough so that $\mathcal{E}(\varepsilon)>0$, \eqref{eq: proof gluing 2.1 - pre condition massless 1} is equivalent to 
    \begin{equation}
        |\alpha|^2>\underline{\delta}\doteq\frac{2}{3}\frac{1}{C'_{\mathrm{low}}} \frac{\varepsilon}{r_0}\frac{\mathfrak{c}}{\mathcal{E}(\varepsilon)}. 
        \label{eq: proof gluing 2.1 - condition massless 1}
    \end{equation}
    We now need to make sure this condition is compatible with $|\alpha|^2<\delta$. Given that
    \begin{equation}
        \delta= \frac{1}{C'_{\mathrm{upp}}}\frac{\varepsilon}{r_0+\varepsilon},
    \end{equation}
    it suffices to show that we can choose $\varepsilon$ small enough so that $\mathcal{E}(\varepsilon)>0$ and moreover
    \begin{equation}
        \delta-\underline{\delta}=\frac{1}{C'_{\mathrm{upp}}}\frac{\varepsilon}{r_0+\varepsilon}-\frac{2}{3}\frac{1}{C'_{\mathrm{low}}} \frac{\varepsilon}{r_0}\frac{\mathfrak{c}}{\mathcal{E}(\varepsilon)}>0,
    \end{equation}
    or equivalently 
    \begin{equation}
        F(\varepsilon)\doteq \frac{3}{2}\frac{r_0}{r_0+\varepsilon}\mathcal{E}(\varepsilon)\frac{C'_{\mathrm{low}}}{C'_{\mathrm{upp}}}-\mathfrak{c}>0.
    \end{equation}

    Note that 
    \begin{equation}
        F(0)= \frac{3}{2}r_0\Big(1-\frac{r_0^2}{3}\Big)\frac{C'_{\mathrm{low}}}{C'_{\mathrm{upp}}}-\mathfrak{c}.
    \end{equation}
    By choosing the scalar field profile so that $C'_{\mathrm{low}}=C'_{\mathrm{upp}}$ (recall Remark \ref{rmk:norm scalar field}), we can ensure
    \begin{equation}
        F(0)=\frac{3}{2}r_0\Big(1-\frac{r_0^2}{3}\Big)-\mathfrak{c}> 3\cos\bigg(\frac{\cos^{-1}(-\mathfrak{c})}{3}\bigg)\bigg(1-\frac43\cos^2\bigg(\frac{\cos^{-1}(-\mathfrak{c})}{3}\bigg)\bigg)-\mathfrak{c}=0,
    \end{equation}
    where we used that $r_0\Big(1-\frac{r_0^2}{3}\Big)$ is a decreasing function  of $r_0$ for $1\leq r_0 < 2\cos\big(\frac{\cos^{-1}(-\mathfrak{c})}{3}\big)$, and the same rearrangement of the triple angle identity that we used in Remark \ref{rmk:mass horizon spheres}. This ensures that, as long as $\varepsilon$ is small enough, we can choose $|\alpha|^2$ such that $\underline{\delta}<|\alpha|^2<\delta$, hence satisfying \eqref{eq: proof gluing 2.1 - aim first statement} too. 

    \newpage

    \noindent\underline{Case $\mathfrak{m}>0$} 

    \noindent Rewriting \eqref{eq: proof gluing 2.1 - aim first statement} in terms of the prescribed values, and setting $\mathfrak{m}^2=\beta\Lambda$, we find
    \begin{align}
        \frac{r_0}{2}\Bigg(\bigg(1+\frac{r_0}{\varepsilon}\Big(1&-\frac{r_0^2}{3}\Big)-\big(r_0+\varepsilon\big)^2\bigg)C'_{\mathrm{low}}+\beta r_0 \varepsilon C_{\mathrm{low}}\Bigg)|\alpha|^2
        \\[0.1em]-&\frac{\beta}{2}\bigg(r_0\big(r_0+\varepsilon\big)^2C'_{\mathrm{low}}C_{\mathrm{upp}}+r_0^2\big(r_0+\varepsilon\big)C_{\mathrm{low}}C'_{\mathrm{upp}}\bigg)|\alpha|^4> \frac{\mathfrak{c}}{3}.
        \label{eq: proof gluing 2.1 - pre condition massive 1}
    \end{align}
    Thus, in order to prove the Lemma and ensure \eqref{eq: proof gluing 2.1 - mass lower statement}, it suffices to show that we can find $0<|\alpha|^2<\delta$ such that 
    \begin{equation}
        \hspace{-0.15em}\mathfrak{q}\big(|\alpha|^2\big)\doteq \frac32\beta\varepsilon\bigg(r_0\big(r_0+\varepsilon\big)^2C'_{\mathrm{low}}C_{\mathrm{upp}}+r_0^2\big(r_0+\varepsilon\big)C_{\mathrm{low}}C'_{\mathrm{upp}}\bigg)|\alpha|^4-\frac32\bigg(r_0\mathcal{E}(\varepsilon)C'_{\mathrm{low}}+\beta r_0^2 \varepsilon^2 C_{\mathrm{low}}\bigg)|\alpha|^2+\varepsilon\mathfrak{c}<0,
        \label{eq: proof gluing 2.1 - condition massive 1}
    \end{equation}
    with $\mathfrak{q}$ a quadratic polynomial whose discriminant is given by 
    \begin{equation}
        \Delta(\mathfrak{q})=\frac94\bigg(r_0\mathcal{E}(\varepsilon)C'_{\mathrm{low}}+\beta r_0^2 \varepsilon^2 C_{\mathrm{low}}\bigg)^2-6\mathfrak{c}\beta\varepsilon^2\bigg(r_0\big(r_0+\varepsilon\big)^2C'_{\mathrm{low}}C_{\mathrm{upp}}+r_0^2\big(r_0+\varepsilon\big)C_{\mathrm{low}}C'_{\mathrm{upp}}\bigg).
        \label{eq: proof gluing 2.1 - discriminant}
    \end{equation}

    Remember that $\varepsilon$ is small enough so that $\mathcal{E}(\varepsilon)>0$. Now, by making $\varepsilon$ possibly even smaller, we can ensure $\Delta(\mathfrak{q})$ is positive, since the dominant term in \eqref{eq: proof gluing 2.1 - discriminant} if $\varepsilon$ is very small will be
    \begin{equation}
        \frac94r_0^2\mathcal{E}(\varepsilon)^2 \big(C'_{\mathrm{low}}\big)^2,
    \end{equation}
    which is positive and finite as $\varepsilon\rightarrow0$. This means that for $\varepsilon$ small enough, $\mathfrak{q}$ has two real roots, and analysing further its coefficients, we conclude they are both positive. Since $\mathfrak{q}(0)=\varepsilon \mathfrak{c}>0$, in order to have \eqref{eq: proof gluing 2.1 - condition massive 1}, we will need $|\alpha|^2$ to be greater than the smallest of the roots of $\mathfrak{q}$. Consequently, we will define $\underline{\delta}$ as the smallest root of $\mathfrak{q}$ in this case. We can write $\mathfrak{q}$ schematically as
    \begin{equation}
        \mathfrak{q}\big(|\alpha|^2\big)=a_{\varepsilon}|\alpha|^4+b_{\varepsilon}|\alpha|^2+c_{\varepsilon}, 
    \end{equation}
    and hence
    \begin{equation}
        \underline{\delta}\doteq-\frac{b_{\varepsilon}+\sqrt{b_{\varepsilon}^2-4a_{\varepsilon}c_{\varepsilon}}}{2a_{\varepsilon}}. 
    \end{equation}

    Summarising, in order to complete the proof of the lemma (in the massive case), we need to show that we can find $|\alpha|^2$ satisfying
    \begin{equation}
        -\frac{b_{\varepsilon}+\sqrt{b_{\varepsilon}^2-4a_{\varepsilon}c_{\varepsilon}}}{2a_{\varepsilon}}=\underline{\delta}<|\alpha|^2<\delta=\frac{1}{C'_{\mathrm{upp}}}\frac{\varepsilon}{r_0+\varepsilon}.
    \end{equation}
    by choosing $\varepsilon$ appropriately small.  Equivalently, we need to show that  
    \begin{equation}
        \bar{F}(\varepsilon)\doteq2a_{\varepsilon}\varepsilon+C'_{\mathrm{upp}}(r_0+\varepsilon)b_{\varepsilon}+C'_{\mathrm{upp}}(r_0+\varepsilon)\sqrt{b_{\varepsilon}^2-4a_{\varepsilon}c_{\varepsilon}}>0
        \label{eq: proof gluing 2.1 - condition massive 2}
    \end{equation}
    for $\varepsilon$ appropriately small. 

    Expanding the expression for $\bar{F}$ we get 
    \begin{align}
        \bar{F}(\varepsilon)&=3\beta\varepsilon^2\Big(r_0\big(r_0+\varepsilon\big)^2C'_{\mathrm{low}}C_{\mathrm{upp}}+r_0^2\big(r_0+\varepsilon\big)C_{\mathrm{low}}C'_{\mathrm{upp}}\Big)-\frac32C'_{\mathrm{upp}}(r_0+\varepsilon)\Big(r_0\mathcal{E}(\varepsilon)C'_{\mathrm{low}}+\beta r_0^2 \varepsilon^2 C_{\mathrm{low}}\Big)
        \\[0.5em]&+C'_{\mathrm{upp}}(r_0+\varepsilon)\sqrt{\frac94\Big(r_0\mathcal{E}(\varepsilon)C'_{\mathrm{low}}+\beta r_0^2 \varepsilon^2 C_{\mathrm{low}}\Big)^2-6\mathfrak{c}\beta\varepsilon^2\Big(r_0\big(r_0+\varepsilon\big)^2C'_{\mathrm{low}}C_{\mathrm{upp}}+r_0^2\big(r_0+\varepsilon\big)C_{\mathrm{low}}C'_{\mathrm{upp}}\Big)}.
    \end{align}
    In order to be able to say something about the sign of $\bar{F}(\varepsilon)$ for $\varepsilon$ small, we would like to know the sign of $\bar{F}$ at $\varepsilon=0$ and its local monotonicity properties around that point. Given that
    \begin{equation}
        \bar{F}(0)=-\frac32C'_{\mathrm{upp}}r_0^2\mathcal{E}(0)C'_{\mathrm{low}}+C'_{\mathrm{upp}}r_0\sqrt{\frac94\big(r_0\mathcal{E}(0)C'_{\mathrm{low}}\big)^2}=0,
    \end{equation}
    we compute $\bar{F}'(0)$, obtaining:
    \begin{align}
        \bar{F}'(0)=-\frac32C'_{\mathrm{low}}C'_{\mathrm{upp}}r_0\mathcal{E}(0)-\frac32C'_{\mathrm{low}}C'_{\mathrm{upp}}r_0^2\mathcal{E}'(0)+\frac32C'_{\mathrm{low}}C'_{\mathrm{upp}}r_0\mathcal{E}(0)+\frac32C'_{\mathrm{low}}C'_{\mathrm{upp}}r_0^2\mathcal{E}'(0)=0.
    \end{align}
    Because the first derivative at $\varepsilon=0$ also vanishes, we differentiate once more and evaluate at $\varepsilon=0$, finding
    \begin{align}
        \bar{F}''(0) = 6\beta r_0^3\big(C'_{\mathrm{low}}C_{\mathrm{upp}}+C_{\mathrm{low}}C'_{\mathrm{upp}}\big)\bigg(1-\frac23\frac{\mathfrak{c}}{\mathcal{E}(0)}\frac{C'_{\mathrm{upp}}}{C'_{\mathrm{low}}}\bigg). 
    \end{align}
    
    Hence, as in the massless case, by choosing the scalar field profile so that $C'_{\mathrm{low}}=C'_{\mathrm{upp}}=C'$, and given that $r_0$ satisfies
    \begin{equation}
        \frac{3}{2}r_0\Big(1-\frac{r_0^2}{3}\Big)> 3\cos\bigg(\frac{\cos^{-1}(-\mathfrak{c})}{3}\bigg)\bigg(1-\frac43\cos^2\bigg(\frac{\cos^{-1}(-\mathfrak{c})}{3}\bigg)\bigg)=\mathfrak{c},
    \end{equation}
    we get
    \begin{equation}
        \bar{F}''(0)=6\beta r_0^3C'\big(C_{\mathrm{upp}}+C_{\mathrm{low}}\big)\left(1-\frac{\mathfrak{c}}{\frac32r_0\Big(1-\frac{r_0^2}{3}\Big)}\right)>0. 
    \end{equation}
    Therefore, for $\varepsilon$ small enough, we can ensure $\bar{F}(\varepsilon)>0$, since for $\varepsilon$ close to $0$ we have
    \begin{equation}
        \bar{F}(\varepsilon)=\bar{F}(0)+\bar{F}'(0)\varepsilon+\frac{\bar{F}''(0)}{2}\varepsilon^2+\mathcal{O}\big(\varepsilon^3\big)=\frac{\bar{F}''(0)}{2}\varepsilon^2+\mathcal{O}\big(\varepsilon^3\big),
    \end{equation}
    and the term involving $\varepsilon^2$ will dominate. 
    Therefore, by choosing $\varepsilon$ appropriately small, we can find $|\alpha|^2$ such that $\underline{\delta}<|\alpha|^2<\delta$, and thus satisfying \eqref{eq: proof gluing 2.1 - aim first statement} too.
\end{proof}

In what follows we will rewrite 
\begin{equation}
    \widetilde{m}(1;\alpha)=\frac{\tilde{\mathfrak{c}}}{3\sqrt{\Lambda}}.
\end{equation}
If $\varepsilon$ is small enough and $\underline{\delta}<|\alpha|^2<\delta$, then $\mathfrak{c}<\widetilde{\mathfrak{c}}<1$, as a consequence of Lemma \ref{lemma:proof gluing 2.1 - upper bound mass} and Lemma \ref{lem:proof gluing 2.1 - lower bound mass}.
\begin{remark}
    Note that, if $\partial_vr(1;\alpha)>0$ and $\partial_ur(1;\alpha)<0$, then
    \begin{equation}
        f_{\tilde{\mathfrak{c}}}\big(r(1;\alpha)\big)=1-\frac{2}{r(1;\alpha)}\frac{\tilde{\mathfrak{c}}}{3\sqrt{\Lambda}}-\frac13\Lambda r(1;\alpha)^2=-4\partial_vr(1;\alpha)\partial_ur(1;\alpha)>0, 
    \end{equation}
    and thus $r(1;\alpha)$ must lie between the two roots of $f_{\tilde{\mathfrak{c}}}(r)$, so, in particular, $r(1;\alpha)<r_{\overline{\mathcal{H}}}(\tilde{\mathfrak{c}})$.
    \label{rmk: proof gluing 2.1 - estimate final radius}
\end{remark}

Given $\varepsilon$ small enough so that Lemma \ref{lem:proof gluing 2.1 - negativity} and Lemma \ref{lem:proof gluing 2.1 - lower bound mass} hold, and $\alpha\in\mathbb S^k_{\sqrt{\delta_*}}$, for some $\delta_*>0$ such that $\underline{\delta}<\delta_*<\delta$, we define $D_{\alpha}(0)$ as the sphere data set satisfying: 
\begin{itemize}
    \item $\varrho=R_i$, with $\frac{1}{\sqrt{\Lambda}}\leq R_i< r_{\overline{\mathcal{H}}}({\mathfrak{c}})$,
    \item $\varrho^1_v=\frac{\varepsilon}{\sqrt{\Lambda}}>0$,
    \item $\varrho^1_u=-\frac{\sqrt{\Lambda}}{4\varepsilon}\big(1-\frac13\Lambda R_i^2\big)<0$,
    \item $\omega =1$,
    \item all other components are zero.
\end{itemize}

\vspace{0.5em}

Therefore, $D_{\alpha}(0)$ is gauge equivalent to the de Sitter static sphere $_{\mathcal{S}}D^{\mathrm{dS}}_{R_i, k}$ with $\varrho^1_u<0$. Now, with $D_{\alpha}(0)$ and using $\phi_{\alpha}$ as a seed, we generate a cone data set $D_{\alpha}:[0,1]\longrightarrow \mathcal{D}_k$. Then the sphere data set $D_{\alpha}(1)$ will satisfy
\begin{itemize}
    \item $\varrho=R_1\doteq r(1;\alpha)$, with $R_1<r_{\overline{\mathcal{H}}}(\tilde{\mathfrak{c}})$ (see Remark \ref{rmk: proof gluing 2.1 - estimate final radius}),
    \item $\varrho^1_v=\partial_vr(1;\alpha)>0$ (see Lemma \ref{lem:proof gluing 2.1 - positivity}),
    \item $\varrho^1_u=\partial_ur(1;\alpha)<0$ (see Lemma \ref{lem:proof gluing 2.1 - negativity}),
    \item $\widetilde{m}\big(D_{\alpha}(1)\big)=\frac{\tilde{\mathfrak{c}}}{3\sqrt{\Lambda}}$, with $\mathfrak{c}<\widetilde{\mathfrak{c}}<1$ (see Lemma \ref{lemma:proof gluing 2.1 - upper bound mass} and Lemma \ref{lem:proof gluing 2.1 - lower bound mass}),
    \item $\omega =1$,
    \item $\varphi^i_v=0$ for all $0\leq i \leq k$.
\end{itemize}

Since this is true for any $\alpha\in\mathbb S^k_{\sqrt{\delta_*}}$, and the parities of $r(v;\alpha)$, $\Omega^2(v;\alpha)$, $\phi(v;\alpha)$ and their $u$- and $v$-derivatives as functions of $\alpha$ are the same as the ones discussed in Lemma \ref{lem:parity lemma}, we can once again apply the Borsuk--Ulam theorem to conclude that there exists $\alpha_*\in\mathbb S^k_{\sqrt{\delta_*}}$ such that $\varphi^i_u(D_{\alpha_*}(1))=0$ for all $1\leq i \leq k$, and therefore  $D_{\alpha_*}(1)$ is gauge equivalent to the Schwarzschild--de Sitter static sphere $_\mathcal{S}D^{\mathrm{SdS}}_{\tilde{\mathfrak{c}},R_1, k}$ with $\varrho^1_u<0$.

\subsubsection*{Step 3: gluing $_{\mathcal{C}}D^{\mathrm{SdS}}_{\tilde{\mathfrak{c}}, R_2, k}$ to $_{\overline{\mathcal{H}}^+}D^{\mathrm{SdS}}_{\mathfrak{c}, k}$}

We start prescribing
\begin{equation}
    \begin{aligned}
        r(1;\alpha&)=  r_{\overline{\mathcal{H}}}(\mathfrak{c}),\\
        \partial_vr(1;\alpha&)= 0.
        \label{eq: proof gluing 2.3 - initialisation}
    \end{aligned}
\end{equation}
We will moreover fix a radius $R_2$, with $r_{\overline{\mathcal{H}}}(\tilde{\mathfrak{c}})<R_2<r_{\overline{\mathcal{H}}}(\mathfrak{c})$.  Repeating the bootstrap argument laid out in Lemma \ref{lem:proof gluing 1 - positivity}, we conclude that there exists a constant $\delta >0$, depending on the baseline scalar field profile, $R_2$ and $\mathfrak{c}$ such that, if $0<|\alpha|^2\leq\delta$, then, for $v\in[0,1]$, 
\begin{equation}
    \begin{aligned}
        r(v;\alpha&)\geq R_2, \\
        \partial_vr(v;\alpha&)\geq  0,\\
        \partial_vr(0;\alpha&)>0.
    \label{eq: proof gluing 2.3 - positivity}
    \end{aligned}
\end{equation}
For $\alpha\in \mathbb S^k_{\sqrt{\delta}}$, we define $D_{\alpha}(0)$ to be the sphere data set with: 
\begin{itemize}
    \item $\varrho=r(0;\alpha)$,
    \item $\varrho^1_v=\partial_vr(0;\alpha)>0$,
    \item $\varrho^1_u=-\frac{1}{4\partial_vr(0;\alpha)}\Big(1-\frac{2\tilde{\mathfrak{c}}}{3\sqrt{\Lambda}r(0;\alpha)}-\frac13\Lambda r^2(0;\alpha)\Big)$,
    \item $\omega =1$,
    \item all other components are zero. 
\end{itemize}
Note that 
\begin{equation}
    1-\frac{2\tilde{\mathfrak{c}}}{3\sqrt{\Lambda}r(0;\alpha)}-\frac13\Lambda r^2(0;\alpha)=f_{\tilde{\mathfrak{c}}}\big(r(0;\alpha)\big)<0,
\end{equation}
since $ r(0;\alpha)\geq R_2>r_{\overline{\mathcal{H}}}(\tilde{\mathfrak{c}})$, and thus $\varrho^1_u\big(D_{\alpha}(0)\big)>0$. Moreover, 
\begin{equation}
    \widetilde{m}\left(D_{\alpha}(0)\right)=\frac{\varrho}{2}\left(1+4\varrho^1_v\varrho^1_u-\frac13\Lambda\varrho^2\right)=\frac{\tilde{\mathfrak{c}}}{3\sqrt{\Lambda}}.
\end{equation}
Hence $D_{\alpha}(0)$ is gauge equivalent to the Schwarzschild--de Sitter cosmological sphere $_{\mathcal{C}}D^{SdS}_{\tilde{\mathfrak{c}},r(0;\alpha), k}$ with $\varrho^1_u>0$. With $D_{\alpha}(0)$ and using $\phi_{\alpha}$ as a seed, we generate a cone data set $D_{\alpha}:[0,1]\longrightarrow \mathcal{D}_k$. 

Given that $r(v;\alpha)\geq r(0;\alpha) > r_{\overline{\mathcal{H}}}(\tilde{\mathfrak{c}})>1/{\sqrt{\Lambda}}$ for all $v\in[0,1]$, it is clear that 
\begin{equation}
    \partial_v(r\partial_ur)(v;\alpha)=\frac14\left( r^2(v;\alpha)\big(\Lambda+\mathfrak{m}^2|\alpha|^2\phi_{\hat{\alpha}}^2\big)-1\right)>0.
\end{equation}
This, together with the fact that $\partial_ur(0;\alpha)=\varrho^1_u\big(D_{\alpha}(0)\big)>0$, guarantees that  $\partial_ur(1;\alpha)>0$. Therefore, $D_{\alpha}(1)$ has:
\begin{itemize}
    \item $\varrho=r(1;\alpha)=r_{\overline{\mathcal{H}}}(\mathfrak{c})$,
    \item $\varrho^1_v=\partial_vr(1;\alpha)=0$,
    \item $\varrho^1_u=\partial_ur(1;\alpha)>0$,
    \item $\omega =1$, 
    \item $\varphi^i_v=0$ for all $0\leq i \leq k$.
\end{itemize}

This is true for any $\alpha\in\mathbb S^k_{\sqrt{\delta}}$, and given that the parities of the metric and scalar field quantities are the same as the ones discussed in Lemma \ref{lem:parity lemma}, we can once again apply the Borsuk--Ulam theorem to conclude that there exists $\alpha_*\in\mathbb S^k_{\sqrt{\delta}}$ such that $\varphi^i_u(D_{\alpha_*}(1))=0$ for all $1\leq i \leq k$, and therefore  $D_{\alpha_*}(1)$ is gauge equivalent to the Schwarzschild--de Sitter cosmological horizon sphere $_{\overline{\mathcal{H}}^+}D^{\mathrm{SdS}}_{\mathfrak{c}, k}$, with $\varrho^1_u>0$. 

In the case that $r(0;\alpha_*)>R_2$, we trivially glue the sphere $_{\mathcal{C}}D^{\mathrm{SdS}}_{\tilde{\mathfrak{c}}, R_2, k}$ with $\varrho^1_u>0$ to the sphere $_{\mathcal{C}}D^{\mathrm{SdS}}_{\tilde{\mathfrak{c}}, r(0;\alpha_*), k}$ with $\varrho^1_u>0$, to complete the Step 3 gluing.

\subsubsection*{Step 2: gluing $_{\mathcal{S}}D^{\mathrm{SdS}}_{\tilde{\mathfrak{c}}, R_1, k}$ to $_{\mathcal{C}}D^{\mathrm{SdS}}_{\tilde{\mathfrak{c}}, R_2, k}$}

Now that the gluings in both Step 1 and Step 3 are complete, we simply need to make sure that $R_1<R_2$ to justify that this intermediate trivial gluing is possible. Indeed, we will have 
\begin{equation}
    R_1<r_{\overline{\mathcal{H}}}(\tilde{\mathfrak{c}})<R_2,
\end{equation}
where the first inequality was argued in Remark \ref{rmk: proof gluing 2.1 - estimate final radius}, and the second one follows from the definition we made for $R_2$. 

Note that in this gluing with trivial scalar field, as it follows from Raychaudhuri's equation \eqref{eq:gluing - Raychaudhuri's v}, we will have $\partial_vr$ constant. One could be concerned that the value of $\varrho^1_v$ at $_{\mathcal{S}}D^{\mathrm{SdS}}_{\tilde{\mathfrak{c}}, R_1, k}$ and at $_{\mathcal{C}}D^{\mathrm{SdS}}_{\tilde{\mathfrak{c}}, R_2, k}$ may not coincide if we take them directly from the previous gluing constructions. Recall that the remaining gauge freedom allows us to rescale\footnote{Recall that our definition of characteristic gluing is up to gauge transformations.} $\varrho^1_v$ in any of these two spheres to make it coincide with the value at the other sphere, making possible then the characteristic gluing with trivial scalar field.

\subsection{Proof of Theorem \ref{thm:impossible gluing}}
\label{subsec:proof impossible gluing}
Recall that Theorem \ref{thm:impossible gluing} is not a gluing result, and consequently the set-up to prove it will be different to the one in previous sections. We start considering a spherically symmetric solution to the Einstein-massless scalar field system with $\Lambda>0$, $\big(\mathcal{Q},r,\Omega^2,\phi\big)$, endowed with double null coordinates $(u,v)$, and ${C}\doteq\{u=u_0\}\cap\mathcal{Q}$ a null cone in $\mathcal{Q}$. We set $\Omega^2\equiv 1$ along ${C}$ (cf. Remark \ref{rmk:lapse-normalised characteristic initial data}), and we see $r$, $\phi$ and their $u$- and $v$- derivatives as functions of $v$ along ${C}$. Following the statement of Theorem \ref{thm:impossible gluing}, we assume 
\begin{equation}
    \begin{aligned}
        r(v_1) &= \frac{1}{\sqrt{\Lambda}},\\
        \partial_vr(v_1) &= 0,\\[0.7em]
        \partial_ur(v_1) &\geq 0,
        \label{eq: proof no gluing - initialisation}
    \end{aligned}
\end{equation}
and we want to show that it is not possible for any of the spheres on ${C}\cap\{v<v_1\}$ to be gauge equivalent to a subextremal Schwarzschild--de Sitter or de Sitter static sphere with $\partial_ur<0$.

From Raychaudhuri's equation we deduce that $\partial_vr$ is a decreasing function of $v$, since
\begin{equation}
    \partial_v^2 r=-r\big(\partial_v\phi\big)^2\leq0.
    \label{eq:proof no gluing - Raychaudhuri's v}
\end{equation}
Given that $\partial_vr(v_1)= 0$, we must have $\partial_vr(v)\geq 0$ for $v< v_1$, and hence $r(v)\leq r(v_1)$ for $v< v_1$. This also implies, looking at the wave equation for $r$ rewritten as in \eqref{eq:EE-massive scalar field - eqn for r partial_ur} with $\mathfrak{m}=0$:
\begin{equation}
    \partial_v\big(r\partial_u r \big)=\frac14\big(\Lambda r^2-1)\leq0.
    \label{eq:proof no gluing - r partial_ur}
\end{equation}
As $r\partial_ur (v_1)\geq0$, we must have $r\partial_ur (v)\geq0$ for $v<v_1$, and thus $\partial_ur(v)\geq0$ for all $v<v_1$. In particular, this rules out the possibility of having, along ${C}\cap\{v<v_1\}$, spheres with $\partial_vr>0$ and $\partial_ur<0$, that is, spheres which are gauge equivalent to static spheres (in the subextremal Schwarzschild--de Sitter or de Sitter spacetimes).

Note that the fact that $\mathfrak{m}=0$ is crucial for this argument. If $\mathfrak{m}>0$, then the contribution of the  term involving $\mathfrak{m}^2$ which is not present in \eqref{eq:proof no gluing - r partial_ur} could make $\partial_v\big(r\partial_ur)$ positive even for $r<1/{\sqrt{\Lambda}}$. And, indeed, that is precisely what Theorem \ref{thm:subext SdS to Nariai} and Theorem \ref{thm:subext SdS to ext COSMhorizon} reflect, that this gluing is possible if $\mathfrak{m}$ is positive.

\subsection{Proof of gluing to a Nariai sphere (Theorem \ref{thm:subext SdS to Nariai})}
\label{subsec:proof subext SdS to Nariai}

It is clear that the 3-gluing strategy used to prove Theorem \ref{thm:dS to subext COSMhorizon} cannot be used to prove the gluing to the extremal cosmological horizon or to Nariai spheres, as it is not possible to find an intermediate subextremal spacetime with greater mass than the extremal solution. Our approach to prove Theorem \ref{thm:subext SdS to ext COSMhorizon} and Theorem~\ref{thm:subext SdS to Nariai} shall therefore be different. Since the strategy in both cases will be similar, but algebraically less involved in the case of Nariai, we first present the proof of gluing to Nariai spheres for clarity. 

We start prescribing
\begin{equation}
    \begin{aligned}
        r(1;\alpha&)= \frac1{\sqrt{\Lambda}},\\
        \partial_vr(1;\alpha&)= 0,
        \label{eq: proof gluing 3 - initialisation}
    \end{aligned}
\end{equation}
Repeating the bootstrap argument laid out in Lemma \ref{lem:proof gluing 1 - positivity}, we conclude that there exists a constant $\delta$, explicitly given by\footnote{Note that this first part of the proof is identical to the proof for the gluing to the extremal black hole horizon in Section \ref{subsec:proof subext SdS to BHhorizon}, since $r_{\overline{\mathcal{H}}}(1)=r_{\mathcal{H}}(1)$ and we have not yet introduced any assumptions about $\partial_ur$.}
\begin{equation}
    \delta \doteq \frac{1}{2C'_{\mathrm{upp}}},
    \label{eq: proof gluing 3 - def delta}
\end{equation} 
such that, if $0<|\alpha|^2\leq\delta$, then 
\begin{align}
    r(v;\alpha&)\geq \frac{1}{2\sqrt{\Lambda}},\label{eq: proof gluing 3 - positivity r}  \\
    \partial_vr(v;\alpha&)\geq  0, \label{eq: proof gluing 3 - positivity partial_vr}\\
    \partial_vr(0;\alpha&)>0. \label{eq: proof gluing 3 - positivity partial_vr initial}
\end{align}
for all $v\in[0,1]$. 

In this case, instead of prescribing the value of $\partial_ur$ at $v=0$, as we did in the gluing to the black hole horizon, we prescribe its value at $v=1$:
\begin{equation}
    (r\partial_ur)(1;\alpha)=0, 
    \label{eq: proof gluing 3 - initialization 2}
\end{equation}
We show now that, given $\mathfrak{m}>\sqrt{\Lambda}/2$, we can choose an appropriate baseline scalar field profile to achieve $\partial_ur(0;\alpha)<0$: 

\begin{lemma}
    \label{lem: proof gluing 3 - negativity partial_u r initial}
    If $\mathfrak{m}^2> \Lambda/4$, we can choose an appropriate dissection of $[0,1]$ and associated bump functions $\chi_1,\dots\chi_{k+1}$ so that: 
    \begin{equation}
        \partial_ur(0;\alpha)<0
        \label{eq: proof gluing 3 - initial partial_u r statement}
    \end{equation}
    for any $\alpha$ satisfying $0<|\alpha|^2< \delta$, with $\delta$ possibly smaller than \eqref{eq: proof gluing 3 - def delta} depending on $\mathfrak{m}$ and the baseline scalar field profile. 
\end{lemma}

\begin{proof}
    We assume $0<|\alpha|^2\leq\delta$, with $\delta$ given by \eqref{eq: proof gluing 3 - def delta}. Hence $r$ is monotone increasing and $\partial_vr$ monotone decreasing as functions of $v$. Integrating Raychaudhuri's equation and using the monotonicity of $r$ we find:
    \begin{equation} 
        \partial_vr(0;\alpha)=\int_0^1r|\alpha|^2 \big(\partial_v\phi_{\hat{\alpha}})^2dv \,\leq r(1;\alpha)|\alpha|^2C'_{\mathrm{upp}},
        \label{eq: proof gluing 3 - partial_v upper}
     \end{equation}
    Integrating in $v$ once more and making use of the monotonicity of $\partial_vr$, we find
    \begin{align}
        r(0;\alpha)&=r(1;\alpha)-\int_0^1\partial_vr\,dv\geq r(1;\alpha)-\partial_vr(0;\alpha)\geq r(1;\alpha)\big(1-|\alpha|^2C'_{\mathrm{upp}} \big).
        \label{eq: proof gluing 3 - r lower pre}
     \end{align}
    On the other hand, integrating equation \eqref{eq:gluing - eqn for r partial_ur}, we get
    \begin{align}
         -(r\partial_ur)(0;\alpha) =& \frac14\int_0^1 \Big(r^2\big(\Lambda + \mathfrak{m}^2|\alpha|^2\phi_{\hat{\alpha}}^2\big)-1\Big) dv= \frac14\bigg(-\int_0^1 \big(1- \Lambda r^2\big) dv+\int_0^1 r^2\mathfrak{m}^2|\alpha|^2\phi_{\hat{\alpha}}^2 dv\bigg). 
         \label{eq: proof gluing 3 - partial_u integral}
    \end{align}
    Note that 
    \begin{align}
         \int_0^1 \big(1-\Lambda r^2\big) dv \leq 2\int_0^1 \big(1-\sqrt{\Lambda} r\big) dv = 2\sqrt{\Lambda}\int_0^1 v\partial_vr\,dv=-2\sqrt{\Lambda} \int_0^1 \frac{v^2}{2}\partial^2_vr\,dv\leq|\alpha|^2\int_0^1 v^2\big(\partial_v\phi_{\hat{\alpha}}\big)^2dv.
    \end{align}
    where we have integrated by parts twice, substituted Raychaudhuri's equation in $v$, and used the fact that $r(v;\alpha)\leq 1/\sqrt{\Lambda}$ for all $v\in[0,1]$. Plugging this in \eqref{eq: proof gluing 3 - partial_u integral}, together with \eqref{eq: proof gluing 3 - r lower pre}, we obtain: 
    \begin{equation}
        -(r\partial_ur)(0;\alpha)\geq\frac{|\alpha|^2}4\bigg(\beta \big(1-|\alpha|^2C'_{\mathrm{upp}} \big)^2\int_0^1\phi_{\hat{\alpha}}^2 dv - \int_0^1 v^2\big(\partial_v\phi_{\hat{\alpha}}\big)^2dv\bigg),
    \end{equation}
    where we have once again introduced $\beta = \mathfrak{m}^2/\Lambda$. Hence in order to ensure $-(r\partial_ur)(0;\alpha)>0$ we need 
    \begin{equation}
         \beta \big(1-|\alpha|^2C'_{\mathrm{upp}} \big)^2\int_0^1\phi_{\hat{\alpha}}^2 dv - \int_0^1 v^2\big(\partial_v\phi_{\hat{\alpha}}\big)^2dv>0
    \end{equation}
    or equivalently
    \begin{equation}
         \big(1-|\alpha|^2C'_{\mathrm{upp}} \big)^2 -\frac{1}{\beta} \frac{\int_0^1 v^2\big(\partial_v\phi_{\hat{\alpha}}\big)^2dv}{\int_0^1\phi_{\hat{\alpha}}^2 dv}>0. 
         \label{eq: proof gluing 3 - initial partial_u r pre}
    \end{equation}
    Notice that 
    \begin{align}
        \int_{0}^{1}\hspace{-0.2em}\phi_{\hat{\alpha}}^2 \, dv &=\sum_{j=1}^{k+1} \hat{\alpha}_j^2\int_{v_{j-1}}^{v_j}\hspace{-0.2em}\chi_j^2 \, dv \\[0.5em]
        \int_{0}^{1}\hspace{-0.2em}v^2\big(\partial_v\phi_{\hat{\alpha}}\big)^2 dv &=\sum_{j=1}^{k+1} \hat{\alpha}_j^2\int_{v_{j-1}}^{v_j}\hspace{-0.2em}v^2\big(\chi'_j\big)^2 dv,
    \end{align}
    for a dissection 
    \begin{equation}
        0=v_0<v_1<\dots<v_{k+1}=1, 
    \end{equation}
    and associated bump functions $\chi_j \in C^{\infty}_c\big((v_{j-1},v_j)\big)$, $j=1,\dots,k+1$. Note that, on every interval of the form $(v_{j-1},v_j)$, for $j=1,\dots,k+1$, by the sharpness of the Hardy inequality (see, for example, \cite[Thm.~3.1]{Gesztesy_Pang_2025}), we have:  
    \begin{equation}
    \inf_{\substack{f \in C^{\infty}_c((v_{j-1},v_j)) \\ f \not\equiv 0}} \frac{\int_{v_{j-1}}^{v_j} v^2 (f')^2 \, dv}{\int_{v_{j-1}}^{v_j} f^2 \, dv} =\frac{1}{4} + \bigg(\frac{\pi}{\ln(\frac{v_j}{v_{j-1}})}\bigg)^2. 
    \end{equation}

    Since $\mathfrak{m}^2>\Lambda/4$, and hence $\beta>1/4$, we can start choosing $0<v_k<v_{k+1}=1$ so that 
    \begin{equation}
        \frac{1}{4} + \bigg(\frac{\pi}{\ln(\frac{1}{v_k})}\bigg)^2<\beta
    \end{equation}
    and $\chi_{k+1}\in C^{\infty}_c\big((v_k,1)\big)$ satisfying 
    \begin{equation}
        \int_{v_k}^{1} v^2 \big(\chi_{k+1}'\big)^2 \, dv<\beta \int_{v_k}^1 \chi_{k+1}^2 \, dv.
    \end{equation}
    We iterate the process, choosing $v_{j-1}\in(0,1)$ so that 
    \begin{equation}
        \frac{1}{4} + \bigg(\frac{\pi}{\ln(\frac{v_{j}}{v_{j-1}})}\bigg)^2<\beta
    \end{equation}
    for $j=k, \dots,2$. In this way we obtain $v_{k-1}, \dots, v_1$, and we can choose associated bump functions so that 
    \begin{equation}
        \int_{v_{j-1}}^{v_j} v^2 \big(\chi_{j}'\big)^2 \, dv<\beta \int_{v_{j-1}}^{v_j} \chi_{j}^2 \, dv
    \end{equation}
    holds for $j=2, \dots, k$. For $j=1$, the sharp Hardy constant is in fact
    \begin{equation}
    \inf_{\substack{f \in C^{\infty}_c((0,v_1)) \\ f \not\equiv 0}} \frac{\int_{0}^{v_1} v^2 (f')^2 \, dv}{\int_{0}^{v_1} f^2 \, dv} =\frac{1}{4},  
    \end{equation}
    irrespective of the value of $v_1$, as long as $v_1>0$ (cf. \cite[Cor.~1.2.2]{Balinsky_Evans_Lewis_2015}), and hence we can choose $\chi_1$ satisfying 
    \begin{equation}
        \int_{0}^{v_1} v^2 \big(\chi_{1}'\big)^2 \, dv<\beta \int_{0}^{v_1} \chi_{1}^2 \, dv.
    \end{equation}

    Therefore, for this choice of bump functions $\chi_1,\dots,\chi_{k+1}$, we have
    \begin{equation}
        \mathfrak{C}\doteq \max\limits_{j=1,\dots,k+1} \frac{1}{\beta}\frac{\int_{v_{j-1}}^{v_j} v^2 \big(\chi_{j}'\big)^2 \, dv}{\int_{v_{j-1}}^{v_j} \chi_{j}^2 \, dv}<1,
    \end{equation}
    and hence
    \begin{equation}
        \int_{0}^{1}\hspace{-0.2em}v^2\big(\partial_v\phi_{\hat{\alpha}}\big)^2 dv =\sum_{j=1}^{k+1}\hat{\alpha}_j^2\int_{v_{j-1}}^{v_j} v^2 \big(\chi_{j}'\big)^2 \, dv\leq\mathfrak{C}\beta \sum_{j=1}^{k+1}\hat{\alpha}_j^2\int_{v_{j-1}}^{v_j} \chi_{j}^2 \, dv=\mathfrak{C}\beta  \int_{0}^{1}\hspace{-0.2em}\phi_{\hat{\alpha}}^2 \, dv. 
    \end{equation}
    It is clear now that by choosing 
    \begin{equation}
        |\alpha|^2< \frac{1}{C'_{\mathrm{upp}}}\big(1-\sqrt{\mathfrak{C}}\,\big)
        \label{eq: proof gluing 3 - new cond alpha}
    \end{equation}
    we can ensure \eqref{eq: proof gluing 3 - initial partial_u r pre}, and hence \eqref{eq: proof gluing 3 - initial partial_u r statement} (and note that $\mathfrak{C}$ here is independent of $\hat{\alpha}$). By possibly taking $\delta$ smaller, we can guarantee that \eqref{eq: proof gluing 3 - new cond alpha} holds for all $0<|\alpha|^2< \delta$. 
\end{proof}

\begin{remark} 
    Note that, if we choose $\delta$ so that Lemma \ref{lem: proof gluing 3 - negativity partial_u r initial} holds, then 
    \begin{equation}
        f_{\tilde{\mathfrak{c}}}\big(r(0;\alpha)\big)=-4\partial_vr(0;\alpha)\partial_ur(0;\alpha)>0, 
    \end{equation}
    and thus $$0\leq\widetilde{m}(0;\alpha)<\frac1{3\sqrt{\Lambda}}$$ for all $\alpha$ with $0<|\alpha|^2< \delta$.
    \label{rmk: proof gluing 3 - estimate initial mass}
\end{remark}

\begin{lemma}
    By possibly making $\delta$ even smaller (depending on $\mathfrak{m}$ and the scalar field profile), we can ensure 
    \begin{equation}
        \widetilde{m}(0;\alpha)>0
        \label{eq: proof gluing 3 - positive mass}
    \end{equation}
    for all $\alpha$ satisfying $0<|\alpha|^2<\delta$. 
    \label{lem: proof gluing 3 - positive mass}
\end{lemma}

\begin{proof}
    We assume $0<|\alpha|^2<\delta$, with $\delta$ small enough so that  Lemma \ref{lem: proof gluing 3 - negativity partial_u r initial} holds. Therefore $r$ is monotone increasing and $\partial_vr$ monotone decreasing on $[0,1]$. 

    Note that \eqref{eq: proof gluing 3 - positive mass} is equivalent to
    \begin{equation}
        -(r\partial_ur)(0;\alpha)<\frac{r(0;\alpha)}{4\partial_vr(0;\alpha)}\bigg(1-\frac13\Lambda r^2(0;\alpha)\bigg). 
        \label{eq: proof gluing 3 - positive mass equiv} 
    \end{equation}
    Integrating equation \eqref{eq:gluing - eqn for r partial_ur}, and using the monotonicity of $r$, we find: 
    \begin{equation}
        -(r\partial_ur)(0;\alpha) \leq \frac14\bigg(r(1;\alpha)^2\left(\Lambda+\mathfrak{m}^2|\alpha|^2C_{\mathrm{upp}}\right)-1\bigg)=\frac14\beta|\alpha|^2C_{\mathrm{upp}}.
    \end{equation}
    On the other hand, we have 
    \begin{equation}
        \frac{r(0;\alpha)}{4\partial_vr(0;\alpha)}\bigg(1-\frac13\Lambda r^2(0;\alpha)\bigg)\geq \frac{1}{12C'_{\mathrm{upp}}|\alpha|^2},
    \end{equation}
    where we have used estimate \eqref{eq: proof gluing 3 - partial_v upper} and the fact that $r(0;\alpha)<1/\sqrt{\Lambda}$. Hence, it suffices to choose 
    \begin{equation}
        |\alpha|^2<\sqrt{\frac{1}{3\beta}\frac1{C'_{\mathrm{upp}}C_{\mathrm{upp}}}}
        \label{eq: proof gluing 3 - new cond alpha 2}
    \end{equation}
    to ensure \eqref{eq: proof gluing 3 - positive mass equiv} and hence \eqref{eq: proof gluing 3 - positive mass}. 

    By possibly taking $\delta$ smaller, we can guarantee that \eqref{eq: proof gluing 3 - new cond alpha 2} holds for all $0<|\alpha|^2< \delta$. 
\end{proof}

\begin{remark}
    If we choose $\delta$ so that Lemma \ref{lem: proof gluing 3 - positive mass} holds, we know that, for any $\alpha$ with $0<|\alpha|^2<\delta$, there exists $\tilde{\mathfrak{c}}\in(0,1)$ such that $$\widetilde{m}(0;\alpha)=\frac{\tilde{\mathfrak{c}}}{3\sqrt{\Lambda}},$$ and also, given that 
    \begin{equation}
        f_{\tilde{\mathfrak{c}}}\big(r(0;\alpha)\big)=-4\partial_vr(0;\alpha)\partial_ur(0;\alpha)>0, 
    \end{equation}
    we conclude that $r_{\mathcal{H}}(\tilde{\mathfrak{c}})<r(0;\alpha)<r_{\overline{\mathcal{H}}}(\tilde{\mathfrak{c}})$.
    \label{rmk: proof gluing 3 - initial radius}
\end{remark}

Take $\delta_*>0$ satisfying $\delta_*<\delta$, so that Lemma \ref{lem: proof gluing 3 - positive mass} holds (and, in particular, Lemma \ref{lem: proof gluing 3 - negativity partial_u r initial} holds as well). For $\alpha\in \mathbb S^k_{\sqrt{\delta_*}}$, we define $D_{\alpha}(0)$ to be the sphere data set with: 
\begin{itemize}
    \item $\varrho=R\doteq r(0;\alpha)$ with $R>r_{\mathcal{H}}(\tilde{\mathfrak{c}})$ for some $\tilde{\mathfrak{c}}\in(0,1)$ (see Remark \ref{rmk: proof gluing 3 - initial radius}),
    \item $\varrho^1_v=\partial_vr(0;\alpha)>0$ (see \eqref{eq: proof gluing 3 - positivity partial_vr initial}),
    \item $\varrho^1_u=\partial_ur(0;\alpha)<0$ (see Lemma \ref{lem: proof gluing 3 - negativity partial_u r initial}),
    \item $\widetilde{m}\big(D_{\alpha}(0)\big)=\frac{\tilde{\mathfrak{c}}}{3\sqrt{\Lambda}}$ (see Remark \ref{rmk: proof gluing 3 - initial radius}),
    \item $\omega =1$,
    \item all other components are zero,
\end{itemize}
and hence $D_{\alpha}(0)$ is gauge equivalent to the Schwarzschild--de Sitter static sphere $_{\mathcal{S}}D^{\mathrm{SdS}}_{\tilde{\mathfrak{c}},R, k}$ with $\varrho^1_u<0$. 

Now, with $D_{\alpha}(0)$ and using $\phi_{\alpha}$ as seed, we generate the cone data set $D_{\alpha}:[0,1]\longrightarrow \mathcal{D}_k$. Therefore, $D_{\alpha}(1)$ has:
\begin{itemize}
    \item $\varrho=r(1;\alpha)=\frac{1}{\sqrt{\Lambda}}$,
    \item $\varrho^1_v=\partial_vr(1;\alpha)=0$ (see \eqref{eq: proof gluing 3 - initialisation}),
    \item $\varrho^1_u=\partial_ur(1;\alpha)=0$ (see \eqref{eq: proof gluing 3 - initialization 2}),
    \item $\omega =1$, 
    \item $\varphi^i_v=0$ for all $0\leq i \leq k$.
\end{itemize}

Since this is true for all $\alpha\in\mathbb S^k_{\sqrt{\delta_*}}$, and the parities of the metric and scalar field quantities are the same as the ones discussed in Lemma \ref{lem:parity lemma}, we can once again apply the Borsuk--Ulam theorem to conclude that there exists $\alpha_*\in\mathbb S^k_{\sqrt{\delta_*}}$ such that $\varphi^i_u(D_{\alpha_*}(1))=0$ for all $1\leq i \leq k$, and therefore  $D_{\alpha_*}(1)$ is gauge equivalent to a Nariai sphere $D^{\mathrm{Nar}}_{k}$.

\subsection{Proof of gluing to the extremal cosmological horizon (Theorem \ref{thm:subext SdS to ext COSMhorizon})}
\label{subsec:proof subext SdS to ext COSMhorizon}

The first part of this proof is completely identical to the proof of Theorem \ref{thm:subext SdS to Nariai} presented in the previous section. We choose the same dissection of $[0,1]$ and bump functions $\chi_1,\dots,\chi_{k+1}$ that we chose in the proof of Theorem \ref{thm:subext SdS to Nariai}. Therefore, as before, we have
\begin{equation}
    \mathfrak{C}\doteq \max\limits_{j=1,\dots,k+1} \frac{1}{\beta}\frac{\int_{v_{j-1}}^{v_j} v^2 \big(\chi_{j}'\big)^2 \, dv}{\int_{v_{j-1}}^{v_j} \chi_{j}^2 \, dv}<1,
\end{equation}
and thus
\begin{equation}
    \int_{0}^{1}\hspace{-0.2em}v^2\big(\partial_v\phi_{\hat{\alpha}}\big)^2 dv \leq\mathfrak{C}\beta  \int_{0}^{1}\hspace{-0.2em}\phi_{\hat{\alpha}}^2 \, dv. 
\end{equation}
where $\beta=\mathfrak{m}^2/\Lambda>1/4$. We also prescribe the same initial values at $v=1$ for $r$ and $\partial_vr$ as in the proof of Theorem \ref{thm:subext SdS to Nariai}. Hence for $0< |\alpha|^2\leq \hat{\delta}$, with 
\begin{equation}
    0<|\alpha|^2\leq\hat{\delta} \doteq \frac{1}{2C'_{\mathrm{upp}}},
    \label{eq: proof gluing 4 - def delta}
\end{equation} 
we have that $r$ is a positive and monotonically increasing function on $v\in[0,1]$, $\partial_vr$ is non-negative and monotone decreasing and moreover 
\begin{align}
    r(0;\alpha&)\geq \frac{1}{2\sqrt{\Lambda}}, \label{eq: proof gluing 4 - r at 0}\\
    \partial_vr(0;\alpha&)>0.\label{eq: proof gluing 4 - derivative at 0}
\end{align}

The key difference with respect to the proof in Section \ref{subsec:proof subext SdS to Nariai} however, is the prescribed value for $\partial_u r$ at the final sphere 
\begin{equation}
    \partial_ur(1;\alpha)=\varepsilon\sqrt{\Lambda}, 
    \label{eq: proof gluing 4 - initialisation 2}
\end{equation}
with $\varepsilon>0$ possibly very small and to be determined later. We show now that we can choose $\varepsilon$ small enough so that $\partial_ur$ is in the desired range at $v=0$: 
\begin{lemma}
    \label{lem: proof gluing 4 - negativity partial_u r initial}
    For any $\alpha$ satisfying $0<|\alpha|^2<\delta$, with $\delta$ depending on $\beta$ and the baseline scalar field profile, and satisfying $\delta\leq\hat{\delta}$, we have 
    \begin{equation}
        \partial_ur(0;\alpha)>-\frac{1}{4\partial_vr(0;\alpha)}\Big(1-\frac13\Lambda r^2(0;\alpha)\Big),
        \label{eq: proof gluing 4 - positive mass statement}
    \end{equation}
    and moreover we can choose $\varepsilon$ sufficiently small (depending on $|\alpha|^2$) so that
    \begin{equation}
        \partial_ur(0;\alpha)<0.\\
        \label{eq: proof gluing 4 - initial partial_u r statement}
    \end{equation}
\end{lemma}

\begin{proof}
    We assume $0<|\alpha|^2\leq\hat{\delta}$. Proceeding as in the proof of Lemma \ref{lem: proof gluing 3 - negativity partial_u r initial}, we find
    \begin{align}
         -(r\partial_ur)(0;\alpha)\geq&\frac{|\alpha|^2}4\bigg(\beta \big(1-|\alpha|^2C'_{\mathrm{upp}} \big)^2\int_0^1\phi_{\hat{\alpha}}^2 dv - \int_0^1 v^2\big(\partial_v\phi_{\hat{\alpha}}\big)^2dv\bigg)-\varepsilon
    \end{align}
    and thus, in order to prove \eqref{eq: proof gluing 4 - initial partial_u r statement}, it suffices to show that we can choose $|\alpha|^2$ and $\varepsilon$ satisfying 
    \begin{equation}
        |\alpha|^2\Big(\big(1-|\alpha|^2C'_{\mathrm{upp}} \big)^2 - \mathfrak{C}\Big)-\frac{4\varepsilon}{\beta C_{\mathrm{low}}}>0. 
        \label{eq: proof gluing 4 - condition}
    \end{equation}
    We know (recall Section \ref{subsec:proof subext SdS to Nariai}) that the term in brackets is positive for any $\alpha$ satisfying
    \begin{equation}
        0<|\alpha|^2<\check{\delta}\doteq\frac{1}{C'_{\mathrm{upp}}}\big(1-\sqrt{\mathfrak{C}}\,\big). 
    \end{equation}
    
    Additionally, we also want to show \eqref{eq: proof gluing 4 - positive mass statement}, or equivalently
    \begin{equation}
        -(r\partial_ur)(0;\alpha)<\frac{r(0;\alpha)}{4\partial_vr(0;\alpha)}\Big(1-\frac13\Lambda r^2(0;\alpha)\Big).
        \label{eq: proof gluing 4 - condition 2}
    \end{equation}
    Integrating equation \eqref{eq:gluing - eqn for r partial_ur}, and using the monotonicity of $r$, we find: 
    \begin{equation}
        -(r\partial_ur)(0;\alpha) \leq \frac14\beta|\alpha|^2C_{\mathrm{upp}}-\varepsilon<\frac14\beta|\alpha|^2C_{\mathrm{upp}},
    \end{equation}
    and, as for the gluing to Nariai spheres, we also have
    \begin{equation}
        \frac{r(0;\alpha)}{4\partial_vr(0;\alpha)}\bigg(1-\frac13\Lambda r^2(0;\alpha)\bigg)\geq \frac{1}{12C'_{\mathrm{upp}}|\alpha|^2}.
    \end{equation}
    Therefore, \eqref{eq: proof gluing 4 - condition 2} holds for any $\alpha$ satisfying
    \begin{equation}
         0<|\alpha|^2<\mathring{\delta}\doteq\sqrt{\frac{1}{3\beta}\frac1{C'_{\mathrm{upp}}C_{\mathrm{upp}}}}
    \end{equation}
    
    Now we can define
    \begin{equation}
        \delta\doteq \min\big\{\hat{\delta}, \check{\delta}, \mathring{\delta}\big\}. 
    \end{equation}
    For any $0<|\alpha|^2<\delta$, \eqref{eq: proof gluing 4 - condition 2} (thus \eqref{eq: proof gluing 4 - positive mass statement}) holds, and also
    \begin{equation}
        |\alpha|^2\Big(\big(1-|\alpha|^2C'_{\mathrm{upp}} \big)^2 - \mathfrak{C}\Big)>0. 
    \end{equation}
    Hence for that $|\alpha|^2$ we can choose $\varepsilon$ sufficiently small so that \eqref{eq: proof gluing 4 - condition} (thus \eqref{eq: proof gluing 4 - initial partial_u r statement}) holds. 
\end{proof}

\begin{remark}
    Choosing $\varepsilon$ so that Lemma \ref{lem: proof gluing 4 - negativity partial_u r initial} holds, we know that, for any $\alpha$ with $0<|\alpha|^2<\delta$, there exists $\tilde{\mathfrak{c}}\in(0,1)$ such that $$\widetilde{m}(0;\alpha)=\frac{\tilde{\mathfrak{c}}}{3\sqrt{\Lambda}},$$ and also, given that 
    \begin{equation}
        f_{\tilde{\mathfrak{c}}}\big(r(0;\alpha)\big)=-4\partial_vr(0;\alpha)\partial_ur(0;\alpha)>0, 
    \end{equation}
    we conclude that $r_{\mathcal{H}}(\tilde{\mathfrak{c}})<r(0;\alpha)<r_{\overline{\mathcal{H}}}(\tilde{\mathfrak{c}})$.
    \label{rmk: proof gluing 4 - initial radius}
\end{remark}

Now, for $\alpha\in \mathbb S^k_{\sqrt{\delta_*}}$ with $0<\delta_*<\delta$, we choose $\varepsilon$ appropriately small according to Lemma \ref{lem: proof gluing 4 - negativity partial_u r initial}, and we define $D_{\alpha}(0)$ to be the sphere data set with:
\begin{itemize}
    \item  $\varrho=R\doteq r(0;\alpha)$ with $R>r_{\mathcal{H}}(\tilde{\mathfrak{c}})$ for some $\tilde{\mathfrak{c}}\in(0,1)$ (see Remark \ref{rmk: proof gluing 4 - initial radius}),
    \item $\varrho^1_v=\partial_vr(0;\alpha)>0$,
    \item $\varrho^1_u=\partial_ur(0;\alpha)<0$ (see Lemma \ref{lem: proof gluing 4 - negativity partial_u r initial}),
    \item $\widetilde{m}\big(D_{\alpha}(0)\big)=\tfrac{\tilde{\mathfrak{c}}}{3\sqrt{\Lambda}}$ (see Lemma \ref{lem: proof gluing 4 - negativity partial_u r initial} and Remark \ref{rmk: proof gluing 4 - initial radius}),
    \item $\omega =1$,
    \item all other components are zero.
\end{itemize}
Hence $D_{\alpha}(0)$ is gauge equivalent to the Schwarzschild--de Sitter static sphere $_{\mathcal{S}}D^{\mathrm{SdS}}_{\tilde{\mathfrak{c}},R, k}$ with $\varrho^1_u<0$. 

Now, with $D_{\alpha}(0)$ and using $\phi_{\alpha}$ as seed, we generate the cone data set $D_{\alpha}:[0,1]\longrightarrow \mathcal{D}_k$. Therefore, $D_{\alpha}(1)$ has:
\begin{itemize}
    \item $\varrho=r(1;\alpha)=\frac{1}{\sqrt{\Lambda}}$,
    \item $\varrho^1_v=\partial_vr(1;\alpha)=0$,
    \item $\varrho^1_u=\partial_ur(1;\alpha)=\varepsilon \sqrt{\Lambda}>0$ ,
    \item $\omega =1$, 
    \item $\varphi^i_v=0$ for all $0\leq i \leq k$.
\end{itemize}

Since this argument works for all $\alpha\in\mathbb S^k_{\sqrt{\delta_*}}$, and the parities of the metric and scalar field quantities are the same as the ones discussed in Lemma \ref{lem:parity lemma}, we can once again apply the Borsuk--Ulam theorem to conclude that there exists $\alpha_*\in\mathbb S^k_{\sqrt{\delta_*}}$ such that $\varphi^i_u(D_{\alpha_*}(1))=0$ for all $1\leq i \leq k$, and therefore  $D_{\alpha_*}(1)$ is gauge equivalent to the Schwarzschild--de Sitter extremal cosmological horizon sphere $_{\overline{\mathcal{H}}^+}D^{\mathrm{SdS}}_{1, k}$, with $\varrho^1_u>0$. 

\section{Applications of the characteristic gluing results}
\label{sec:applications}

In this section we show how the characteristic gluing results stated in Section \ref{sec:gluing results} and proved in Section \ref{sec:proofs} can be used to construct appropriate spacetimes and, ultimately, identify suitable Cauchy data satisfying the gravitational collapse results presented in Section \ref{subsec:main theorems intro}. 

For the set-up considered here, an admissible Cauchy data set for the Einstein-scalar field system consists of a quintuple $\big(\Sigma, g_0, k_0, \phi_0, \phi_1\big)$, where $\Sigma$ is a 3-dimensional manifold diffeomorphic to $\mathbb S^3$, $g_0$ is a Riemannian metric on $\Sigma$, $k_0$ is a symmetric 2-tensor on $\Sigma$, and $\phi_0$ and $\phi_1$ are scalar fields on $\Sigma$. We recall that the existence of a future maximal globally hyperbolic development associated to such initial data follows from classical results \cite{Foures-Bruhat_1952, Choquet-Bruhat_Geroch_1969}. See also \cite{Ringstrom_2009} for an explicit proof covering the $\Lambda\neq0$ case coupled to a massive scalar field. Moreover, if the initial data are spherically symmetric, the future maximal globally hyperbolic development $(\mathcal{M}, g, \phi)$ is also spherically symmetric by uniqueness, and its quotient $(\mathcal{Q},g_{\mathcal{Q}}, \phi)$ admits global double null coordinates.

Unlike in the asymptotically flat setting, where it is standard to require the initial data to be free of antitrapped surfaces, no such condition will be imposed here. In the presence of a positive cosmological constant, the spacetime naturally contains cosmological regions foliated by antitrapped spheres. The potential presence of antitrapped spheres in $\Sigma$ should therefore be seen as an expected and physically reasonable feature of our set-up.  

For reasons of clarity, we will follow a slightly different order to that used to present these results in Section \ref{subsec:main theorems intro}. We start by establishing the dynamical formation of the subextremal Schwarzschild--de Sitter black hole (Theorem \ref{thm:grav collapse to subext SdS}). Next, we present the results related to the dynamical formation of the extremal Schwarzschild--de Sitter black hole spacetime (Theorem \ref{thm:grav collapse to ext SdS BH}), and we analyse some of the geometric properties of the resulting spacetime, namely the lack of other event horizons and the connectedness of the apparent horizon(s). Finally, we present the results concerning the formation in gravitational collapse of a portion of the extremal Schwarzschild--de Sitter white hole spacetime (Theorem \ref{thm:grav collapse to ext SdS WH}) and of the Nariai spacetime (Theorem \ref{thm:grav collapse to Nariai}). 

\subsection{Dynamical formation of Schwarzschild--de Sitter black holes}
\subsubsection{Gravitational collapse to the subextremal Schwarzschild--de Sitter black hole}

\begin{customthm}{2}
    For any $k\in \mathbb N$, $\mathfrak{m}\geq0$ and $\mathfrak{c}\in(0,1)$, there exist spherically symmetric Cauchy data for the Einstein-scalar field system with scalar field mass $\mathfrak{m}$, $\big(\Sigma, g_0, k_0, \phi_0, \phi_1\big)$, with $\Sigma \cong \mathbb S^3$ and two regular centres, such that the future maximal globally hyperbolic development $\big(\mathcal{M},g,\phi\big)$ satisfies: 
    \begin{itemize}
        \item All dynamical quantities are at least $C^k$-regular.
        \item The spacetime contains a region of the form 
        \begin{equation}
            \mathcal{R}\doteq \{u_1 \leq u \leq u_2\} \cap \{v_1 \leq v \leq v_2\}
        \end{equation}
        for some $u_1<u_2$, $v_1<v_2$, that is isometric to a double null rectangle in the Schwarzschild--de Sitter spacetime of extremality index $\mathfrak{c}$. In particular, the truncated null cone
        \begin{equation}
            H^+\doteq\{u=u_2\}\cap \mathcal{R}
        \end{equation} 
        can be identified with an appropriate portion of the constant-$u$ component of the black hole horizon $\mathcal{H}^+$ of the Schwarzschild--de Sitter spacetime of extremality index $\mathfrak{c}$. Similarly, the truncated null cone  
        \begin{equation}
            \overline{H}^+\doteq\{v=v_2\}\cap\mathcal{R}
        \end{equation} 
        can be identified with an appropriate portion of the constant-$v$ component of the future cosmological horizon $\overline{\mathcal{H}}^+$ of the Schwarzschild--de Sitter spacetime of extremality index $\mathfrak{c}$.   

        \item There are no other event horizons in the region $\{u>u_2\}\cap\{v<v_2\}$. 

        \item The hypersurface $\Sigma$ does not intersect the null cones ${C}\doteq\{u=u_2\}\cap\mathcal{M}$ and $\underline{{C}}\doteq\{v=v_2\}\cap\mathcal{M}$.
    \end{itemize}
    \label{thm:grav collapse to subext SdS}
\end{customthm}

\begin{proof}
    We refer to Figure \ref{fig:grav collapse to subext SdS} for a visual summary of the proof. Given $\mathfrak{c}\in(0,1)$, we fix $r_0<r_{\mathcal{H}}(\mathfrak{c})$ and $\varepsilon>0$ such that $r_0-\varepsilon>0$. In view of Theorem \ref{thm:subext SdS to BHhorizon}, we can apply Proposition \ref{prop:spacetime gluing} taking $\big(\mathcal{Q}_1,r_1, \Omega^2_1 \big)$ to be a double null rectangle in the static region $\mathcal{S}$ of de Sitter spacetime with
    \begin{equation}
        \begin{aligned}
        r_0-\varepsilon &\leq t+r\leq r_0, \\
        -r_0 +\varepsilon &\geq t-r\geq -r_0, 
        \end{aligned}
    \end{equation}
    and $\big(\mathcal{Q}_2, r_2, \Omega^2_2\big)$ a double null rectangle in the static region $\mathcal{S}$ of the subextremal Schwarzschild--de Sitter spacetime of extremality index $\mathfrak{c}$, extending up to and including a portion of the constant-$u$ component of the black hole event horizon. We glue these two solutions along the null cone ${C}=\{u=u_2\}$, hence obtaining a solution to the Einstein-scalar field system on thin slabs to the future and the past of ${C}$. 

    Note that we can extend $\big(\mathcal{Q}_1,r_1, \Omega^2_1 \big)$ so that it reaches up to the left centre of de Sitter spacetime. The thin slab to the past of ${C}$ on which the solution was obtained by local existence can be extended all the way up to the centre by the same arguments used in \cite{Kehle_Unger_2025} (see also \cite{Christodoulou_2009_blackholes,Luk_Oh_Yang_2017} for arguments specific to the spherically symmetric setting). Given the loss of derivatives that spherically symmetric solutions to wave equations suffer at the centre of symmetry, the fact that we have carried out the gluing up to order $k+1$ ensures that the solution ultimately obtained is of order $C^k$ as desired. 
    
    We also extend $\big(\mathcal{Q}_2, r_2, \Omega^2_2\big)$ so that it reaches the constant-$v$ component of the future cosmological event horizon of the Schwarzschild--de Sitter spacetime of extremality index $\mathfrak{c}$ and the boundary point $i^+$. We can subsequently extend the thin slab to the future of ${C}$ on which the solution is obtained by local existence, but in this case we may not be able to solve in a full double null neighbourhood near $i^+$, as depicted in Figure \ref{fig:grav collapse to subext SdS}.

    \begin{figure}[H]
        \centering
        \includegraphics[width=0.65\textwidth]{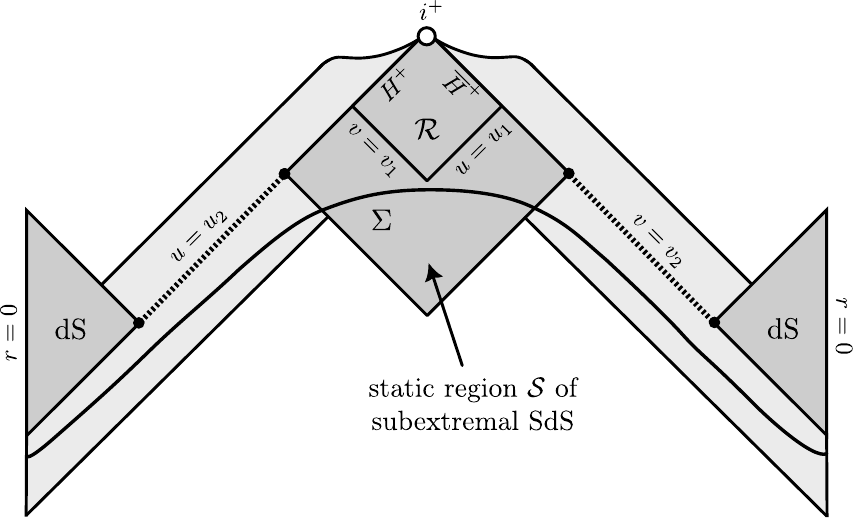}
        \caption{Penrose diagram for the proof of Theorem \ref{thm:grav collapse to subext SdS}.} 
        \label{fig:grav collapse to subext SdS}
    \end{figure}
    
    Note that $\mathcal{Q}_2$ will be of the form $\{u_0 \leq u \leq u_2\} \cap \{v_0 \leq v \leq v_2\}$ for some $u_0<u_2$ and $v_0<v_2$. Using the analogue of Theorem \ref{thm:dS to subext COSMhorizon} and Proposition \ref{prop:spacetime gluing} along a constant-$v$ cone, we can repeat the procedure to glue, along the cone $\underline{{C}}=\{v=v_2\}$, the subextremal Schwarzschild--de Sitter portion from a cosmological horizon sphere to a portion in the static region $\mathcal{S}'$ of de Sitter spacetime with 
    \begin{equation}
        \begin{aligned}
        t+r\leq r_0, \\
        t-r\geq -r_0, 
        \end{aligned}
    \end{equation}
    that thus extends up to the right centre of de Sitter. The Penrose diagram of the resulting spacetime, $\big(\mathcal{M},g,\phi\big)$, looks as shown in Figure \ref{fig:grav collapse to subext SdS}. 

    Now that we have constructed the spacetime, we can identify an appropriate Cauchy hypersurface $\Sigma$ in it, reaching the left and right regular centres and with $\Sigma\cong\mathbb S^3$  (in the diagram in Figure \ref{fig:grav collapse to subext SdS}, this simply corresponds to a spacelike curve connecting the two centres). Note that $\Sigma$ can be chosen so that it does not intersect the cones ${C}$, $\underline{{C}}$. We consider then the Cauchy data induced by the solution $\big(\mathcal{M},g,\phi\big)$ on $\Sigma$, $\big(\Sigma, g_0, k_0, \phi_0, \phi_1\big)$. The future maximal globally hyperbolic development of $\big(\Sigma, g_0, k_0, \phi_0, \phi_1\big)$, by uniqueness, will include the domain of dependence of $\Sigma$ in $\big(\mathcal{M},g,\phi\big)$, and thus, in particular, will contain a region 
    \begin{equation}
        \mathcal{R}= \{u_1 \leq u \leq u_2\} \cap \{v_1 \leq v \leq v_2\}
    \end{equation}
    for $u_1$ with $u_0<u_1<u_2$ and sufficiently close to $u_2$, and $v_1$ with $v_0<v_1<v_2$ and sufficiently close to $v_2$, lying to the future of $\Sigma$ and satisfying the desired properties.

    Finally, the claim about no other horizons in the region $\{u>u_2\}\cap\{v<v_2\}$ follows from Theorem~\ref{thm:no horizon points} and Corollary~\ref{cor:no hor points subextremal} below.
\end{proof}

\subsubsection{Gravitational collapse to the extremal Schwarzschild--de Sitter black hole}
\label{subsec:grav collapse extremal BH}

Proceeding as in Theorem \ref{thm:grav collapse to subext SdS}, but applying only Theorem~\ref{thm:subext SdS to BHhorizon}, once along a constant-$u$ cone and once along a constant-$v$ cone, one can readily construct spacetimes whose Penrose diagram looks like the one shown in Figure \ref{fig:grav collapse to subext SdS} but with the central region $\mathcal{R}$ replaced by a double null rectangle in the extremal Schwarzschild--de Sitter spacetime, bounded to the future by portions of the extremal black hole horizons.  

In order to provide explicit counterexamples to the third law, however, it is convenient to combine several gluings in such a way that portions of the subextremal black hole horizons can be clearly identified. This is precisely what is done in Theorem~\ref{thm:grav collapse to ext SdS BH}. 

\begin{customthm}{1.1}
    For any $k\in \mathbb N$, $\mathfrak{m}\geq0$, and $\mathfrak{c}, \mathfrak{c}' \in(0,1)$, there exist spherically symmetric Cauchy data for the Einstein-scalar field system with scalar field mass $\mathfrak{m}$, $\big(\Sigma, g_0, k_0, \phi_0, \phi_1\big)$, with $\Sigma \cong \mathbb S^3$ and two regular centres, such that the future maximal globally hyperbolic development $\big(\mathcal{M},g,\phi\big)$ satisfies: 
    \begin{itemize}
        \item All dynamical quantities are at least $C^k$-regular.
        \item Moreover, the spacetime contains:
        \begin{enumerate}[(i)]
            \item A region 
            \begin{equation}
                \hat{\mathcal{R}}\doteq \{u_1 \leq u \leq u_2\} \cap \{v_1 \leq v \leq v_2\}
            \end{equation}
            for some $u_1<u_2$, $v_1<v_2$, that is isometric to a double null rectangle in the extremal Schwarzschild--de Sitter black hole spacetime. In particular, the truncated null cones
            \begin{equation}
                \begin{aligned}
                    H_{l}^+ &\doteq \{u=u_2\}\cap \hat{\mathcal{R}},\\
                    H_{\mathrm{r}}^+ &\doteq \{v=v_2\}\cap \hat{\mathcal{R}},
                \end{aligned}
            \end{equation} 
            can be identified with appropriate portions of the constant-$u$ and constant-$v$ components, respectively, of the black hole horizon $\mathcal{H}^+$ of the extremal Schwarzschild--de Sitter black hole spacetime.
            \item A region
            \begin{equation}
                \mathcal{R}\doteq \{u_2 \leq u \leq u_3\} \cap \{{v_{-2} \leq v \leq v_{-1}}\},
            \end{equation}
            for some $u_3$ with $u_3>u_2$, and $v_{-1}$, $v_{-2}$ with $v_1>v_{-1}>v_{-2}$, that is isometric to a double null rectangle in the static region $\mathcal{S}$ of the Schwarzschild--de Sitter spacetime of extremality index $\mathfrak{c}$. In particular, the truncated null cone $\{u=u_3\}\cap \mathcal{R}$ can be identified with an appropriate portion of the constant-$u$ component of the subextremal black hole horizon. 
            \item A region 
            \begin{equation}
                \mathcal{R}'\doteq \{u_{-2} \leq u \leq u_{-1}\} \cap \{v_2 \leq v \leq v_3\},
            \end{equation}
            for some $u_{-1}$, $u_{-2}$ with $u_1>u_{-1}>u_{-2}$, and $v_3$ with $v_3>v_2$, that is isometric to a double null rectangle in the static region $\mathcal{S}'$ of the Schwarzschild--de Sitter spacetime of extremality index $\mathfrak{c}'$. In particular, the truncated null cone $\{v=v_3\}\cap \mathcal{R}'$ can be identified with an appropriate portion of the constant-$v$ component of the subextremal black hole horizon. 
        \end{enumerate} 
        \item The spacetime contains no event horizons other than the two bounding the extremal region.  
        \item The hypersurface $\Sigma$ does not intersect the null cones $\{u=u_2\}\cap\mathcal{M}$ and $\{v=v_2\}\cap\mathcal{M}$.
    \item Additionally, if the scalar field is \myuline{massless} (i.e. $\mathfrak{m}=0$), the left  and right apparent horizons, defined as 
\begin{equation}
        \mathcal{A}_{l}\doteq\big\{(u,v)\in \mathcal{M}: u\geq u_2, \, \partial_vr(u,v)=0\big\},
        \label{eq:def left apparent horizon}
    \end{equation}
    \begin{equation}
        \mathcal{A}_{\mathrm{r}}\doteq\big\{(u,v)\in \mathcal{M}: v \geq v_2, \, \partial_ur(u,v)=0\big\},
        \label{eq:def right apparent horizon}
    \end{equation}
    are connected, achronal curves in the (1+1)-dimensional reduction of the spherically symmetric spacetime.
        \end{itemize} 
    \label{thm:grav collapse to ext SdS BH}
\end{customthm}

\begin{figure}[H]
        \centering
        \includegraphics[width=0.80\textwidth]{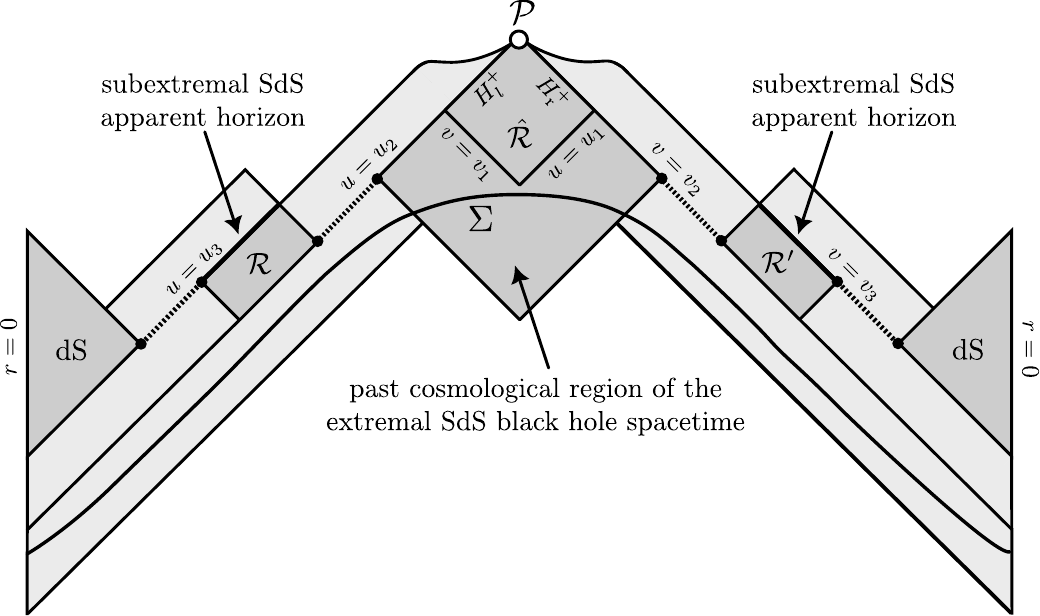}
        \caption{Penrose diagram illustrating the proof of Theorem \ref{thm:grav collapse to ext SdS BH}.} 
        \label{fig:grav collapse to ext SdS BH}
\end{figure}  

\begin{proof}
    We refer to Figure \ref{fig:grav collapse to ext SdS BH} for a visual summary of the proof. We start as in the proof of Theorem \ref{thm:grav collapse to subext SdS}, gluing to order $k+1$ a portion of the static patch $\mathcal{S}$ of de Sitter spacetime to a double null rectangle in the Schwarzschild--de Sitter spacetime of extremality index $\mathfrak{c}$, along the cone $\{u=u_3\}$, for some $u_3\in \mathbb R$. In this case we do not take the double null rectangle in the subextremal Schwarzschild--de Sitter spacetime extending to the boundary point $i^+$ and the cosmological horizon, but rather we arrange it so that its easternmost sphere has $r<r_{\mathcal{H}}(1)$. We obtain then a solution to the Einstein-scalar field system in double null neighbourhoods to the future and the past of the cone $\{u=u_3\}$, and the thin slab to the past can be extended to reach up to the centre by the same arguments as in the proof of Theorem \ref{thm:grav collapse to subext SdS}. 

    Now we glue to order $k$ from the easternmost sphere in the double null rectangle of the subextremal spacetime to a black hole horizon sphere in a double null rectangle in the extremal Schwarzschild--de Sitter black hole spacetime, making use again of Proposition \ref{prop:spacetime gluing} and Theorem \ref{thm:subext SdS to BHhorizon}. Note that the subextremal double null rectangle will be of the form $\{u_2 \leq u \leq u_3\} \cap \{{v_{-2} \leq v \leq v_{-1}}\}$, for some $u_3$ with $u_3>u_2$, and $v_{-1}$, $v_{-2}$ with $v_1>v_{-1}>v_{-2}$, so we carry out this second gluing along the cone $\{u=u_2\}$, thus obtaining a $C^k$ solution in a neighbourhood of $\{u=u_2\}$ which in this case may not be a full double null neighbourhood near the boundary point $\mathcal{P}$, and that, using again the same arguments as in Theorem \ref{thm:grav collapse to subext SdS}, can be extended to reach up to the centre. 

    Now we repeat the exact same construction using the analogues of Proposition \ref{prop:spacetime gluing} and Theorem \ref{thm:subext SdS to BHhorizon} along constant-$v$ cones. The double null rectangle of the extremal Schwarzschild--de Sitter spacetime will be of the form 
    $\{u_0 \leq u \leq u_2\} \cap \{v_0 \leq v \leq v_2\}$, for some $u_0$ with $u_0<u_2$ and $v_0, v_2$ with $v_{-1}<v_0<v_2$. Hence we start gluing this double null rectangle to an appropriate portion of the static patch $\mathcal{S}'$ of the subextremal Schwarzschild--de Sitter spacetime of extremality index\footnote{Note that, although we could have $\mathfrak{c}=\mathfrak{c}'$, in principle nothing forces the two extremality indices to be equal, since the gluing constructions along the constant-$u$ cone and the constant-$v$ cone are completely independent.} $\mathfrak{c}'$ along the cone $\{v=v_2\}$. Next, we glue this portion of the subextremal spacetime, which will be of the form $\{u_{-2} \leq u \leq u_{-1}\} \cap \{v_2 \leq v \leq v_3\}$ for some $u_{-2}<u_{-1}<u_0$ and $v_3>v_2$, to a portion of the static region $\mathcal{S}'$ of de Sitter spacetime, extending up to the centre.  
    
    We have constructed the spacetime shown in Figure \ref{fig:grav collapse to ext SdS BH}. As in the previous constructions, we can now identify an appropriate Cauchy hypersurface $\Sigma$, which can be chosen so that it does not intersect the cones $\{u=u_2\}$ and $\{v=v_2\}$. We consider then the Cauchy data induced by the constructed spacetime on $\Sigma$, $\big(\Sigma, g_0, k_0, \phi_0, \phi_1\big)$, whose future maximal globally hyperbolic development will include a region 
    \begin{equation}
        \hat{\mathcal{R}}= \{u_1 \leq u \leq u_2\} \cap \{v_1 \leq v \leq v_2\}
    \end{equation}
    for $u_1$ with $u_0<u_1<u_2$ and sufficiently close to $u_2$, and $v_1$ with $v_0<v_1<v_2$ and sufficiently close to $v_2$, lying to the future of $\Sigma$ and satisfying the desired properties. Note that the two double null rectangles that are isometric to regions of the subextremal Schwarzschild--de Sitter spacetimes and that satisfy the claims in parts $(ii)$ and $(iii)$ in the statement of the theorem will also be contained in the future maximal globally hyperbolic development of $\Sigma$. 
    
    The presence of additional event horizons in $\big(\mathcal{M}, g,\phi\big)$ is ruled out by Theorem~\ref{thm:no horizon points} and Corollary~\ref{cor:no hor points extremal}, and the connectedness of the apparent horizons in the massless case follows from Theorem~\ref{thm:connectedness}, all proved below. 
\end{proof}

We emphasise that the absence of additional event horizons in these spacetimes is what allows us to regard them as genuine counterexamples to the third law for $\Lambda>0$, as discussed in Section~\ref{subsec:intro third law}: it guarantees that the black hole region is unique, so the extremal region belongs to the same black hole that was subextremal at earlier times. These solutions therefore describe a subextremal black hole apparent horizon that becomes extremal in finite time. In fact, whenever the scalar field is massless, this is even more evident, since the portion of the apparent horizon lying between the subextremal and extremal pieces is connected.

\subsubsection{Ruling out the presence of additional event horizons}

We now examine whether the maximal developments of the data considered above could include additional event horizons. More precisely, for the future maximal globally hyperbolic development of Cauchy data satisfying the conclusions of Theorem \ref{thm:grav collapse to ext SdS BH} or Theorem~\ref{thm:grav collapse to subext SdS}, we aim to rule out the existence of boundary points that correspond neither to a central singularity (where $r\rightarrow 0$) nor to a component of future null infinity $\mathcal{I}^+$ (where $r\rightarrow \infty$), since, as will be discussed below, any additional event horizon would have to end at such a point.  

In order to achieve this, we seek a suitable extension principle that characterises when a solution to the Einstein-scalar field system \eqref{eq:EE massive scalar field - eqn for Omega}--\eqref{eq:EE massive scalar field - Raychaudhuri's u} can be continued, thereby allowing us to classify the boundary of the maximal globally hyperbolic development.

Extension principles have played a decisive role in understanding the global structure of dynamical solutions to the Einstein equations, starting with Christodoulou's seminal study on the gravitational collapse of a massless scalar field in the spherically symmetric, asymptotically flat setting \cite{Christodoulou_1993, Christodoulou_1999}. In the presence of a positive cosmological constant, the systematic study of the boundary decomposition of dynamical spacetimes via extension principles was initiated by Dafermos and Rendall in their work on surface-symmetric cosmological spacetimes with Vlasov matter \cite{Dafermos_Rendall_2016}. 

Our analysis builds on the work of Kommemi \cite{Kommemi_2013}, who obtained a classification of the possible boundary components of the maximal development, together with a generalised extension principle \cite[Thm.~1.8]{Kommemi_2013}, for solutions to the Einstein--Maxwell-charged scalar field system. Although Kommemi's analysis assumes $\Lambda=0$, one easily sees that the proof carries over to the $\Lambda>0$ case, yielding the following: 

\begin{extension}
    \label{thm:extension principle}
    Let $k\in \mathbb N_{\geq2}\cup\{\infty\}$ and $\mathfrak{m}\geq0$. Consider $\big(\mathcal{Q},r, \Omega^2, \phi\big)$ the $C^k$ future maximal globally hyperbolic development of a spherically symmetric Cauchy data set for the Einstein-scalar field system \eqref{eq:EE massive scalar field - eqn for Omega}--\eqref{eq:EE massive scalar field - Raychaudhuri's u} with scalar field mass $\mathfrak{m}$. Let $\mathcal{Y}\subset\mathcal{Q}$ be a region of the form 
    \begin{equation}
        \mathcal{Y}\doteq\big(\big[U, U'\big]\times\big[V, V'\big]\big)\big \backslash\big\{\big(U',V'\big)\big\},
    \end{equation}
    for some $U<U'$, $V<V'$. If there exist constants $R_1, R_2>0$ such that 
    \begin{equation}
        0<R_1\leq r(u,v)\leq R_2<\infty
        \label{eq:extension princ 1}
    \end{equation}
    for all $(u,v)\in\mathcal{Y}$, and moreover
    \begin{equation}
        \mathrm{Vol}(\mathcal{Y})=\int_U^{U'}\int_V^{V'}\Omega^2(u,v)dudv<\infty,
        \label{eq:extension princ 2}
    \end{equation}
    then $(U',V')\in \mathcal{Q}$. 
\end{extension}

\vspace{0.7em}

Put in other words, the generalised extension principle asserts the following. Let $(U',V')\in \overline{\mathcal{Q}}\backslash\mathcal{Q}$\footnote{We denote by $\overline{\mathcal{Q}}$ the closure of $\mathcal{Q}$.} be a \textit{(non-central) first singularity}, that is, a point not belonging to $\mathcal{Q}$ for which there exists a region of the form
\begin{equation}
    \mathcal{Y}\doteq\big(\big[U, U'\big]\times\big[V, V'\big]\big)\big\backslash\big\{\big(U',V'\big)\big\}\subset\mathcal{Q}
\end{equation}
with $U<U'$, $V<V'$. If $r$ remains bounded away from $0$ and $\infty$ on such a region $\mathcal{Y}$, then $\mathcal{Y}$ must have infinite volume. 

Motivated by this, we introduce the notion of \textit{horizon point}, which precisely identifies non-central first singularities on which $r\not\to0$, $r\not\to\infty$ but the generalised extension principle cannot be applied:

\begin{definition}[Horizon point]
    \label{def:horizon point}
    Let $k\in \mathbb N_{\geq2}\cup\{\infty\}$ and $\mathfrak{m}\geq0$. Consider $\big(\mathcal{Q},r, \Omega^2, \phi\big)$ the $C^k$ future maximal globally hyperbolic development of a spherically symmetric Cauchy data set for the Einstein-scalar field system \eqref{eq:EE massive scalar field - eqn for Omega}--\eqref{eq:EE massive scalar field - Raychaudhuri's u} with scalar field mass $\mathfrak{m}$. We will say that $(U',V')\in \overline{\mathcal{Q}}\backslash\mathcal{Q}$ is a \textit{horizon point} if 
    \begin{enumerate}[(i)]
        \item there exists a causal diamond 
        \begin{equation}
            \mathcal{Y}\doteq\big(\big[U, U'\big]\times\big[V, V'\big]\big)\big \backslash\big\{\big(U',V'\big)\big\}\subset \mathcal{Q}, 
        \end{equation}
        with $U<U'$, $V<V'$, and positive real numbers $R_1, R_2$ such that 
        \begin{equation}
            0<R_1\leq r(u,v)\leq R_2<\infty
            \label{eq:horizon pt 1}
        \end{equation}
        for all $(u,v)\in\mathcal{Y}$; 
        \item every causal diamond $\mathcal{Y}$ satisfying (i) has infinite volume, that is 
        \begin{equation}
            \mathrm{Vol}(\mathcal{Y})=\displaystyle \int_U^{U'}\int_V^{V'}\Omega^2(u,v)dudv=\infty. \\[0.8em]
        \end{equation}
    \end{enumerate}
\end{definition}

This notion of horizon point is adopted from \cite{Dafermos_Rendall_2016,Kommemi_2013}. In the asymptotically flat setting, spherical symmetry rules out the presence of multiple black hole event horizons, and the future endpoint of the unique event horizon already fails to satisfy condition (i) of Definition~\ref{def:horizon point}. In the presence of $\Lambda>0$, however, Definition~\ref{def:horizon point} identifies the future endpoints of event horizons (black hole or cosmological, extremal or subextremal), which in this setting satisfy both conditions (i) and (ii), the latter by the generalised extension principle. All the points on the future boundary of Nariai spacetime are also horizon points.

One can in fact easily show that if $(U',V')$ is a non-central first singularity that is not a horizon point, then $r$ extends ``continuously'' to zero or to $\infty$ at $(U',V')$ (see \cite{Kommemi_2013,Dafermos_Rendall_2016}). Moreover, if $(U',V')$ is a non-central first singularity at which $r$ extends continuously to zero, then $r$ also extends continuously to zero on any null component emanating from $(U',V')$. 

In view of the discussion above, the future endpoint of any additional event horizon is a horizon point, so to rule out additional event horizons it suffices to rule out horizon points. To this end, we establish the following general result:

\begin{theorem}
    Let $k\in \mathbb N_{\geq2}$, $\mathfrak{m}\geq0$, and consider $\big(\mathcal{Q},r, \Omega^2, \phi\big)$ the $C^k$ future maximal globally hyperbolic development of a spherically symmetric Cauchy data set for the Einstein-scalar field system \eqref{eq:EE massive scalar field - eqn for Omega}--\eqref{eq:EE massive scalar field - Raychaudhuri's u} with scalar field mass $\mathfrak{m}$, endowed with global double null coordinates $(u,v)$. Assume there exists a null cone $${C}\doteq\{u=u_*\}\cap\mathcal{Q}$$ such that $\partial_ur<0$ along ${C}$. Then there are no horizon points (in the sense of Definition \ref{def:horizon point}) in the region $$\widetilde{\mathcal{R}}\doteq\big\{(u,v)\in\overline{\mathcal{Q}}: u> u_*\, , \, (u_*,v)\in {C}  \big\}.$$ 
    \label{thm:no horizon points}
\end{theorem}
\begin{proof}
    Since $\partial_ur<0$ along ${C}$, and as a consequence of Raychaudhuri's equation,
    \begin{equation}
        \partial_u\left(\frac{\partial_ur}{\Omega^2}\right)=-\frac{r}{\Omega^2}\big(\partial_u\phi\big)^2\leq0,
        \label{eq:proof no other horizon points Raychaudhuri}
    \end{equation}
    we deduce that $\partial_ur<0$ everywhere in the region $\widetilde{\mathcal{R}}$. Moreover, for all $(u,v)\in\widetilde{\mathcal{R}}$, we have 
    \begin{equation}
        \frac{\partial_ur}{\Omega^2}(u,v)\leq \frac{\partial_ur}{\Omega^2}(u_*,v), 
    \end{equation}
    or equivalently
    \begin{equation}
        -\frac{\Omega^2}{\partial_ur}(u,v)\leq -\frac{\Omega^2}{\partial_ur}(u_*,v). 
        \label{eq:proof no other horizon points ineq}
    \end{equation}

    Now let $(U_1,V_1)\in\widetilde{\mathcal{R}}$ be a point satisfying condition (i) in Definition \ref{def:horizon point}. Then there exist $(U,V)\in \mathcal{Q}$, a causal diamond
    \begin{equation}
        \mathcal{Y}\doteq\big([U,U_1]\times[V,V_1]\big)\backslash\{(U_1,V_1)\}\subset \mathcal{Q},
    \end{equation}
    and real constants $R_1, R_2 >0$, such that 
    \begin{equation}
        0<R_1\leq r(u,v)\leq R_2<\infty, \quad \text{ for all } (u,v)\in\mathcal{Y}.
        \label{eq:proof no other horizon points condition r}
    \end{equation}
    Without loss of generality, we assume that $\mathcal{Y}\subset\widetilde{\mathcal{R}}$: otherwise, we may replace $\mathcal{Y}$ by a causal diamond contained in $\mathcal{Y}\cap\widetilde{\mathcal{R}}$ with future tip $(U_1,V_1)$, which still satisfies \eqref{eq:proof no other horizon points condition r}. Hence we have:
    \begin{align}
        \int_V^{V_1}\int_U^{U_1}\Omega^2(u,v)dudv&= \int_V^{V_1}\int_U^{U_1}\bigg(-\frac{\Omega^2}{\partial_ur}\bigg)(-\partial_ur)(u,v)dudv\\ 
        & \leq \int_V^{V_1} -\frac{\Omega^2}{\partial_ur}(u_*,v)\bigg(\int_U^{U_1}(-\partial_ur)(u,v)du\bigg)dv\\
        & \leq \int_V^{V_1} -\frac{\Omega^2}{\partial_ur}(u_*,v)\big(r(U,v)-r(U_1,v))dv\\ 
        & \leq \sup_{v\in[V,V_1]}\bigg(-\frac{\Omega^2}{\partial_ur}(u_*, v)\bigg)(R_2-R_1)  (V_1-V)<\infty,
    \end{align}
    where we have used \eqref{eq:proof no other horizon points ineq} and the fact that $\Omega^2(-\partial_ur)^{-1}$ is continuous as a function of $v$ on the truncated null cone $\{u_*\}\times[V,V_1]$ and therefore bounded. We have shown that any points in the region $\widetilde{\mathcal{R}}$ satisfying condition (i) in Definition \ref{def:horizon point} necessarily fail to satisfy (ii), and thus we conclude there are no horizon points in region $\widetilde{\mathcal{R}}$. 
\end{proof}

\begin{remark}
    \label{rmk:no horizon points symmetry}
    Note that, by the symmetry of the Einstein-scalar field system \eqref{eq:EE massive scalar field - eqn for Omega}--\eqref{eq:EE massive scalar field - Raychaudhuri's u} under $u\mapsto v$, $v\mapsto u$, it also follows from Theorem~\ref{thm:no horizon points} that if $\partial_v r <0$ along a null cone $\underline{{C}}\doteq\{v=v_*\}\cap\mathcal{Q}$, then there are no horizon points in the region $$\big\{(u,v)\in\overline{\mathcal{Q}}: v>v_*\, , \, (u,v_*)\in \underline{{C}} \big\}.$$ 
\end{remark}

\vspace{0.3em}

As shown in Corollary~\ref{cor:no hor points extremal} and Corollary~\ref{cor:no hor points subextremal} below, Theorem~\ref{thm:no horizon points} rules out horizon points in the black hole regions of the solutions of Theorem~\ref{thm:grav collapse to ext SdS BH} and Theorem~\ref{thm:grav collapse to subext SdS}, and hence, assuming the generalised extension principle, additional event horizons. 

\begin{corollary}
    Let $k\in \mathbb N_{\geq2}$, $\mathfrak{m}\geq0$, and consider $\big(\mathcal{Q},r, \Omega^2, \phi\big)$ the future maximal globally hyperbolic development of a Cauchy data set satisfying the conclusions of Theorem~\ref{thm:grav collapse to ext SdS BH} with regularity index $k$ and scalar field mass $\mathfrak{m}$. There are no horizon points in $\big(\mathcal{Q},r, \Omega^2, \phi\big)$ other than the one located at $(u_2, v_2)$ and $r$ extends continuously to zero on all other non-central first singularities.
    \label{cor:no hor points extremal}
\end{corollary}

\begin{proof}
    Let us consider the regions
    \begin{equation}
        \begin{aligned}
            \mathcal{R}_l&\doteq\Big(\big(\{u>u_2\}\cap\{v_{-1}<v< v_{2}\}\big)\cup\big(\{u>u_3\}\cap\{v_{-3}<v\leq v_{-1}\}\big)\Big)\cap\overline{\mathcal{Q}},\\
            \mathcal{R}_{\mathrm{r}}&\doteq\Big(\big(\{u_{-1}<u< u_{2}\}\cap\{v>v_2\}\big)\cup\big(\{u_{-3}<u\leq u_{-1}\}\cap\{v>v_3\}\big)\Big)\cap\overline{\mathcal{Q}}.
            \label{eq:no other horizons regions}
        \end{aligned}
    \end{equation}
    Here, $v_{-3}$ and $u_{-3}$ are such that the region isometric to a portion of the static patch $\mathcal{S}$ of de Sitter spacetime is $\big(\{u\geq u_3\}\cap\{v\leq v_{-3}\}\big)\cap \overline{\mathcal{Q}}$, whereas the region isometric to a portion of the static patch $\mathcal{S}'$ of de Sitter spacetime is $\big(\{u\leq u_{-3}\}\cap\{v\geq v_3\}\big)\cap\overline{\mathcal{Q}}$.

    Note that, by construction, there are no horizon points in $\overline{\mathcal{Q}}\backslash(\mathcal{R}_l\cup\mathcal{R}_{\mathrm{r}})$ other than the one at  $(u_2, v_2)$, which corresponds to the extremal Schwarzschild--de Sitter horizon endpoint $\mathcal{P}$. Therefore we need to show that there are no horizon points in the regions $\mathcal{R}_l, \mathcal{R}_{\mathrm{r}}$.

    Focusing first on $\mathcal{R}_l$, we consider the null cone ${C}_1\doteq\{u=u_2\}\cap\{v_{-1}<v< v_{2}\}$, along which $\partial_ur<0$ by construction. Then, by Theorem~\ref{thm:no horizon points}, we conclude that there are no horizon points in the region  
    \begin{equation}
        \mathcal{R}^1_l=\big(\{u>u_2\}\cap\{v_{-1}<v< v_{2}\}\big)\cap\overline{\mathcal{Q}}.
    \end{equation}
    Similarly, applying Theorem~\ref{thm:no horizon points} along the cone ${C}_2\doteq\{u=u_3\}\cap\{v_{-3}<v\leq v_{-1}\}$, which also has $\partial_ur<0$ by construction, we can rule out the existence of horizon points in the region 
    \begin{equation}
        \mathcal{R}^2_l=\big(\{u>u_3\}\cap\{v_{-3}<v\leq v_{-1}\}\big)\cap\overline{\mathcal{Q}}.
    \end{equation}
    
    \vspace{-0.5em}

    Consequently, we deduce that there are no horizon points in the region $\mathcal{R}^1_l\cup\mathcal{R}^2_l=\mathcal{R}_l$. Moreover $r$ extends continuously to zero on all other non-central first singularities in $\mathcal{R}_l$ (see \cite{Kommemi_2013,Dafermos_Rendall_2016}). By Remark \ref{rmk:no horizon points symmetry}, an analogous argument shows the claim for region $\mathcal{R}_{\mathrm{r}}$.
\end{proof}

\begin{corollary}
    Let $k\in \mathbb N_{\geq2}$, $\mathfrak{m}\geq0$, and consider $\big(\mathcal{Q},r, \Omega^2, \phi\big)$ the future maximal globally hyperbolic development of a Cauchy data set satisfying the conclusions of Theorem~\ref{thm:grav collapse to subext SdS} with regularity index $k$ and scalar field mass $\mathfrak{m}$. There are no horizon points in the region $\{u>u_2\}\cap\{v<v_2\}$, and moreover $r$ extends continuously to zero on all other non-central first singularities in this region. 
    \label{cor:no hor points subextremal}
\end{corollary}

\begin{proof}
    Consider the cone ${C}=\{u=u_2\}\cap\{v<v_2\}\subset\mathcal{Q}$, along which $\partial_ur<0$ by construction. It follows from Theorem~\ref{thm:no horizon points} that there are no horizon points in the region $\{u>u_2\}\cap\{v<v_2\}$. As before, $r$ extends continuously to zero on all other non-central first singularities in this region \cite{Kommemi_2013,Dafermos_Rendall_2016}. 
\end{proof}

\begin{figure}[H]
        \centering
        \hspace{-0.7em}
        \begin{subfigure}[c]{0.4\textwidth}
            \centering
            \includegraphics[width=1\textwidth]{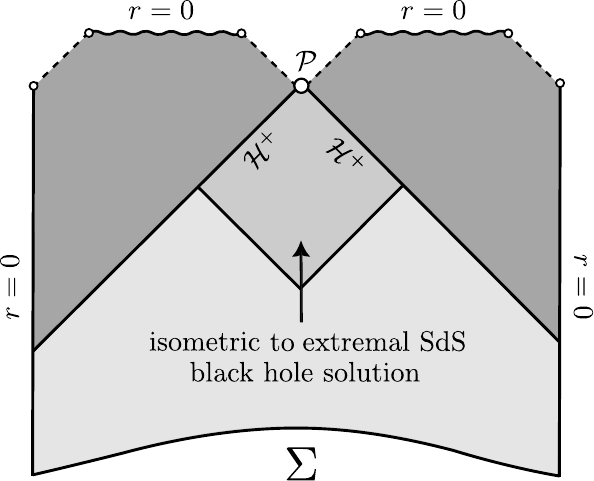}
            \caption{}
            \label{subfig:global no other horizons ext}
        \end{subfigure}
        \hspace{4.0em}
        \begin{subfigure}[c]{0.4\textwidth}
            \centering
            \includegraphics[width=1\textwidth]{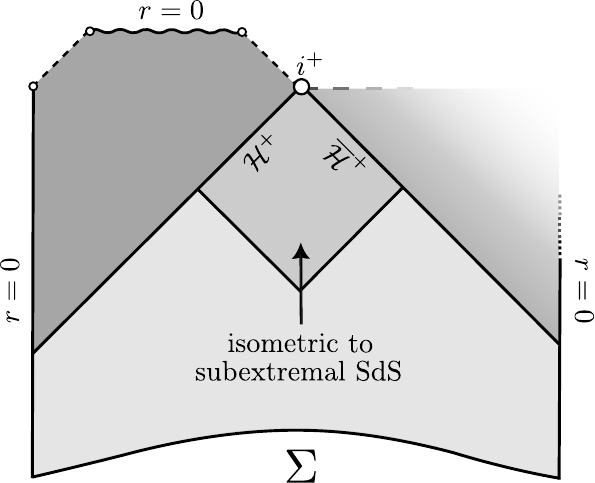}
            \caption{}
            \label{subfig:global no other horizons subext}
        \end{subfigure}
        \caption{General Penrose diagrams for the solutions described by (a) Theorem~\ref{thm:grav collapse to ext SdS BH} and (b) Theorem~\ref{thm:grav collapse to subext SdS}, taking into account the absence of additional horizon points in the black hole regions (as shown in Corollary~\ref{cor:no hor points extremal} and Corollary~\ref{cor:no hor points subextremal}, respectively). In both cases, the null components of the future boundary could potentially be non-empty.} 
        \label{fig:no other horizons}
\end{figure}

In view of Corollary~\ref{cor:no hor points extremal}, the solutions described by Theorem~\ref{thm:grav collapse to ext SdS BH} have, in general, the Penrose diagram depicted in Figure \ref{subfig:global no other horizons ext}, where the null components of the future boundary could potentially be non-empty. Similarly, as a consequence of Corollary~\ref{cor:no hor points subextremal}, the solutions described by Theorem~\ref{thm:grav collapse to subext SdS} look in general as shown in Figure \ref{subfig:global no other horizons subext}, where again the null components of the future boundary could potentially be non-empty. 

\vspace{1em}

As already mentioned in Section \ref{subsec:global structure}, it remains to be seen in which cases the existence of null components emanating from the centres or from the horizon points $i^+$ (in the subextremal case) or $\mathcal{P}$ (in the extremal case) can be ruled out. We also point out that, for both the solutions described by Theorem~\ref{thm:grav collapse to ext SdS BH} and the ones described by Theorem~\ref{thm:grav collapse to subext SdS}, the $r=0$ boundary is spacelike (as depicted) for $\mathfrak{m}=0$; if $\mathfrak{m}>0$, however, the $r=0$ boundary may be merely achronal and have null pieces.

We emphasise that, in the subextremal case (Figure \ref{subfig:global no other horizons subext}), the argument does not carry over to the null cone containing the portion of the subextremal cosmological horizon, so the existence of additional horizon points beyond that cone cannot, a priori, be ruled out.

\subsubsection{Connectedness of the apparent horizon}

As argued in Section~\ref{subsec:intro third law}, it is also interesting to examine the connectedness of the apparent horizon in the third-law-violating spacetimes constructed in Theorem \ref{thm:grav collapse to ext SdS BH}, a feature that cannot arise in the asymptotically flat setting, as shown by Israel~\cite{Israel_1986}. In this regard, we show the following:

\begin{theorem}
    In the future maximal globally hyperbolic development of any Cauchy data set for the Einstein-massless scalar field system satisfying the conclusions of Theorem \ref{thm:grav collapse to ext SdS BH}, the left and right apparent horizons, $\mathcal{A}_{l}$ and $\mathcal{A}_{\mathrm{r}}$, defined by \eqref{eq:def left apparent horizon} and \eqref{eq:def right apparent horizon},  are connected, achronal curves in the (1+1)-dimensional reduction of the spherically symmetric spacetime.
    \label{thm:connectedness}
\end{theorem}

\begin{proof}
    Let us consider a Cauchy data set for the Einstein-massless scalar field satisfying Theorem \ref{thm:grav collapse to ext SdS BH} for some $k \in \mathbb N$, $\mathfrak{c}, \mathfrak{c}' \in(0,1)$. Then its future maximal globally hyperbolic development $\big(\mathcal{Q},r,\Omega^2, \phi\big)$ will satisfy the conclusions of Theorem \ref{thm:grav collapse to ext SdS BH}. By the $u\leftrightarrow v$ symmetry of the Einstein-scalar field system \eqref{eq:EE massive scalar field - eqn for Omega}--\eqref{eq:EE massive scalar field - Raychaudhuri's u}, it suffices to prove the result for the left apparent horizon $\mathcal{A}_{l}$, so we restrict our attention to the region
    \begin{equation}
    \widetilde{\mathcal{R}}_l\doteq \{u\geq u_2\}\cap \mathcal{Q}
    \end{equation} 
    and hence $\mathcal{A}_l$ as defined in \eqref{eq:def left apparent horizon} is simply
    \begin{equation}
        \mathcal{A}_{\mathrm{l}}=\big\{(u,v)\in \widetilde{\mathcal{R}}_l: \partial_vr(u,v)=0\big\}.
    \end{equation}
    For convenience, we will write $\widetilde{\mathcal{R}}_l=\widetilde{\mathcal{R}}^1_l\cup\widetilde{\mathcal{R}}^2_l$, with
    \begin{equation}
        \widetilde{\mathcal{R}}^1_l\doteq \big(\{u\geq u_2\}\cap\{v_{-2}\leq v < v_2\}\big)\cap \mathcal{Q},
    \end{equation}
    \begin{equation}
        \widetilde{\mathcal{R}}^2_l\doteq \big(\{u\geq u_2\}\cap\{v\leq v_{-1}\}\big)\cap \mathcal{Q},
    \end{equation}
    and study separately the connectedness of $\mathcal{A}_l$ in the regions $\widetilde{\mathcal{R}}^1_l$ and $\widetilde{\mathcal{R}}^2_l$. 

    \vspace{0.7em}

    \textbf{\myuline{Connectedness and causal character of $\mathcal{A}_l\cap\widetilde{\mathcal{R}}^1_l$}} 

    \vspace{0.3em}
    
    \noindent Instead of working with the whole region $\widetilde{\mathcal{R}}^1_l$, for certain parts of our argument it will be useful to consider separately the region 
      \begin{equation}
        \underline{\mathcal{R}}\doteq \{u_2\leq u \leq u_3\}\cap\{v_{-1}<v<v_0\}, 
    \end{equation}
    and also 
    \begin{equation}
        \begin{aligned}
            \mathcal{R}_{\mathrm{in}}&\doteq \big(\{u>u_3\}\cap \{v_{-2}\leq v \leq v_{-1}\}\big)\cap\mathcal{Q},\\
            \underline{\mathcal{R}}_{\mathrm{in}}&\doteq\big(\{u>u_3\}\cap \{v_{-1}<v< v_{0}\}\big)\cap\mathcal{Q},\\
            \hat{\mathcal{R}}_{\mathrm{in}}&\doteq \big(\{u>u_2\}\cap \{v_0\leq v< v_2\}\big)\cap\mathcal{Q}.\\[0.5em]
        \end{aligned}
    \end{equation}
    These are all depicted in Figure~\ref{fig:setup proof connected apparent horizon}. 

    An immediate observation is that $\partial_ur<0$ everywhere in $\widetilde{\mathcal{R}}^1_l$. To see this, recall that $\partial_ur<0$ along the cone $\{u_2\}\times[v_{-2},v_2)$ by construction. From Raychaudhuri's equation \eqref{eq:EE massive scalar field - Raychaudhuri's u}, we deduce
    \begin{equation}
        \frac{\partial_ur}{\Omega^2}(u,v)\leq \frac{\partial_ur}{\Omega^2}(u_2,v)
    \end{equation}
    for all $(u,v)$ with $u\geq u_2$ and $v\in[v_{-2},v_{2})$. As a consequence of this, for all $u>u_2$, $v\in[v_{-2},v_2)$, we also have   
    \begin{equation}
        r(u,v)<r(u_2,v)\leq\frac{1}{\sqrt{\Lambda}}. 
    \end{equation}

    \vspace{0.4em}

    \begin{figure}[H]
        \centering
        \includegraphics[width=0.64\textwidth]{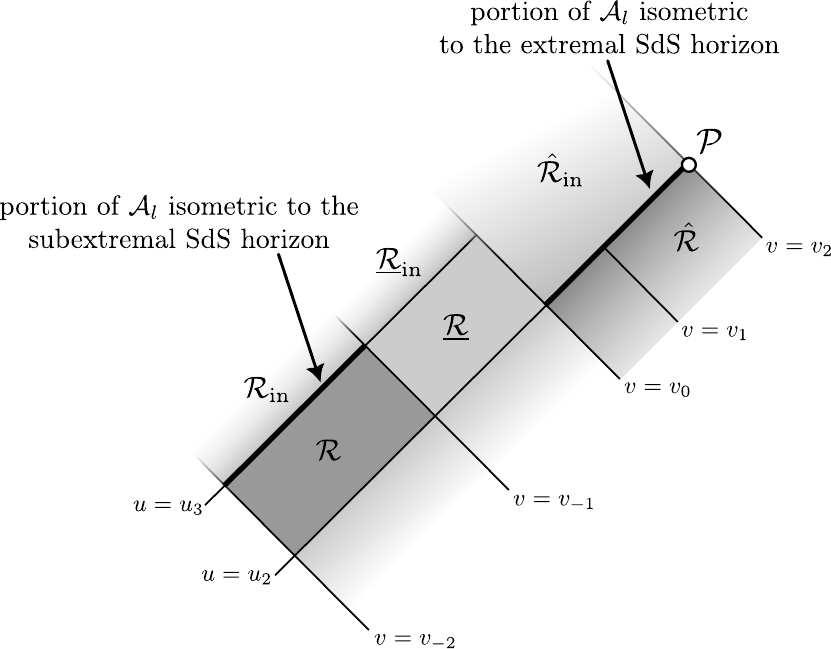}
        \caption{Set-up for the proof that the portion of the left apparent horizon $\mathcal{A}_l$ lying in the region $\widetilde{\mathcal{R}}^1_l$ is connected, showing the regions that will be relevant for the analysis.} 
        \label{fig:setup proof connected apparent horizon}
    \end{figure}  

    For our discussion here it will be useful to consider $s\doteq\partial_vr^2$ instead of $\partial_vr$. The points on the apparent horizon will then have $s=0$. Note that, in view of Raychaudhuri's equation \eqref{eq:EE massive scalar field - Raychaudhuri's v}, $s$ satisfies the equation
    \begin{equation}
        \partial_v\Big(\frac{s}{2r\Omega^2}\Big)=-\frac{r}{\Omega^2}(\partial_v\phi)^2, 
        \label{eq:partial v s}
    \end{equation}
    and, considering the wave equation for $r$ rewritten in the form \eqref{eq:EE-massive scalar field - eqn for r partial_ur}, we also have
    \begin{equation}
        \partial_us=\frac{\Omega^2}{2}\big(r^2\Lambda-1\big),
        \label{eq:partial u s}
    \end{equation}
    where we have already taken into account that we are considering a solution involving a massless scalar field.

    In particular, we have $\partial_ur<0$ and $r<1/\sqrt{\Lambda}$ everywhere in $\mathcal{R}_{\mathrm{in}}\cup\underline{\mathcal{R}}_{\mathrm{in}}\cup\hat{\mathcal{R}}_{\mathrm{in}}$. Moreover, $s<0$ in these regions, so they are foliated by trapped spheres, as shown in the following lemma.
    \begin{lemma}
        Any symmetry sphere $(u,v)\in \mathcal{R}_{\mathrm{in}} \cup \underline{\mathcal{R}}_{\mathrm{in}}\cup \hat{\mathcal{R}}_{\mathrm{in}}$ has $s(u,v)<0$. 
        \label{lem:connectedness trapped regions}
    \end{lemma}
    \begin{proof}[Proof of Lemma \ref{lem:connectedness trapped regions}]
        We start with region $\mathcal{R}_{\mathrm{in}}$. Recall that the truncated null cone $\{u_3\}\times[v_{-2},v_{-1}]$ can be identified with an appropriate portion of the right component of the subextremal Schwarzschild--de Sitter black hole event horizon, and hence $s(u_3,v)=0$ for $ v\in[v_{-2},v_{-1}]$. We now consider \eqref{eq:partial u s} along every constant-$v$ cone with $ v\in[v_{-2},v_{-1}]$. Since the right-hand side of \eqref{eq:partial u s} is strictly negative in $\mathcal{R}_{\mathrm{in}}\cup\big(\{u_3\}\times[v_{-2},v_{-1}]\big)$, we conclude
        \begin{equation}
            s(u,v)<s(u_3,v) 
        \end{equation}
        for all $u>u_3$, $v\in [v_{-2},v_{-1}]$, and hence $s<0$ everywhere in $\mathcal{R}_{\mathrm{in}}$. 

        In order to show the claim for region $\underline{\mathcal{R}}_{\mathrm{in}}$, it suffices to consider \eqref{eq:partial v s}, which along every constant-$u$ cone with $u>u_3$ implies
        \begin{equation}
            \frac{s}{2r\Omega^2}(u,v)\leq \frac{s}{2r\Omega^2}(u,v_{-1})<0, 
        \end{equation}
        for any $v>v_{-1}$, where we have used the negativity of $s$ in $\mathcal{R}_{\mathrm{in}}$. 

        Finally we prove the claim for region $\hat{\mathcal{R}}_{\mathrm{in}}$. In this case the truncated cone $\{u_2\}\times[v_0,v_2)$ can be identified with an appropriate portion of the right component of the extremal Schwarzschild--de Sitter black hole event horizon, and hence $s(u_2,v)=0$ for $v\in[v_0,v_2)$. However this horizon also has $r=1/\sqrt{\Lambda}$, and thus the right-hand side of \eqref{eq:partial u s} vanishes identically for all $(u_2,v)$ with  $v\in[v_0,v_2)$. Differentiating \eqref{eq:partial u s} in $u$ we obtain 
        \begin{equation}
            \partial_u^2s (u_2,v) = \bigg(\frac{\partial_u\Omega^2}{2}\big(r^2\Lambda -1\big)+\frac{\Omega^2}{2}(2r\partial_ur\Lambda)\bigg)(u_2,v)=\sqrt{\Lambda}\big(\Omega^2\partial_ur)(u_2,v)<0,
        \end{equation}
        for all $v\in[v_0,v_2)$. This, combined with the fact that $\partial_u s(u,v)<0$ for all $(u,v)\in \hat{\mathcal{R}}_{\mathrm{in}}$ (since $r<1/\sqrt{\Lambda}$ in $\hat{\mathcal{R}}_{\mathrm{in}}$), allows us to conclude that $s<0$ in $\hat{\mathcal{R}}_{\mathrm{in}}$.
    \end{proof}
    As a consequence of this, we can establish the following lemma.  

    \begin{lemma}
        \label{lem:spheres on the apparent horizon}
        In region $\widetilde{\mathcal{R}}^1_l$, the following hold:
        \begin{enumerate}[(a)]
            \item For each $v\in[v_{-2}, v_2)$, there exists a unique $u_*\in[u_2,u_3]$ with $s(u_*,v)=0$. That is, each constant-$v$ cone in $\widetilde{\mathcal{R}}^1_l$ contains exactly one sphere belonging to $\mathcal{A}_l$. 
            \item For each $u\in[u_2, u_3]$, there exists a non-empty interval $I\subset [v_{-2}, v_2)$ (possibly a single point, $I=\{v_*\}$) such that $s(u,v)=0$ for all $v\in I$. That is, each constant-$u$ cone in $\widetilde{\mathcal{R}}^1_l$ with $u\in[u_2,u_3]$ contains a non-empty, connected set of spheres belonging to $\mathcal{A}_l$, which may reduce to a single sphere.
        \end{enumerate}
    \end{lemma}

    \begin{proof}[Proof of Lemma \ref{lem:spheres on the apparent horizon}]
        We start with claim \textit{(a)}. Note that, in $\mathcal{R}\cup\mathcal{R}_{\mathrm{in}}\cup\underline{\mathcal{R}}\cup\underline{\mathcal{R}}_{\mathrm{in}}$, we have $r<1/\sqrt{\Lambda}$ everywhere. In view of \eqref{eq:partial u s}, we then deduce that along constant-$v$ cones with $v\in[v_{-2}, v_0)$, $s(\cdot,v)$ is a strictly decreasing function of $u$, and hence, for each $v\in[v_{-2}, v_0)$, there exists at most one $u_*\geq u_2$ such that $s(u_*,v)=0$. 
        
        For all $v\in [v_{-2}, v_{-1}]$, $u_*=u_3$ by construction. For all $v\in (v_{-1}, v_0)$, recall that $s(u_2,v)>0$ by construction, since the truncated null cone $\{u_2\}\times(v_{-1},v_0)$ is the one along which we carried out the gluing from the subextremal Schwarzschild--de Sitter static region to the extremal Schwarzschild--de Sitter black hole horizon. Let us fix $\tilde{v}\in(v_{-1},v_0)$. We know from Lemma \ref{lem:connectedness trapped regions} that $s(u,\tilde{v})<0$ for all $u>u_3$, whereas $s(u_2,\tilde{v})>0$. Since $s(\cdot,\tilde{v})$ is a $C^k$ function of $u$, there must exist $u_*\in(u_2,u_3]$ such that $s(u_*, \tilde{v})=0$. Given that $\tilde v$ was arbitrary, this proves the claim for all $v\in(v_{-1},v_0)$. 
        
        For $v\in[v_0,v_2)$, we have by construction that $s(u_2, v)=0$, and by Lemma \ref{lem:connectedness trapped regions}, we know that $s(u,v)<0$ for all $u>u_2$. Hence claim \textit{(a)} holds for all $v\in[v_{-2},v_2)$. 

        For the proof of claim \textit{(b)}, we argue in an analogous manner. Let us first consider $\tilde{u}\in(u_2,u_3)$. From Lemma \ref{lem:connectedness trapped regions} we have $s(\tilde{u},v)<0$ for $v\geq v_0$, and $s(\tilde{u}, v)>0$ for $v\leq v_{-1}$, since $(\tilde{u},v)$ is in this case a sphere in the static region $\mathcal{S}$ of a subextremal Schwarzschild--de Sitter spacetime. Then there exists $v_*\in (v_{-1},v_0)$ such that $s(\tilde{u},v_*)=0$. Note that this $v_*$ may not be unique, but, in view of \eqref{eq:partial v s}, $s(\tilde{u},v)\leq0$ for all $v\geq v_*$, and moreover if $s(\tilde{u},\mathring{v})<0$, then $s(\tilde{u},v)<0$ for all $v\geq \mathring{v}$. From this we conclude claim \textit{(b)} for all $\tilde{u}\in(u_2,u_3)$. For $u=u_3$, by construction $s(u_3, v)=0$ for all $v\in[v_{-2},v_{-1}]$, and as before there could be more values of $v$ for which $s(u_3,v)=0$, but they must form a subinterval of $[v_{-2},v_0)$ in view of \eqref{eq:partial v s}. For $u=u_2$ we know that $s(u_2,v)=0$ for all $v\in[v_0,v_2)$ by construction, and $s(u_2,v)>0$ for $v<v_0$, thus claim \textit{(b)} holds for all $u\in[u_2,u_3]$.
    \end{proof}

    The last ingredient needed to establish the connectedness of the apparent horizon relies on its causal character. It is clear that the portions of $\mathcal{A}_l$ that are isometric to portions of the subextremal and extremal Schwarzschild--de Sitter apparent horizons are null. The portion of $\mathcal{A}_l$ lying in $\underline{\mathcal{R}}$ is in fact everywhere spacelike or null, as shown in the following lemma.

    \begin{lemma}
        The set $\mathcal{A}_l\cap\underline{\mathcal{R}}$ is a $C^k$ hypersurface whose normal vector field is everywhere causal, and hence it is spacelike or null.
        \label{lem:causal character apparent horizon}
    \end{lemma}

    \begin{proof}[Proof of Lemma \ref{lem:causal character apparent horizon}]
        Note that $\mathcal{A}_l\cap\underline{\mathcal{R}}=\big(s|_{\underline{\mathcal{R}}}\big)^{-1}(\{0\})$, where $0$ is a regular value of the restriction $s|_{\underline{\mathcal{R}}}$. In order to see this, we compute $ds$ and evaluate at a sphere $(u,v)\in \mathcal{A}_{l}\cap\underline{\mathcal{R}}$. Since $r<{1}/{\sqrt{\Lambda}}$ in $\underline{\mathcal{R}}$, we have $\partial_us<0$ everywhere on $\underline{\mathcal{R}}$ so, in particular, 
        \begin{equation}
            \partial_us\vert_{\mathcal{A}_l\cap\underline{\mathcal{R}}}<0. 
        \end{equation}
        On the other hand, from \eqref{eq:partial v s} we deduce
        \begin{equation}
            \partial_vs\vert_{\mathcal{A}_l\cap\underline{\mathcal{R}}}= \Big(-2r^2(\partial_v\phi)^2+s\, \partial_v \ln \big(r\Omega^2\big)\Big)\Big\vert_{\mathcal{A}_l\cap\underline{\mathcal{R}}}=-2r^2(\partial_v\phi)^2\big\vert_{\mathcal{A}_l\cap\underline{\mathcal{R}}}\leq 0.
        \end{equation}
        More precisely, $\partial_vs(u,v)<0$ if $\partial_v\phi (u,v)\neq0$, and  $\partial_vs(u,v)=0$ if  $\partial_v\phi (u,v)=0$. 
        Thus we conclude that for every sphere $(u,v)\in \mathcal{A}_l\cap\underline{\mathcal{R}}$, we have
        \begin{equation}
            ds\vert_{(u,v)}=\partial_us\vert_{(u,v)}du+\partial_vs\vert_{(u,v)}dv\neq 0. 
        \end{equation}

        Now it is straightforward to compute the normal to $\mathcal{A}_l\cap\underline{\mathcal{R}}$, which will simply be given by
        \begin{equation}
            N\doteq\mathrm{grad} \, s= -\frac{2}{\Omega^2}\partial_vs \,\partial_u -\frac{2}{\Omega^2}\partial_us \,  \partial_v 
        \end{equation}
        and hence 
        \begin{equation}
            g(N,N)\big \vert_{\mathcal{A}_l\cap\underline{\mathcal{R}}} = -\frac{4}{\Omega^2}\partial_vs\partial_us \Big \vert_{\mathcal{A}_l\cap\underline{\mathcal{R}}} 
            \begin{cases} 
                <0 &\, \text{ and }N \text{ is timelike} \text{ if } \, \partial_v\phi\neq 0, \\
                =0 & \, \text{ and }N \text{ is null} \text{ if } \, \partial_v\phi= 0.
            \end{cases}
        \end{equation}
    \end{proof}

    We then conclude that $\mathcal{A}_l\cap\widetilde{\mathcal{R}}^1_l$ is an achronal curve in the (1+1)-dimensional reduction of the spherically symmetric spacetime. In order to see that it is connected, we consider the map $u_{\mathcal{A}}: [v_{-2}, v_2)\rightarrow[u_2, u_3]$, defined by the condition $s\big(u_{\mathcal{A}}(v),v\big)=0$, for every $v\in[v_{-2}, v_2)$. This is a well-defined function in view of part \textit{(a)} in Lemma \ref{lem:spheres on the apparent horizon} and, by part \textit{(b)} of the same lemma, it takes all values in $[u_2,u_3]$. Now $\mathcal{A}_{l}\cap\widetilde{\mathcal{R}}^1_l$ can also be expressed as the graph of the function $u_{\mathcal{A}}$ over $[v_{-2}, v_2)$:
    \begin{equation}
        \mathcal{A}_{l}\cap\widetilde{\mathcal{R}}^1_l=\big\{(u_{\mathcal{A}}(v),v): v\in[v_{-2}, v_2)\big\}. 
        \label{eq:def apparent horizon proof connectedness}
    \end{equation}

    Finally, we show that:
    \begin{lemma}
        The function $u_{\mathcal{A}}: [v_{-2}, v_2)\rightarrow[u_2, u_3]$ is monotone decreasing. 
        \label{lem:decreasing}
    \end{lemma}
    \begin{proof}[Proof of Lemma \ref{lem:decreasing}]
        On $(v_{-1},v_0)$, $u_{\mathcal{A}}$ is differentiable by the implicit function theorem, since $\partial_us<0$ everywhere in $\underline{\mathcal{R}}$. In this region, the tangent vector field to its graph, that is $\mathcal{A}_{l}\cap\underline{\mathcal{R}}$, can thus be expressed as
        \begin{equation}
            T=\bigg(\frac{du_{\mathcal{A}}}{dv}, 1\bigg),
        \end{equation} 
        and, in view of Lemma \ref{lem:causal character apparent horizon}, we get
        \begin{equation}
            g(T,T) = -\Omega^2 \frac{du_{\mathcal{A}}}{dv}\geq 0
        \end{equation}
        We deduce that $u_{\mathcal{A}}$ is a monotone decreasing function of $v$ on $(v_{-1},v_0)$. 

        On $[v_{-2},v_{-1}]$, $u_{\mathcal{A}}$ is constant and equal to $u_3$ (its maximum value). On $[v_0,v_2)$, $u_{\mathcal{A}}$ is constant and equal to $u_2$ (its minimum value). Hence $u_{\mathcal{A}}$ is decreasing on $[v_{-2},v_2)$.
    \end{proof}
    
    Since $u_{\mathcal{A}}:[v_{-2}, v_2)\rightarrow[u_2, u_3]$ is monotone decreasing and takes all values $u_2\leq u \leq u_3$, it is a continuous function of $v$ on $[v_{-2}, v_2)$, and hence its graph $\mathcal{A}_{l}\cap\widetilde{\mathcal{R}}^1_l$ is connected as a subset of $\widetilde{\mathcal{R}}^1_l$.

    \vspace{0.8em}

    \textbf{\myuline{Connectedness and causal character of $\mathcal{A}_l\cap\widetilde{\mathcal{R}}^2_l$}} 

    \vspace{0.3em}
    
    \noindent In this case, the region we analyse extends up to the regular left centre, $\Gamma_l$, the boundary point $b_{\Gamma_l}$, a future boundary null component emanating from $b_{\Gamma_l}$ that could be non-empty, and the $r=0$ singularity.  In what follows we assume that $u\to u_b$ and $v\to v_b$ as we approach $b_{\Gamma_l}$, for some finite values $u_b,v_b$. The set-up for this case is illustrated in Figure~\ref{fig:setup proof connected apparent horizon 2}.  

    As before we will consider $s=\partial_vr^2$ instead of $\partial_vr$. We start noting that all spheres in the region 
    \begin{equation}
        \{u_2\leq u < u_3\}\cap \widetilde{\mathcal{R}}^2_l
    \end{equation} 
    have $s>0$, given that this is true for the static spheres in $\{u_2\leq u < u_3\}\cap\{v_{-2}\leq v \leq v_{-1}\}$, and therefore true for spheres in $\{u_2\leq u < u_3\}\cap\{v<v_{-2}\}$ in view of \eqref{eq:partial v s}. Hence the portion of $\mathcal{A}_l$ lying in $\widetilde{\mathcal{R}}^2_l$ must be contained in the region $\{u\geq u_3\}\cap \widetilde{\mathcal{R}}^2_l$.

    Along $\{u=u_3\}\cap \widetilde{\mathcal{R}}^2_l$ we have, by construction, $\partial_ur<0$ and $r<1/{\sqrt{\Lambda}}$, and thus, as a consequence of Raychaudhuri's equation in $u$, the same inequalities hold everywhere on $\{u\geq u_3\}\cap \widetilde{\mathcal{R}}^2_l$. Taking into account \eqref{eq:partial u s}, we also conclude that $\partial_us<0$ everywhere on $\{u\geq u_3\}\cap \widetilde{\mathcal{R}}^2_l$.

    \begin{figure}[H]
        \centering
        \hspace{8em}\includegraphics[width=0.66\textwidth]{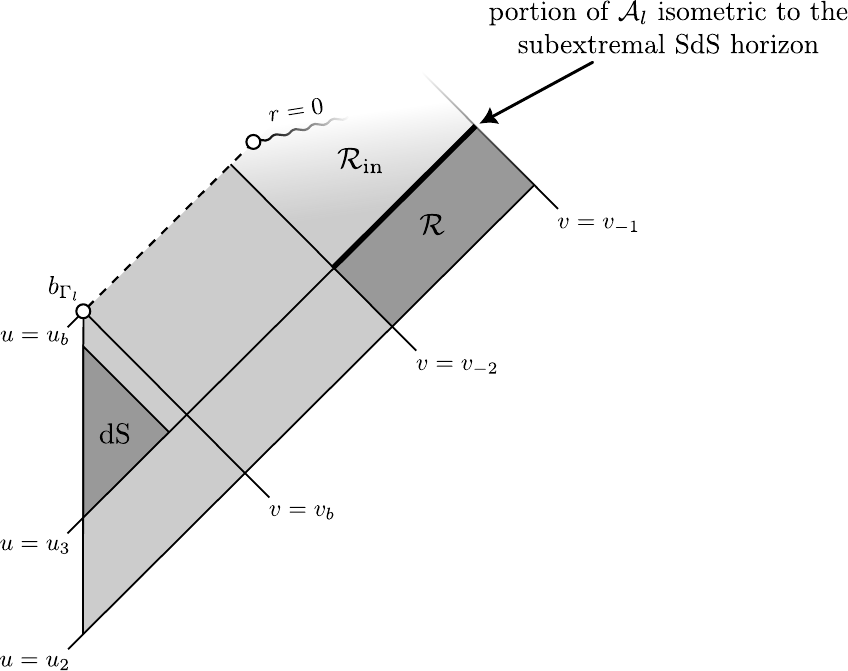}
        \caption{Set-up for the proof that the portion of the left apparent horizon $\mathcal{A}_l$ lying in the region $\widetilde{\mathcal{R}}^2_l$ is connected. Here we are assuming the general global structure already depicted in Figure \ref{subfig:global no other horizons ext}, where the null component of the future boundary could potentially be non-empty.} 
        \label{fig:setup proof connected apparent horizon 2}
    \end{figure} 

    Note that the region $\{u_3\leq u <u_b\}\cap\{v<v_b\}$ also has $s>0$, given that all the spheres along the regular centre have $s>0$ and $s$ is strictly decreasing along constant-$v$ cones, as argued above. Thus we deduce that $\mathcal{A}_l\cap\widetilde{\mathcal{R}}^2_l\subset \mathcal{R}^*$ with $\mathcal{R}^*$ defined as 
    \begin{equation}
        \mathcal{R}^*\doteq\big(\{u_3\leq u<u_b\}\cap\{v_b \leq v \leq v_{-1}\}\big)\cap\mathcal{Q}. 
    \end{equation} 

    We have the following analogue of Lemma \ref{lem:spheres on the apparent horizon}: 
    \begin{lemma}
        \label{lem:spheres on the apparent horizon 2}
        In region $\mathcal{R}^*$, the following hold:
        \begin{enumerate}[(a)]
            \item There exists $v_{\mathrm{inf}}\in [v_b, v_{-2})$ such that for each $v\in(v_{\mathrm{inf}}, v_{-1}]$, there exists a unique $u_*\in[u_3,u_b)$ with $s(u_*,v)=0$, and moreover $s(u,v)>0$ for all $(u,v)\in \{v\leq v_{\mathrm{inf}}\}\cap\mathcal{R}^*$. That is, there exists a threshold value $v_{\mathrm{inf}}$, such that any constant-$v$ cone in $\mathcal{R}^*$ with $v\leq v_{\mathrm{inf}}$ does not contain any points belonging to $\mathcal{A}_l$, and such that any constant-$v$ cone in $\mathcal{R}^*$ with $v>v_{\mathrm{inf}}$ contains exactly one sphere belonging to $\mathcal{A}_l$. 
            \item For each $u\in[u_3, u_b)$, there exists a non-empty interval $I\subset (v_{\mathrm{inf}}, v_{-1}]$ (possibly a single point, $I=\{v_*\}$) such that $s(u,v)=0$ for all $v\in I$. That is, each constant-$u$ cone in $\mathcal{R}^*$ contains a non-empty, connected set of spheres belonging to $\mathcal{A}_l$, which may reduce to a single sphere.
        \end{enumerate}
    \end{lemma}
    \begin{proof}[Proof of Lemma \ref{lem:spheres on the apparent horizon 2}]
        We start proving claim \textit{(a)}. The fact that $s$ is a strictly decreasing function of $u$ along constant-$v$ cones in $\mathcal{R}^*$ guarantees that, for any $v\in[v_b, v_{-1}]$, there exists at most one $u_*\in[u_3,u_b)$ with $s(u_*,v)=0$. Note that for all $v\in[v_{-2}, v_{-1}]$, $u_*=u_3$ by construction. 
        
        We argue now that if a constant-$v$ cone with $v=\tilde{v}_1$ in $\mathcal{R}^*$ contains one sphere belonging to $\mathcal{A}_l$, then any constant-$v$ cone with $v>\tilde{v}_1$ must also contain an apparent horizon sphere. Suppose, for contradiction, that there exists $\tilde{v}_2$, $\tilde{v}_2>\tilde{v}_1$, such that the cone $\{v=\tilde{v}_2\}\cap\mathcal{R}^*$ contains no apparent horizon spheres (hence it must be $\tilde{v}_2<v_{-2}$). Since $\{v=\tilde{v}_1\}\cap\mathcal{R}^*$ contains an apparent horizon sphere, there exists $u_*\in[u_3, u_b)$ such that $s(u_*, \tilde{v}_1)=0$, and hence, by \eqref{eq:partial v s},  $s(u_*, v)\leq0$ for all $v>\tilde{v}_1$. If $(u_*, \tilde{v}_2) \in \mathcal{Q}$, then $s(u_*, \tilde{v}_2) \le 0$; since $s(u_3, \tilde{v}_2) > 0$, continuity of $s(\cdot, \tilde{v}_2)$ gives $u_{**} \in (u_3, u_*]$ with $s(u_{**}, \tilde{v}_2) = 0$, a contradiction. If instead $(u_*, \tilde{v}_2) \notin \mathcal{Q}$, then, by Corollary~\ref{cor:no hor points extremal} and the extension principle, the cone $\{v = \tilde{v}_2\}\cap\mathcal{R}^*$ must terminate at the $r = 0$ singularity at some $u_{\mathrm{max}} \in (u_3, u_*]$. Having no apparent horizon spheres on $\{v = \tilde{v}_2\} \cap \mathcal{R}^*$ forces $s(u, \tilde{v}_2) > 0$ for all $u \in [u_3, u_{\mathrm{max}})$ and hence $\partial_v r > 0$ on $[u_3, u_{\mathrm{max}}) \times [\tilde{v}_1, \tilde{v}_2]$, which implies $r(u, \tilde{v}_2) > r(u, \tilde{v}_1) \ge r(u_*, \tilde{v}_1) > 0$ for all $u\in[u_3, u_{\mathrm{max}})$, contradicting $r(u, \tilde{v}_2) \to 0$ as $u \to u_{\mathrm{max}}^-$. 
        
        Let then $V$ be the set
        \begin{equation}
            V\doteq\big\{\tilde{v}\in[v_b,v_{-1}]: \{v=\tilde{v}\}\cap\mathcal{R}^* \text{ contains a sphere belonging to } \mathcal{A}_l\big\},
        \end{equation}
        and let $v_{\mathrm{inf}}\doteq\inf V$. In view of the above discussion, $V$ is an interval and $v_{\mathrm{inf}}\leq v_{-2}$. To see that $v_{\mathrm{inf}}< v_{-2}$, it suffices to note that there exist horizon spheres with $v<v_{-2}$: given $\tilde{u}\in(u_3, u_b)$ sufficiently close to $u_3$ so that $(\tilde{u},v_{-2})\in \mathcal{Q}$, we have $s(\tilde{u},v_{-2})<0$, whereas $s(\tilde{u},v)>0$ for $v<v_b$, so continuity of $s(\tilde{u},\cdot)$ gives $v_*\in[v_b,v_{-2})$ such that $s(\tilde{u},v_*)=0$. 
        
        In order to see that $v_{\mathrm{inf}}\notin V$, we proceed by contradiction once more: assume there exists $u_*\in[u_3,u_b)$ such that $s(u_*,v_{\mathrm{inf}})=0$ (then necessarily $u_*> u_3$, since $s(u_3,v_{\mathrm{inf}})>0$). Since $s$ is strictly decreasing along constant-$v$ cones, $s(u,v_{\mathrm{inf}})<0$ for all $u>u_*$. Now fix $u'\in(u_*,u_b)$. By continuity of $s$, we have $s(u',\tilde{v})<0$ for $\tilde{v}<v_{\mathrm{inf}}$ close enough to $v_{\mathrm{inf}}$. Given that $s(u_3, \tilde{v})>0$, there must exist $u_{**}\in(u_3,u']$ such that $s(u_{**},\tilde{v})=0$, which contradicts the fact that $\{v=\tilde{v}\}\cap\mathcal{R}^*$ contains no apparent horizon spheres.
        
        Therefore, we have established that there exists $v_{\mathrm{inf}}\in [v_b, v_{-2})$ such that any constant-$v$ cone with $v\in (v_{\mathrm{inf}},v_{-1}]$ contains exactly one apparent horizon sphere, with uniqueness following from the strict monotonicity of $s(\cdot,v)$. For $\tilde{v}\leq v_{\mathrm{inf}}$, we have shown that $\{v=\tilde{v}\}\cap\mathcal{R}^*$ contains no apparent horizon spheres. Thus $s(\cdot,\tilde{v})$ is nowhere zero on $[u_3,u_b)$, and since $s(u_3,\tilde{v})>0$ we conclude that $s(u,\tilde{v})>0$ for all $(u,\tilde{v})\in\{v\leq v_{\mathrm{inf}}\}\cap\mathcal{R}^*$. This completes the proof of claim \textit{(a)}. 

        Regarding claim \textit{(b)}, we start by considering $\tilde{u} \in (u_3, u_b)$. We claim that there exists $v' \in (v_b, v_{-2}]$ such that $(\tilde{u}, v') \in \mathcal{Q}$ and $s(\tilde{u}, v') < 0$. Indeed, if $(\tilde{u}, v_{-2}) \in \mathcal{Q}$, then $s(\tilde{u}, v_{-2}) < 0$ by Lemma~\ref{lem:connectedness trapped regions}. If instead $(\tilde{u}, v_{-2}) \notin \mathcal{Q}$, then, by Corollary~\ref{cor:no hor points extremal} and the generalised extension principle, the cone $\{u = \tilde{u}\}\cap\mathcal{R}^*$ must terminate at the $r = 0$ singularity at some $v_{\mathrm{max}} \le v_{-2}$; because $r(\tilde{u}, \cdot)$ decreases from a positive value at $v_b$ to $0$ as $v \to v_{\mathrm{max}}^-$, there must exist $v' \in (v_b, v_{\mathrm{max}})$ with $\partial_v r(\tilde{u}, v') < 0$ (and hence $s(\tilde{u}, v') < 0$). On the other hand, $s(\tilde{u}, v)>0$ for $v< v_b$. Then there exists $v_*\in [v_b,v_{-2})$ such that $s(\tilde{u},v_*)=0$. This $v_*$ may not be unique but, if there are more values of $v$ for which $s(\tilde{u},v)=0$, they must form an interval  $I\subset[v_b,v_{-2})$ in view of \eqref{eq:partial v s}. Moreover, as a consequence of claim \textit{(a)}, it must be $I\subset(v_{\mathrm{inf}}, v_{-2})$. This establishes the claim for all $\tilde{u}\in(u_3,u_b)$.  
        
        For $u=u_3$, by construction $s(u_3, v)=0$ for all $v\in [v_{-2},v_{-1}]\subset (v_{\mathrm{inf}}, v_{-1}]$ and $s(u_3, v)>0$ for $v<v_{-2}$. Hence claim \textit{(b)} holds for all $u\in[u_3,u_b)$.
    \end{proof}

    Regarding the causal character of the portion of $\mathcal{A}_l$ lying in $\widetilde{\mathcal{R}}^2_l$, the same reasoning as in Lemma~\ref{lem:causal character apparent horizon} gives:  

    \begin{lemma}
        The set $\mathcal{A}_l\cap \widetilde{\mathcal{R}}^2_l$ (or equivalently $\mathcal{A}_l\cap \mathcal{R}^*$) is a $C^k$ hypersurface whose normal vector field is everywhere causal, and hence it is spacelike or null.
        \label{lem:causal character apparent horizon 2}
    \end{lemma}

    \begin{proof}[Proof of Lemma \ref{lem:causal character apparent horizon 2}]
        We have $\mathcal{A}_l\cap\widetilde{\mathcal{R}}^2_l=\big(s|_{\widetilde{\mathcal{R}}^2_l}\big)^{-1}(\{0\})$, where $0$ is a regular value of the restriction $s|_{\widetilde{\mathcal{R}}^2_l}$, given that, in view of the discussion above,
        \begin{equation}
            \partial_us\vert_{\mathcal{A}_l\cap\widetilde{\mathcal{R}}^2_l}<0,
        \end{equation}
        and moreover 
        \begin{equation}
            \partial_vs\vert_{\mathcal{A}_l\cap\widetilde{\mathcal{R}}^2_l}=\partial_vs\vert_{\mathcal{A}_l\cap\mathcal{R}^*}=-2r^2(\partial_v\phi)^2\big\vert_{\mathcal{A}_l\cap\mathcal{R}^*}
            \begin{cases} 
                <0 &\, \text{ if } \, \partial_v\phi\neq 0, \\
                =0 & \, \text{ if } \, \partial_v\phi= 0.
            \end{cases}
        \end{equation} 
        The norm of the normal $N$ to $\mathcal{A}_l\cap\widetilde{\mathcal{R}}^2_l$, will then satisfy
        \begin{equation}
            g(N,N)\big \vert_{\mathcal{A}_l\cap\widetilde{\mathcal{R}}^2_l} = -\frac{4}{\Omega^2}\partial_vs\partial_us \Big \vert_{\mathcal{A}_l\cap\widetilde{\mathcal{R}}^2_l}=-\frac{4}{\Omega^2}\partial_vs\partial_us \Big \vert_{\mathcal{A}_l\cap\mathcal{R}^*}
            \begin{cases} 
                <0 &\, \text{ and }N \text{ is timelike} \text{ if } \, \partial_v\phi\neq 0, \\
                =0 & \, \text{ and }N \text{ is null} \text{ if } \, \partial_v\phi= 0,
            \end{cases}
        \end{equation}
        which concludes the proof.
    \end{proof}
    
    We thus conclude that $\mathcal{A}_l\cap\widetilde{\mathcal{R}}^2_l$ is an achronal curve in the (1+1)-dimensional reduction of the spherically symmetric spacetime. In order to see that it is connected, we use the same strategy as before and introduce the map $\overline{u}_{\mathcal{A}}: (v_{\mathrm{inf}}, v_{-1}]\rightarrow[u_3, u_b)$, with $s\big(\overline{u}_{\mathcal{A}}(v),v\big)=0$, for every $v\in(v_{\mathrm{inf}}, v_{-1}]$. This is a well-defined function as a consequence of part \textit{(a)} in Lemma \ref{lem:spheres on the apparent horizon 2} and it is surjective by part \textit{(b)} of the same lemma. The graph of $\overline{u}_{\mathcal{A}}$ is precisely $\mathcal{A}_{l}\cap\widetilde{\mathcal{R}}^2_l$. 

    Using the causal character of $\mathcal{A}_{l}\cap\widetilde{\mathcal{R}}^2_l$ given by Lemma \ref{lem:causal character apparent horizon 2} and the fact that $\overline{u}_{\mathcal{A}}$ is differentiable (by the implicit function theorem, since $\partial_us<0$ along $\mathcal{A}_l\cap\widetilde{\mathcal{R}}^2_l$), we proceed as in Lemma \ref{lem:decreasing} to conclude that $\overline{u}_{\mathcal{A}}$ is decreasing. Since $\overline{u}_{\mathcal{A}}:(v_{\mathrm{inf}}, v_{-1}]\rightarrow[u_3, u_b)$ also takes all values in $[u_3, u_b)$, it must be continuous, and hence $\mathcal{A}_l\cap\widetilde{\mathcal{R}}^2_l$ is connected.

    Note that $\mathcal{A}_l$ is an achronal curve that has a limit point either at the boundary point $b_{\Gamma_l}$ or on the null component emanating from it. If this null component is empty, the limit point of $\mathcal{A}_l$ must necessarily be $b_{\Gamma_l}$. 
\end{proof}

\subsection{Dynamical formation of the extremal Schwarzschild--de Sitter white hole and of Nariai}
\label{subsec:grav collapse WH and Nariai}

\begin{customthm}{1.2}
    For any $k\in \mathbb N$ and $\mathfrak{m}>\sqrt{\Lambda}/2$, there exist $\mathfrak{c}, \mathfrak{c}' \in(0,1)$ and spherically symmetric Cauchy data for the Einstein-scalar field system with scalar field mass $\mathfrak{m}$, $\big(\Sigma, g_0, k_0, \phi_0, \phi_1\big)$, with $\Sigma \cong \mathbb S^3$ and two regular centres, such that the future maximal globally hyperbolic development $\big(\mathcal{M},g,\phi\big)$ satisfies: 
    \begin{itemize}
        \item All dynamical quantities are at least $C^k$-regular.
        \item The spacetime contains:
        \begin{enumerate}[(i)]
            \item A region $$\check{\mathcal{R}}\doteq \{u_1 \leq u \leq u_2\} \cap \{v_1 \leq v \leq v_2\}$$ for some $u_1<u_2$, $v_1<v_2$, that is isometric to a double null rectangle in the extremal Schwarzschild--de Sitter white hole spacetime. In particular, the truncated null cones
            \begin{equation}
                \begin{aligned}
                    \overline{H}_{l}^+ &\doteq \{u=u_2\}\cap \check{\mathcal{R}},\\
                    \overline{H}_{\mathrm{r}}^+ &\doteq \{v=v_2\}\cap \check{\mathcal{R}},
                \end{aligned}
            \end{equation} 
            can be identified with appropriate portions of the constant-$u$ and constant-$v$ components, respectively, of the cosmological horizon $\overline{\mathcal{H}}^+$ of the extremal Schwarzschild--de Sitter white hole spacetime.
            \item A region
            \begin{equation}
                \mathcal{R}\doteq \{u_2 \leq u \leq u_3\} \cap \{{v_{-2} \leq v \leq v_{-1}}\},
            \end{equation}
            for some $u_3$ with $u_3>u_2$, and $v_{-1}$, $v_{-2}$ with $v_1>v_{-1}>v_{-2}$, that is isometric to a double null rectangle in the static region $\mathcal{S}$ of the Schwarzschild--de Sitter spacetime of extremality index $\mathfrak{c}$. In particular, the truncated null cone $\{u=u_3\}\cap \mathcal{R}$ can be identified with an appropriate portion of the constant-$u$ component of the subextremal black hole horizon.
            \item A region 
            \begin{equation}
                \mathcal{R}'\doteq \{u_{-2} \leq u \leq u_{-1}\} \cap \{v_2 \leq v \leq v_3\},
            \end{equation}
            for some $u_{-1}$, $u_{-2}$ with $u_1>u_{-1}>u_{-2}$, and $v_3$ with $v_3>v_2$, that is isometric to a double null rectangle in the static region $\mathcal{S}'$ of the Schwarzschild--de Sitter spacetime of extremality index $\mathfrak{c}'$. In particular, the truncated null cone $\{v=v_3\}\cap \mathcal{R}'$ can be identified with an appropriate portion of the constant-$v$ component of the subextremal black hole horizon. 
        \end{enumerate} 
        \item The hypersurface $\Sigma$ does not intersect the null cones $\{u=u_2\}\cap\mathcal{M}$ and $\{v=v_2\}\cap\mathcal{M}$.
    \end{itemize} 
    Moreover, if $\mathfrak{m}=0$, no such Cauchy data exist.
    \label{thm:grav collapse to ext SdS WH}
\end{customthm}

\begin{figure}[H]
        \centering
        \includegraphics[width=0.80\textwidth]{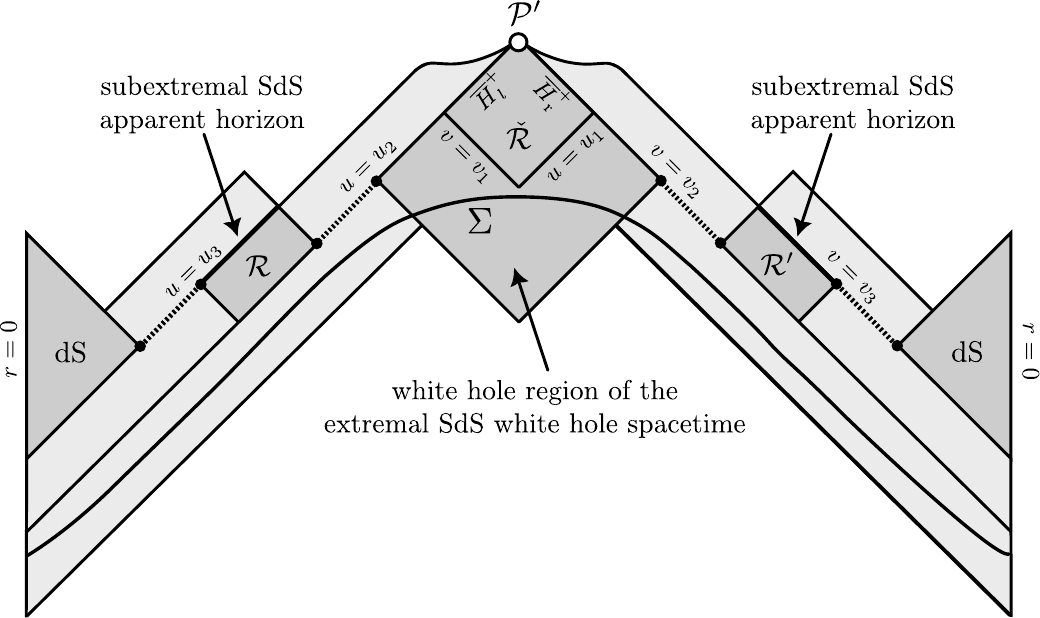} 
        \caption{Penrose diagram depicting the proof of Theorem \ref{thm:grav collapse to ext SdS WH}.}
        \label{fig:grav collapse to ext SdS WH}
\end{figure}

\vspace{1em}

\begin{proof}
    We will carry out this construction in a slightly different way, given that when applying Theorem \ref{thm:subext SdS to ext COSMhorizon}, we cannot freely choose the extremality index of the subextremal Schwarzschild--de Sitter static sphere we glue to the extremal cosmological horizon sphere. 

    \vspace{0.5em}
    
    Therefore, we start considering a double null rectangle in the extremal Schwarzschild--de Sitter white hole solution extending up to the boundary point $\mathcal{P}'$, and hence bounded by portions of the extremal cosmological horizon, which will be of the form  $\{u_0 \leq u \leq u_2\} \cap \{v_0 \leq v \leq v_2\}$ for some $u_0<u_2$, $v_0<v_2$. Then, applying Theorem \ref{thm:subext SdS to ext COSMhorizon}, we can find an extremality index $0<\mathfrak{c}<1$ and glue to order $k$, along the cone $\{u=u_2\}$, a sphere on the extremal cosmological horizon to a sphere in the static region $\mathcal{S}$ of the subextremal Schwarzschild--de Sitter spacetime of extremality index $\mathfrak{c}$. We can now attach to that sphere a double null rectangle extending up to and including a portion of the hypersurface $r=r_{\mathcal{H}}(\mathfrak{c})$ in the subextremal spacetime, and apply Proposition \ref{prop:spacetime gluing} to get a $C^k$ solution in a neighbourhood of the cone $\{u=u_2\}$. 

    \vspace{0.5em}

    Next, we apply Theorem \ref{thm:subext SdS to BHhorizon} and Proposition \ref{prop:spacetime gluing} again to glue to order $k+1$ this double null rectangle in the subextremal Schwarzschild--de Sitter spacetime to a portion of the static region $\mathcal{S}$ of de Sitter spacetime, exactly as in the proof of Theorem \ref{thm:grav collapse to subext SdS}. 

    \vspace{0.5em}

    Performing the analogous construction along constant-$v$ cones, starting from the cone $\{v=v_2\}$, yields the $C^k$ spacetime depicted in Figure \ref{fig:grav collapse to ext SdS WH}. As before, we now identify an appropriate Cauchy hypersurface $\Sigma$, which can be chosen so that it does not intersect the cones $\{u=u_2\}$ and $\{v=v_2\}$. The Cauchy data induced by the constructed spacetime on $\Sigma$, $\big(\Sigma, g_0, k_0, \phi_0, \phi_1\big)$ will satisfy that its future maximal globally hyperbolic development includes a region 

    \begin{equation}
        \check{\mathcal{R}}= \{u_1 \leq u \leq u_2\} \cap \{v_1 \leq v \leq v_2\}
    \end{equation}

    \noindent for $u_1$ with $u_0<u_1<u_2$ and sufficiently close to $u_2$, and $v_1$ with $v_0<v_1<v_2$ and sufficiently close to $v_2$, lying to the future of $\Sigma$ and satisfying $(i)$, and two double null rectangles which are isometric to portions of the subextremal Schwarzschild--de Sitter spacetimes and satisfy the claims in parts $(ii)$ and $(iii)$. Finally, the claim that no such Cauchy data exist when the scalar field is massless follows from Theorem~\ref{thm:impossible gluing}.
\end{proof}

\begin{figure}[H]
        \centering
        \includegraphics[width=0.80\textwidth]{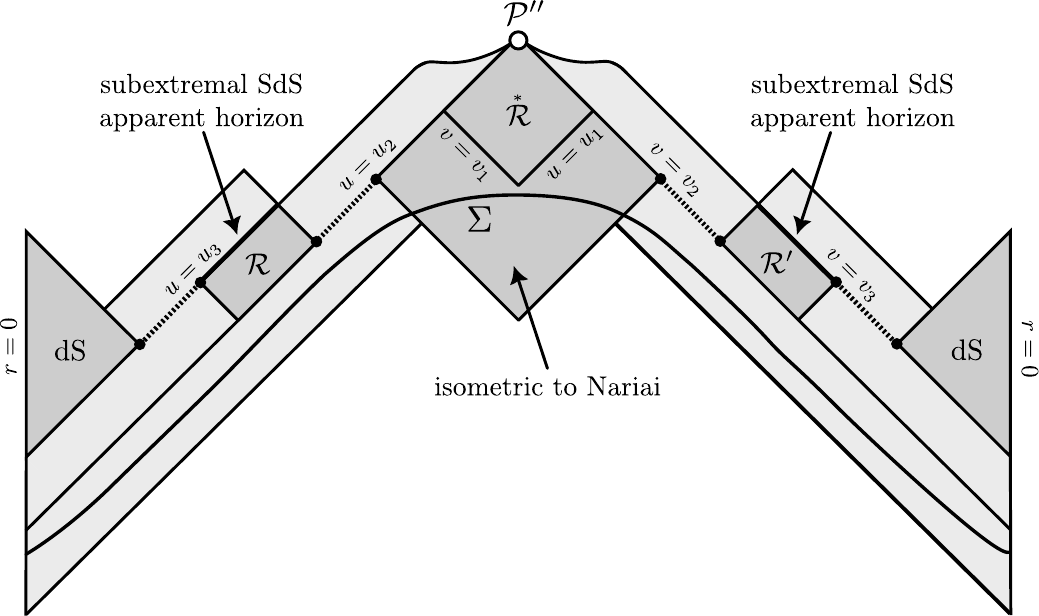}
        \caption{Penrose diagram illustrating the construction in Theorem \ref{thm:grav collapse to Nariai}.} 
        \label{fig:grav collapse to Nariai}
\end{figure}

Following the same procedure, but combining Proposition \ref{prop:spacetime gluing} with Theorem \ref{thm:subext SdS to Nariai} instead of with Theorem \ref{thm:subext SdS to ext COSMhorizon}, we can prove the analogous result for Nariai (see Figure \ref{fig:grav collapse to Nariai} for the diagram of the construction):

\begin{customthm}{1.3}
    For any $k\in \mathbb N$ and $\mathfrak{m}>\sqrt{\Lambda}/2$, there exist $\mathfrak{c}, \mathfrak{c}' \in(0,1)$ and spherically symmetric Cauchy data for the Einstein-scalar field system with scalar field mass $\mathfrak{m}$, $\big(\Sigma, g_0, k_0, \phi_0, \phi_1\big)$, with $\Sigma \cong \mathbb S^3$ and two regular centres, such that the future maximal globally hyperbolic development $\big(\mathcal{M},g,\phi\big)$ satisfies: 
    \begin{itemize}
        \item All dynamical quantities are at least $C^k$-regular.
        \item The spacetime contains:
        \begin{enumerate}[(i)]
            \item A region $\overset{_*}{\mathcal{R}}\doteq \{u_1 \leq u \leq u_2\} \cap \{v_1 \leq v \leq v_2\}$ that is isometric to a double null rectangle in the Nariai spacetime.
            \item A region
            \begin{equation}
                \mathcal{R}\doteq \{u_2 \leq u \leq u_3\} \cap \{{v_{-2} \leq v \leq v_{-1}}\},
            \end{equation}
            for some $u_3$ with $u_3>u_2$, and $v_{-1}$, $v_{-2}$ with $v_1>v_{-1}>v_{-2}$, that is isometric to a double null rectangle in the static region $\mathcal{S}$ of the Schwarzschild--de Sitter spacetime of extremality index $\mathfrak{c}$. In particular, the truncated null cone $\{u=u_3\}\cap \mathcal{R}$ can be identified with an appropriate portion of the constant-$u$ component of the subextremal black hole horizon.
            \item A region 
            \begin{equation}
                \mathcal{R}'\doteq \{u_{-2} \leq u \leq u_{-1}\} \cap \{v_2 \leq v \leq v_3\},
            \end{equation}
            for some $u_{-1}$, $u_{-2}$ with $u_1>u_{-1}>u_{-2}$, and $v_3$ with $v_3>v_2$, that is isometric to a double null rectangle in the static region $\mathcal{S}'$ of the Schwarzschild--de Sitter spacetime of extremality index $\mathfrak{c}'$. In particular, the truncated null cone $\{v=v_3\}\cap \mathcal{R}'$ can be identified with an appropriate portion of the constant-$v$ component of the subextremal black hole horizon. 
        \end{enumerate} 
        \item The hypersurface $\Sigma$ does not intersect the null cones $\{u=u_2\}\cap\mathcal{M}$ and $\{v=v_2\}\cap\mathcal{M}$.
    \end{itemize} 
    Moreover, if $\mathfrak{m}=0$, no such Cauchy data exist.
    \label{thm:grav collapse to Nariai}
\end{customthm}

\vspace{2em}

\newpage

\text{ }

\vspace{-4.5em}

\printbibliography[heading=bibintoc]

\end{document}